\documentclass[aps,prl,twocolumn,preprintnumbers,superscriptaddress,nofootinbib]{revtex4}
\usepackage[utf8]{inputenc}
\usepackage{graphicx}
\usepackage{psfrag}
\usepackage{color}
\usepackage{amssymb}
\usepackage{amsmath}
\usepackage{braket}
\usepackage{epstopdf}
\begin{document}
\title{Study of kaon structure using the light-cone quark model}
\author{Satvir Kaur and Harleen Dahiya}
\affiliation{Dr. B. R. Ambedkar National Institute of Technology,\\ Jalandhar, 144011, India}
\begin{abstract}
We investigate the various distributions explaining multi-dimensional structure of kaon at the level of its constituents ($u$ and $\bar{s}$) using the light-cone quark model. The overlap form of wavefunctions associated with the light-cone quark model is adopted for the calculations. The generalized parton distributions(GPDs)of $u$ and $\bar{s}$ quarks are presented for the case when the momentum transfer in the longitudinal direction is non-zero. The dependence of kaon GPDs is studied in terms of variation of  quark longitudinal momentum fraction,  momentum transfer in longitudinal direction and total momentum transfer to the final state of hadron. The transverse impact-parameter dependent GPDs are also studied by taking the Fourier transformation of general GPDs. Further, the quantum phase-space distributions; Wigner distributions are studied for the case of unpolarized, longitudinally-polarized and transversely-polarized parton in an unpolarized kaon. The Wigner distributions are analysed in the transverse impact-parameter plane, the transverse momentum plane and the mixed plane.  Further, to get a complete picture of kaon in terms of its valence quarks, the variation of longitudinal momentum fraction carried by quark and antiquark in the generalized transverse momentum-dependent parton distributions (GTMDs) is studied for different values of transverse quark and antiquark momentum $({\bf k}_\perp)$ as well as for different values of momentum transferred to the kaon in transverse direction $({\bf \Delta}_\perp)$. This has been done for zero as well as non-zero skewedness representing respectively the absence and presence of momentum transfer to the final state of kaon in longitudinal direction. Furthermore, the possible spin-orbit correlation for $u$ and $\bar{s}$ in kaon is elaborated in context of Wigner distributions and GTMDs.
\end{abstract}
\maketitle
\section{I. Introduction} 
Quantum Chromodynamics (QCD) which is a part of Standard Model (SM) describes the formation of hadron by including the strong interrelation between quarks, antiquarks and gluons. The hadron structure cannot be  derived directly from the structure functions and the non-perturbative effects of QCD are the key to understand the complex internal structure of hadron. This can be attempted through probing where, by knowing the nature of scattering reaction, one can extract the detailed information about the structure. The probe chosen for the interaction with the hadron is a point-like particle such as lepton. The parton distribution functions (PDFs) \cite{pdf1, pdf2} and the form factors (FFs) \cite{ff1, ff2, ff3, ff4} are the basic ingredients to understand the internal hadron structure. The extended information on the internal structure can be explained through the generalized parton distributions (GPDs) \cite{gpd1,gpd2,gpd3} and the transverse momentum-dependent parton distributions (TMDs) \cite{tmd1, tmd2}: the techniques to understand the three-dimensional picture of hadron. The GPDs depend on three variables namely, (a) quark longitudinal momentum fraction $x$, (b) momentum transfer in longitudinal direction $\zeta$ and (c) total momentum transfer to the final state of hadron $t=\Delta^2$. In general, GPDs are responsible in unravelling the longitudinal and transverse distribution of partons inside the hadron. 
The Fourier transformation of GPDs provide us the impact-parameter dependent GPDs carrying the information in terms of transverse distance from the center of hadron. The impact-parameter dependent GPDs (IPDGPD) in the presence of transverse momentum in longitudinal direction have been effectively discussed \cite{ipdgpd1}. In the absence of longitudinal momentum transfer i.e. $\zeta=0$, the impact-parameter dependent GPDs$(x,{\bf b}_\perp)$ end up with a probabilistic interpretation of parton's density in the fast moving hadron, where ${\bf b}_\perp$ is the transverse impact-parameter distance \cite{ipdgpd2}. 
Further, since the GPDs are unable to explain the parton distributions in terms of transverse momentum carried by valence parton, the TMDs can be defined as they provide the information of parton distributions in terms of parton's transverse momentum $({\bf k}_\perp)$. 

FFs enter into the elastic processes to explain the non-perturbative dynamics of partons in terms of photon virtuality. However, the possibility of occurrence of inelastic scattering processes become more as compared to elastic scattering at high photon virtuality. The deep inelastic scattering (DIS) \cite{dis} is helpful in the interpretation of PDFs which explicate the probability of locating the parton carrying the longitudinal momentum fraction $x$ inside the hadron. The GPDs explain through various exclusive processes, 
the presence of the recoil momentum ${\bf \Delta}_\perp$ as the hadron has different initial and final state. Such processes are named as deeply virtual Compton scattering (DVCS) \cite{dvcs1, dvcs2, dvcs3, dvcs4, dvcs5,dvcs6} and hard exclusive meson production (HEMP) \cite{hemp1, hemp2, hemp3}. If the final state has the unpolarized meson with zero spin, one needs to measure the longitudinal cross-section \cite{gpd3}. Further, the TMDs which are compatible in explaining the distribution of partons in transverse momentum space are accessible through semi-inclusive deep inelastic scattering (SIDIS) and Drell-Yan (DY) processes \cite{sidis1, sidis2, sidis3, sidis4, sidis5}. The three-dimensional GPDs are the unification of FFs and PDFs for $\zeta=0$. The $x$-integration of the GPDs leads to the FFs which conceal the location of partons in transverse direction. Whereas in the forward limit, ${\bf \Delta}_\perp=0$, GPDs produce PDFs. 

The quantum phase-space distributions i.e. Wigner distributions \cite{wigner1, wigner2} are explained as the position and momentum space distributions of partons in fast moving hadron. Being limited to the five-dimensional quasi-probabilistic, Wigner distributions interpret in terms of probabilistic quantities i.e. impact-parameter dependent distributions (IPDs) and TMDs. Furthermore, the Fourier transformation of the Wigner distributions are related to generalized transverse momentum-dependent distributions of partons (GTMDs) \cite{gtmd1, gtmd2} for $\zeta=0$. Generally, GTMDs are named as \textit{mother distributions} because GPDs and TMDs can be derived by applying certain integrations and limits. 
The quark GTMDs in a hadron can be measured through the exclusive double Drell-Yan process where two photons are observed in the final state  along with the hadron. The quark GTMDs have been extracted for the case of nucleon by considering all  possible  helicities of hadrons and photons \cite{double-DY}. 
The quark GTMDs can be extracted from the  scattering amplitude for the process where the dominating contribution is taken for the transversely polarized photons. The measured quark GTMDs  lie in ERBL region i.e. $-\zeta<x<\zeta$. The Wigner distributions and GTMDs are also important in calculating the spin-orbital angular momentum (spin-OAM) and spin-spin correlations. In addition to this, quark orbital angular momentum is related to the phase-space average of the Wigner distributions.

The probabilistic distributions i.e. GPDs and TMDs have been studied widely in sense of theories and experiments. The various models which remain successful in explaining the GPDs, both for nucleon and pion are the MIT bag model \cite{gpdmodel1}, the constituent quark model with non-relativistic approach \cite{gpdmodel2, gpdmodel3} as well as with relativistic approach \cite{gpdmodel4, gpdmodel5}, meson cloud model \cite{gpdmodel6}, light-front quark-diquark model \cite{gpdmodel7}, AdS/ QCD inspired light-cone model \cite{gpdmodel8, gpdmodel9} using Bethe-Salpeter approach \cite{gpdmodel10, gpdmodel11}. The GPDs in transverse impact-parameter space have also been studied widely \cite{ipdmodel1, ipdmodel2, ipdmodel3, ipdmodel4, ipdmodel5}. Recently, the unified Wigner distributions have been evaluated extensively using various models by considering different polarization configurations of quarks and gluons in the case of spin$-\frac{1}{2}$ hadron \cite{wdmodel1, wdmodel2, wdmodel3, wdmodel4, wdmodel5, wdmodel6, wdmodel7}. Further, using these distributions, the spin-spin and spin-OAM correlations have also been studied \cite{spinoam1, spinoam2}. Recently, for pion, the Wigner distributions have been successfully investigated \cite{wdmodel8}. Alongwith the Wigner distributions, the GTMDs have also been studied for the spin$-\frac{1}{2}$ as well as spin$-0$ hadrons.

In light of the studies done for the case of pion, we move a step forward to study the internal structure of kaon in context of its valence partons by adopting the motivation towards the spin-0 hadron structure \cite{gtmd1,thesis}. It is important to mention here that even though kaon is a spin$-0$ hadron, it is different from pion  in terms of its constituents that have unequal quark masses (one is light $u$ and other is heavy quark $\bar s$). Further, the dynamics of valence partons in a spin-$0$ meson system are easier to determine as compared to the spin$-\frac{1}{2}$ baryon system as the mesons are composed of a quark-antiquark pair.  However, the experiments are more focused on the internal structure of the lowest lying pion and nucleon and there is no experimental data available for the case of kaon. However, the kaon sea quark distributions can induced by including one charged kaon on an isoscalar target \cite{kaon-DY}. The kaon-nucleus Drell-Yan process demands  several combinations of valence and sea quarks. The combination of differently charged kaons incident on isoscalar-target permits one to evaluate valence-valence interactions separately. In other words, the linear combinations of cross-sections induced by $K^+$ and $K^-$ lead to valence-valence term while the interaction of $K^+$ with deuteron give the sea-valence and sea-sea interaction terms.

One of the important model which can be used to  investigate the kaon at the level of its constituents is the light-cone QCD inspired model. The light-cone framework \cite{lc1} provides a suitable environment for the description of hadron's internal structure when it moves relativistically \cite{lc2, lc3}. The light-cone inspired quark model has been applied to successfully calculate the  electromagnetic form factors and compare them  with the experimental data \cite{kaonmodel}. The mesonic light-cone Fock state wavefunctions are expanded as $| M \rangle=\sum |q \bar{q}\rangle \psi_{q \bar{q}}+ \sum |q \bar{q}g\rangle \psi_{q \bar{q}g} + ...$, where we choose the minimal Fock state description i.e. quark-antiquark state, because we study the leading-twist distributions \cite{kaonmodel1}.

In this work, we study the various distributions of $u$ quark as well as $\bar{s}$ quark in kaon using the light-cone quark model (LCQM). We have used the overlap representation of the light-cone wavefunctions.  While evaluating the $u$ quark distribution, the other quark, i.e. $\bar{s}$ quark is considered as a spectator and vice versa. First of all, we discuss the generalized distributions of quark and antiquark in kaon for $\zeta \neq 0$. We also discuss the GPDs in transverse position space, i.e. transverse impact-parameter GPDs for the case when momentum transfer in longitudinal direction is non zero $\zeta\neq 0$. We also study the case when the momentum transfer in longitudinal direction is zero $\zeta=0$. We discuss the Wigner distributions with the quark and antiquark having different polarizations in the kaon. We take the case where kaon however remains unpolarized throughout the calculations. We further explain the kaon GTMDs for the case of $\zeta=0$ as well as $\zeta\neq 0$. In context of the Wigner distributions and GTMDs, the spin-orbit correlation of $u$ quark and $\bar s$ quark in kaon has also been studied. 

The paper is arranged as follows. We provide the detailed description about the general framework of light-cone quark model and the associated wavefunctions with it in Section II. In Section III, generalized distributions of quark and antiquark in kaon are presented for $\zeta \neq 0$ describing the three-dimensional picture of kaon. Further, the impact-parameter dependent GPDs for non-zero skewedness are evaluated by taking the Fourier transformation from the momentum transferred to impact-parameter distance in Section-IV. We present the definitions of Wigner distributions in terms of polarization configurations and  the calculations for Wigner distributions, the phase-space distribution describing five-dimensional picture of kaon for both $u$ quark and $\bar{s}$ quark in Section V. We present the results in transverse impact-parameter plane, transverse momentum-plane, and in mixed space. In Section VI, GTMDs, the mother distributions of valence partons in kaon are discussed for $\zeta=0$ as well as $\zeta \neq 0$. The possible spin-orbit correlation for $u$ and $\bar{s}$ quarks in kaon are calculated, plotted and discussed in Section VII. At the end, we summarize the results in Section VIII. 
\section{II. Light-cone quark model}
\subsection{General framework}
If the light-cone momentum fractions and relative momentum co-ordinates of the hadronic constituents are denoted by $x_i=k^+_i/P^+$ and ${\bf k}_{\perp i}$, then the light-cone Fock state expansion of hadronic eigenstate $|M (P^+, \textbf{P}_\perp, S_z)\rangle $ in terms of its constituent eigenstates $|n \rangle$ is defined as \cite{kaonmodel1, dvcs}
\begin{eqnarray}
|M (P^+, \mathbf{P}_\perp, S_z) \rangle
   &=&\sum_{n,\lambda_i}\int\prod_{i=1}^n \frac{\mathrm{d} x_i \mathrm{d}^2
        \mathbf{k}_{\perp i}}{\sqrt{x_i}~16\pi^3}
        16\pi^3 \nonumber\\
       && \delta\Big(1-\sum_{i=1}^n x_i\Big)\delta^{(2)}\Big(\sum_{i=1}^n \mathbf{k}_{\perp i}\Big) | n;x_i P^+\nonumber\\
       && x_i \mathbf{P}_\perp+\mathbf{k}_{\perp i},
        \lambda_i \rangle
        \psi_{n/M}^{\lambda_i}(x_i,\mathbf{k}_{\perp i}), \nonumber\\
        \label{meson_eqn}
\end{eqnarray}
where
\begin{eqnarray}
k_i=\bigg[x_i P^+,\frac{(x_i {\bf P}_\perp+{\bf k}_{\perp i})^2+m^2_i}{x_i P^+},x_i {\bf P}_\perp+ {\bf k}_{\perp i}\bigg].
\end{eqnarray}
Here $m_i$ are the masses of $i$ number of constituents of hadron and $\lambda_i$ is the helicity of $ith$ constituent. In Eq. (\ref{meson_eqn}), $x_i {\bf P}_\perp+{\bf k}_{\perp i}={\bf p}_{\perp i}$ is the physical transverse momenta term and $\psi_{n/M}$ gives the probability amplitudes for finding the on-shell mass constituents of meson \cite{pionlc}. The $n$-particle Fock states $|p_i,{\bf p}_{\perp i} \rangle$ are normalized as follows
\begin{eqnarray}
\langle{n;{p'}_i^+, {\textbf{p}'}_{\perp i},\lambda'_i} | {n;{p}_i^+, {\textbf{p}}_{\perp i},\lambda_i}\rangle &=&\prod_{i=1}^n 16 \pi^3 p_i^+\delta({p'}_i^+ -p_i^+) \nonumber\\
&&
\delta^{(2)}({\textbf{p}'}_{\perp i}-{\textbf{p}}_{\perp i})\delta_{\lambda'_i \lambda_i}.\nonumber\\
\end{eqnarray}
We use the frame where the general four-vector $A=[A^+,A^-,{\bf A}_\perp]$ components are described as 
\[A^\pm=A^0 \pm A^3,~~~~ {\bf A}_\perp=(A^1,A^2)~~~~ \rm{and}~~~~ A^2=A^+ A^- - {\bf A}^2_\perp.\]

We choose a symmetric light-cone frame for the calculations with $\Delta \rightarrow -\Delta$ symmetry. The initial and final four-momenta of meson in symmetric  frame are taken as \cite{kaonmodel1, dvcs}
\begin{eqnarray}
P'&=&\bigg[(1+\zeta)P^+, \frac{M^2+{\bf \Delta}^2_\perp/4}{(1+\zeta)P^+},\frac{{\bf \Delta}_\perp}{2} \bigg],\\
P''&=&\bigg[(1-\zeta)P^+, \frac{M^2+{\bf \Delta}^2_\perp/4}{(1-\zeta)P^+},-\frac{{\bf \Delta}_\perp}{2} \bigg],
\end{eqnarray}
respectively.

The average four-vector momentum of meson $P^\mu=\frac{(P'+P'')^\mu}{2}$ and four-vector momentum transfer from the meson $\Delta^\mu={P'}^\mu-{P''}^\mu$ are given as 
\begin{eqnarray}
P&=&\bigg[P^+, \frac{M^2+{\bf \Delta}^2/4}{(1-\zeta^2)P^+},{\bf 0}_\perp \bigg],\\
\Delta &=& \bigg[2 \zeta P^+, -\frac{\zeta {\bf \Delta}_\perp^2+4 \zeta M^2}{2(1-\zeta^2)P^+}, {\bf \Delta}_\perp\bigg],
\end{eqnarray}
where $\zeta=-\frac{\Delta^+}{2P^+}$ is skewedness and $M$ is defined as the meson mass. 
\subsection{Light-cone wavefunctions for kaon}
The two-particle Fock state expansion in Eq. (\ref{meson_eqn}) for meson $(n=2)$ can be reduced to 
\begin{eqnarray}
|M(P,S)\rangle&=&\sum_{\lambda_1, \lambda_2} \int \frac{dx d^2{\bf k}_\perp}{\sqrt{x(1-x)} 16\pi^3} |x,{\bf k}_\perp,\lambda_1,\lambda_2 \rangle \nonumber\\
&&\psi^{\lambda_1,\lambda_2}_{S_z}(x,{\bf k}_\perp).
\label{meson_eqn_kaon}
\end{eqnarray}
Here $\lambda_1$ and $\lambda_2$ describe the helicities of quark and antiquark in meson respectively. 

Since kaon is a pseudoscalar particle with $S=0$, the light-cone wavefunctions $\psi^{\lambda_1,\lambda_2}_{S_z}(x,{\bf k}_\perp)$ in  Eq. (\ref{meson_eqn_kaon}) can be defined for different combinations of helicities of quark and spectator antiquark in kaon as  \cite{kaonmodel1}
\begin{eqnarray}
\psi^{\uparrow,\uparrow}_0(x,{\bf k}_\perp)&=&-\frac{1}{\sqrt{2}}\frac{k_1-i k_2}{\sqrt{{\bf k}^2_\perp+l^2}}\varphi(x,{\bf k}_\perp),\nonumber\\
\psi^{\uparrow,\downarrow}_0(x,{\bf k}_\perp)&=&\frac{1}{\sqrt{2}}\frac{(1-x)m_1+x m_2}{\sqrt{{\bf k}^2_\perp+l^2}}\varphi(x,{\bf k}_\perp),\nonumber\\
\psi^{\downarrow,\uparrow}_0(x,{\bf k}_\perp)&=&-\frac{1}{\sqrt{2}}\frac{(1-x)m_1+x m_2}{\sqrt{{\bf k}^2_\perp+l^2}}\varphi(x,{\bf k}_\perp),\nonumber\\
\psi^{\downarrow,\downarrow}_0(x,{\bf k}_\perp)&=&-\frac{1}{\sqrt{2}}\frac{k_1+i k_2}{\sqrt{{\bf k}^2_\perp+l^2}}\varphi(x,{\bf k}_\perp), \label{wavefunctions}
\end{eqnarray}
with
\begin{eqnarray}
 l^2=(1-x)m_1^2+x m_2^2-x(1-x)(m_1-m_2)^2.
\end{eqnarray}
The positive (negative) helicity of the quark and the antiquark spectator is denoted by $\uparrow(\downarrow)$. Here, the longitudinal momentum fraction of quark and the quark transverse momentum are denoted by $x$ and ${\bf k}_\perp$ respectively. On the other hand, for antiquark spectator, these terms are described by $(1-x)$ and $-{\bf k}_\perp$ respectively.

 The momentum-space wavefunction $\varphi(x,{\bf k}_\perp)$ in Eq. (\ref{wavefunctions}) is described using the Brodsky-Huang-Lepage method \cite{kaonmodel}. We  have
\begin{eqnarray}
\varphi(x,\textbf{k}_\perp)&=& A \ {\rm exp} \Bigg[-\frac{\frac{\textbf{k}^2_\perp+m_1^2}{x}+\frac{\textbf{k}^2_\perp+m_2^2}{1-x}}{8 \beta^2}\nonumber\\
&-&\frac{(m_1^2-m_2^2)^2}{8 \beta^2 \bigg(\frac{\textbf{k}^2_\perp+m_1^2}{x}+\frac{\textbf{k}^2_\perp+m_2^2}{1-x}\bigg)}\Bigg],
\end{eqnarray}
where the parameters $\beta$ and $A$ are defined as harmonic scale and normalization constant respectively. Here $m_1$ and $m_2$ represent the mass of $u$ quark and $\bar{s}$ quark in kaon respectively. The numerical values of parameters used for the calculations are as follows:\\
$m_1=0.25$ $GeV$, $m_2=0.5$ $GeV$ (with the $u$ quark on-shell), $\beta=0.393$ $GeV$ and $A=74.2$. 
\section{III. Generalized quark and antiquark distributions for kaon (GPDs)}
The GPDs are evaluated using the overlap form of wavefunctions in light-cone quark model (LCQM). To carry out with the distributions, the support interval $-1<x<1$ is divided  into three regions: (i) ERBL region $-\zeta<x<\zeta$ where both quark-antiquark pairs are involved, (ii) DGLAP region $\zeta<x<1$ for the distributions of quark and (iii) DGLAP region  $-1<x<-\zeta$ for the distributions of antiquark. In this work, we have focused on the DGLAP regions for evaluating the distributions of quark and antiquark. In these regions, the conserved number of  particle leads to $n \rightarrow n$ overlaps of diagonal elements.  For the case of kaon we have $n=2$ and since it is a spin$-0$ particle, the number of GPDs are less as compared to the higher spin particles. The associated GPD for kaon, defined via the off-diagonal matrix elements of the bilocal field operator, is expressed as \cite{gpd3}
\begin{widetext}
\begin{eqnarray}
H_K(x,\zeta,t)&=&\frac{1}{2}\int \frac{dz^-}{2 \pi} e^{i x P^+ z^-}\bigg\langle{M(P')}\bigg|\bar{q}\bigg(-\frac{z}{2}\bigg)\gamma^+q\bigg(\frac{z}{2}\bigg)\bigg|{M(P)}\bigg\rangle\Bigg|_{z^+=0,{\bf z}_\perp={\bf 0}_\perp}.
\end{eqnarray}
\end{widetext}

The overlap form of wavefunctions for kaon GPD is achieved by operating the quark field operators on the specific state of the kaon. The $2 \rightarrow 2$ overlaps of wavefunctions for $H(x,\zeta,t)$ are given as
\begin{widetext}
\begin{eqnarray}
H_K(x,\zeta,t)&=&\int \frac{d^2 {\bf k}_\perp}{16\pi^3}\big[\psi^{*\uparrow,\uparrow}_0(x'',k'')\psi^{\uparrow,\uparrow}_0(x',k')+\psi^{* \uparrow,\downarrow}_0(x'',k'')\psi^{\uparrow,\downarrow}_0(x',k')\nonumber\\
&&+\psi^{*\downarrow,\uparrow}_0(x'',k'')\psi^{\downarrow,\uparrow}_0(x',k')+\psi^{* \downarrow,\downarrow}_0(x'',k'')\psi^{\downarrow,\downarrow}_0(x',k')\big],
\end{eqnarray}
\end{widetext}
with the initial  and final   momenta (${\bf k}'_{\perp }$ and ${\bf k}''_{\perp }$) and longitudinal momentum fractions ($x'_1$ and $x''_1$) carried by struck quark in symmetric frame in DGLAP domains $\zeta<x<1$ and $-1<x<-\zeta$ are given as
\begin{eqnarray}
{\bf k}'_{\perp }&=& {\bf k}_{\perp 1}+(1-x'_1)\frac{{\bf \Delta}_\perp}{2}; \ \  x'_1=\frac{x_1+\zeta}{1+\zeta}, \label{active_quark_momenta1}\\
{\rm and\ \ } {\bf k}''_{\perp }&=& {\bf k}_{\perp 1}-(1-x''_1)\frac{{\bf \Delta}_\perp}{2}; \ \  x''_1=\frac{x_1-\zeta}{1-\zeta}.
\label{active_quark_momenta2}
\end{eqnarray}
For the spectator, which is an antiquark here, the initial and final states of momenta and corresponding longitudinal momentum fractions are given as
\begin{eqnarray}
{\bf k}'_{\perp s}&=&{\bf k}_{\perp 2}-{x'_2} \frac{{\bf \Delta}_\perp}{2};\ \ x'_2=\frac{x_2}{1+\zeta},\\
{\rm and \ \ }{\bf k}''_{\perp s}&=&{\bf k}_{\perp 2}+{x''_2} \frac{{\bf \Delta}_\perp}{2};\ \ x''_2=\frac{x_2}{1-\zeta}. 
\label{spcetator_momenta}
\end{eqnarray}
It would be important to mention here that for the struck quark, the momentum and longitudinal momentum fraction are taken to be ${\bf k}_{\perp 1}={\bf k}_\perp$ and $x_1=x$ respectively. However, for the antiquark spectator, the respective parameters are ${\bf k}_{\perp 2}=-{\bf k}_\perp$ and  $x_2=1-x$.

The required conditions which should be satisfied for initial and final states are 
\begin{eqnarray}
\sum_{i=1}^2 x'_i&=&1, {\ \ \ \rm and  \ \ } \sum_{i=1}^2 x''_i=1, \\ \sum_{i=1}^2{\bf k}'_{\perp i}&=&{\bf 0}_\perp, {\ \rm and  \ } \sum_{i=1}^2{\bf k}''_{\perp i}={\bf 0}_\perp.
\label{conditions}
\end{eqnarray}

The antiquark GPDs are defined from quark GPDs as \cite{gpd3}
\begin{eqnarray}
H^{q}(x, \zeta, t, m_1,m_2) = - H^{\bar{q}}(-x, \zeta, t,m_2,m_1).
\label{gpd_flavor_decomposition}
\end{eqnarray}
In the above equation, the masses get reversed in $\bar s$ quark case due to the on-shell mass effect. In other words, when the quark (antiquark) case is taken into account $m_1 (m_2)$ is considered to be on-shell, where $m_1 (m_2)$ denotes the mass of $u$ quark ($\bar{s}$ quark). 
On the other hand, if we consider the antiquark as an active parton and quark as the spectator, the conditions described in Eq. ({\ref{conditions}}) should remain the same, however, the antiquark momentum will now be given as ${\bf k}_{\perp 1}=-{\bf k}_\perp$ and longitudinal momentum fraction by active antiquark as $x_1=-x$. In that case, the quark spectator momentum and longitudinal momentum fraction are taken to be ${\bf k}_{\perp 2}={\bf k}_\perp$ and  $x_2=1+x$ respectively.

In the LCQM, the expressions of GPD $H(x,\zeta,t)$ of $u$ and $\bar s$ quarks for kaon in DGLAP regions are expressed as
\begin{widetext}
\begin{eqnarray}
H^{(u)}&=&\int \frac{d^2 {\bf k}_\perp}{16\pi^3}\bigg[{\bf k}_\perp^2-\frac{(1-x)^2}{1-\zeta^2}\frac{{\bf\Delta}_\perp^2}{4}
-\frac{\zeta(1-x)}{1-\zeta^2}(k_x \Delta_x+k_y \Delta_y)+\mathcal{M}_u'\mathcal{M}_u''\bigg]
 \frac{\varphi_u^{*}(x'',{\bf k}''_\perp)\varphi_u(x', {\bf k}'_\perp)}{\sqrt{{\bf k}_\perp''^2+{l}_u''^2}{\sqrt{{\bf k}_\perp'^2+{l}_u'^2}}},\label{gpd_u}\\
 H^{(\bar{s})}&=&-\int \frac{d^2 {\bf k}_\perp}{16\pi^3}\bigg[{\bf k}_\perp^2-\frac{(1+x)^2}{1-\zeta^2}\frac{{\bf\Delta}_\perp^2}{4}-\frac{\zeta (1+x)}{1-\zeta^2}(k_x \Delta_x+k_y \Delta_y)+\mathcal{M}_{\bar{s}}'\mathcal{M}_{\bar{s}}''\bigg]\frac{\varphi_{\bar{s}}^{*}(x'',{\bf k}''_\perp)\varphi_{\bar{s}}(x', {\bf k}'_\perp)}{\sqrt{{\bf k}_\perp''^2+{l}_{\bar{s}}''^2}{\sqrt{{\bf k}_\perp'^2+{l}_{\bar{s}}'^2}}}.\label{gpd_s}
\end{eqnarray}
Here, we have
\begin{eqnarray}
\mathcal{M}_u'&=&\frac{1-x}{1+\zeta}m_1+\frac{x+\zeta}{1+\zeta} m_2, \ \ \ \ \ \ \
\mathcal{M}_u''=\frac{1-x}{1-\zeta}m_1+\frac{x-\zeta}{1-\zeta} m_2,\\
\mathcal{M}_{\bar{s}}'&=&\frac{1+x}{1+\zeta}m_2+\frac{-x+\zeta}{1+\zeta} m_1,\ \ \ \ \
\mathcal{M}_{\bar{s}}''=\frac{1+x}{1-\zeta}m_2+\frac{-x-\zeta}{1-\zeta} m_1,\\
l'^2_u &=&\frac{1-x}{1+\zeta}m_1^2+\frac{x+\zeta}{1+\zeta}m_2^2-\frac{(1-x)(x+\zeta)}{(1+\zeta)^2}(m_1-m_2)^2,\\  
l''^2_u &=& \frac{1-x}{1-\zeta}m_1^2+\frac{x-\zeta}{1-\zeta}m_2^2-\frac{(1-x)(x-\zeta)}{(1-\zeta)^2}(m_1-m_2)^2,
\\
l'^2_{\bar{s}}&=& \frac{1+x}{1+\zeta}m_2^2+\frac{-x+\zeta}{1+\zeta}m_1^2-\frac{(1+x)(\zeta-x)}{(1+\zeta)^2}(m_2-m_1)^2,
\\
l''^2_{\bar{s}}&=& \frac{1+x}{1-\zeta}m_2^2+\frac{-x-\zeta}{1-\zeta}m_1^2+\frac{(1+x)(x+\zeta)}{(1-\zeta)^2}(m_2-m_1)^2.
\end{eqnarray} 
\end{widetext}

\begin{figure*}
\centering
\begin{minipage}[c]{1\textwidth}
(a)\includegraphics[width=.45\textwidth]{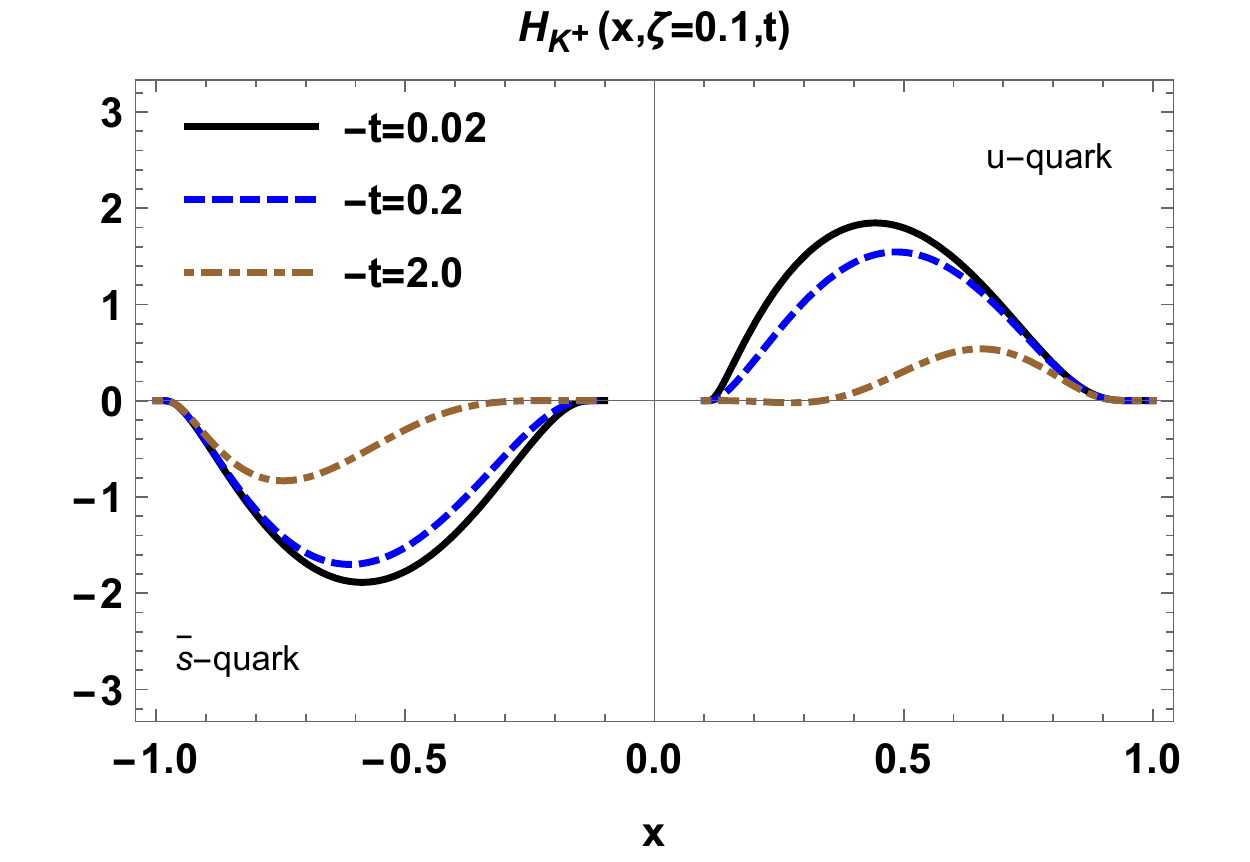}
(b)\includegraphics[width=.45\textwidth]{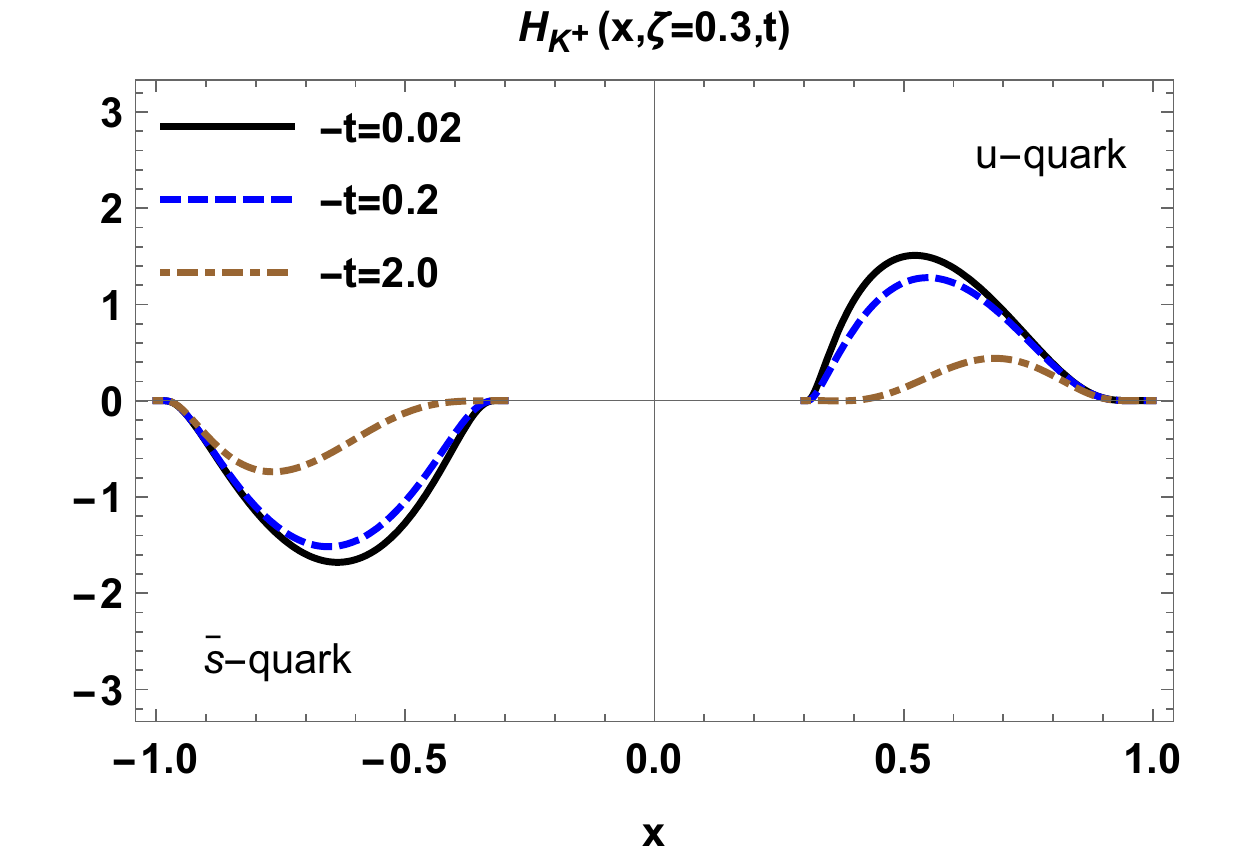}
\end{minipage}
\caption{The variation of unpolarized kaon GPD $H$ for $u$ and $\bar{s}$ quark as a function of $x$ at different values of $-t=0.02, 0.2,2.0$ $GeV^2$ for skewedness (a) $\zeta=0.1$ and (b) $\zeta=0.3$.}
\label{gpd_different_t}
\end{figure*}

There are primarily four parameters used in the calculations: mass of $u$ quark $m_1=0.25$ $GeV$, mass of $\bar s$ $m_2=0.5$ $GeV$ (with the $u$ quark being on-shell),  the harmonic scale constant $\beta=0.393$ $GeV$ and the normalization constant $A=74.2$. Using these parameters, we have calculated the unpolarized GPDs $H(x,\zeta,t)$ of kaon's valence partons corresponding to the $u$ and $\bar s$ quarks for the case when skewedness is non-zero i.e. $\zeta \neq 0$. First of all, we have fixed the value of $\zeta$ as $0.1$ and $0.3$ and have shown the variation of $u$ and $\bar s$ GPDs w.r.t $x$ for different values of momentum transfer $-t$ in Fig. \ref{gpd_different_t}(a) and (b). We observe that the peak of distribution for the $u$ quark  shifts towards higher values of $x$ by when the momentum transfer to the final state $(-t)$ of kaon  increases. The magnitude however ceases down with the increase in the magnitude of momentum transfer. In general, the distribution peaks depend on the momentum transfer to the kaon in the longitudinal direction. Higher the longitudinal momentum transfer, more is the concentration of quark distributions at the higher longitudinal momentum fraction carried by the active quark or antiquark.  Similar effect can be seen for $\bar{s}$ quark case, where the exception lies in the magnitude and polarity. The reason behind the change in magnitude and polarity is the heavier mass of strange quark. The relation between the flavor decomposition of kaon is given in Eq. (\ref{gpd_flavor_decomposition}) and it is clear tha due to heavy mass, the momentum transfer effect is less in case of $\bar s$ quark distribution as compared to that in the $u$ quark. It is observed that the distribution is maximum when the quark longitudinal momentum fraction and total momentum transfer are lower. Further, it can be seen that, at $x \rightarrow 1$, the distribution does not depend on the values of $t$. This is because at $x\rightarrow 1$, the total longitudinal momentum fraction is carried by the struck quark, which was first distributed into all the valence partons in the kaon which makes the contribution from the partons (except the active parton)  negligible. The same happens with the active antiquark in the limit $x\rightarrow -1$.

\begin{figure*}
\includegraphics[width=.45\textwidth]{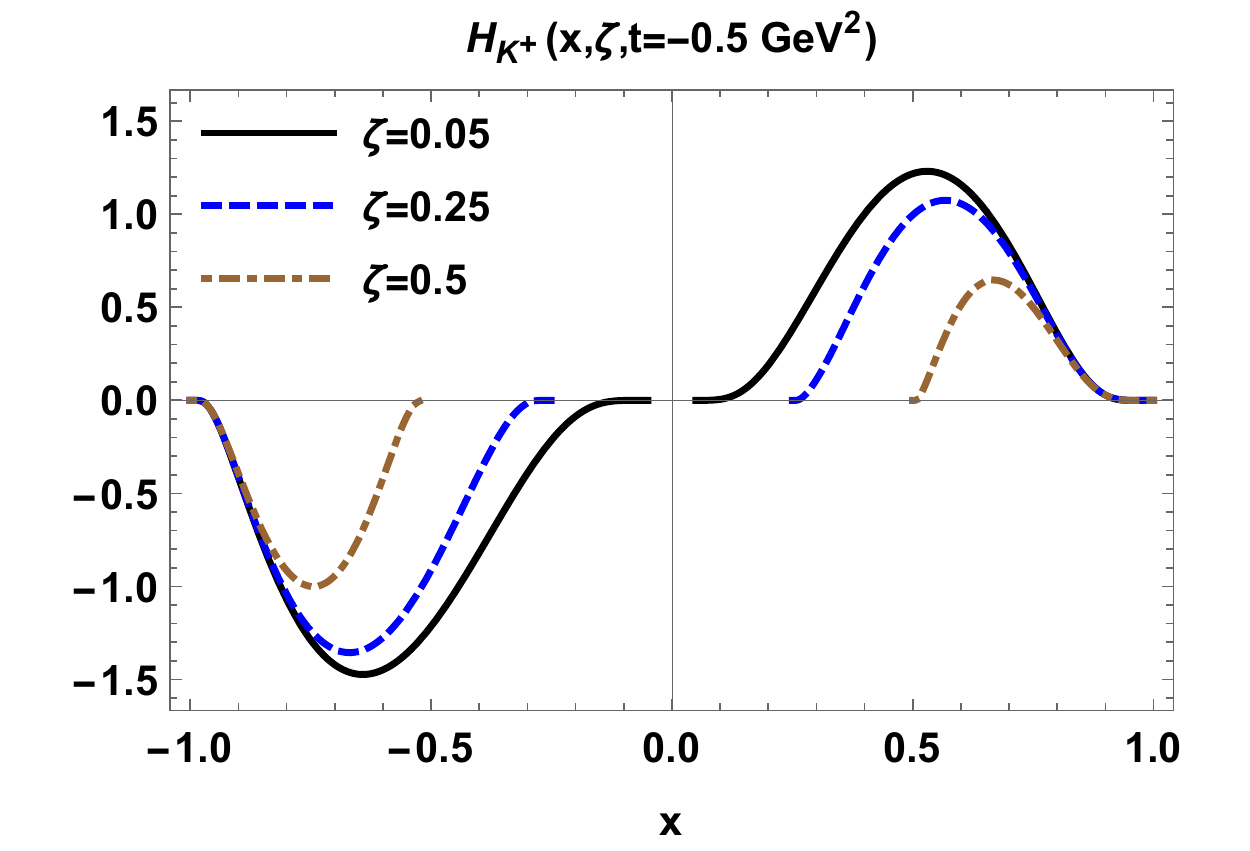}
\caption{The variation of unpolarized kaon GPD $H$ for $u$ and $\bar{s}$ quark as a function of $x$ at constant $-t=0.5$ $GeV^2$ and different values of $\zeta=0.05,0.25,0.5$.}
\label{gpd_different_zeta}
\end{figure*}

In Fig. \ref{gpd_different_zeta}, we have fixed the momentum transfer $-t=0.5$ $GeV^2$ and presented the GPDs w.r.t $x$ for different values of $\zeta$.  We observe that at a lower value of $\zeta$, the distribution shows a maximum peak w.r.t $x$, whereas the magnitude of peak decreases as the value of $\zeta$ increases. The peak also moves towards  higher $|x|$ with the increasing values of $\zeta$. This implies that the  longitudinal momentum fraction carried by the quark depends on the momentum transfer in the longitudinal direction. As the  longitudinal momentum transfer increases, the concentration of quarks move towards higher values longitudinal momentum fraction, however, the overall magnitude of the peak decreases.

\begin{figure*}
\centering
\begin{minipage}[c]{1\textwidth}
(a)\includegraphics[width=.45\textwidth]{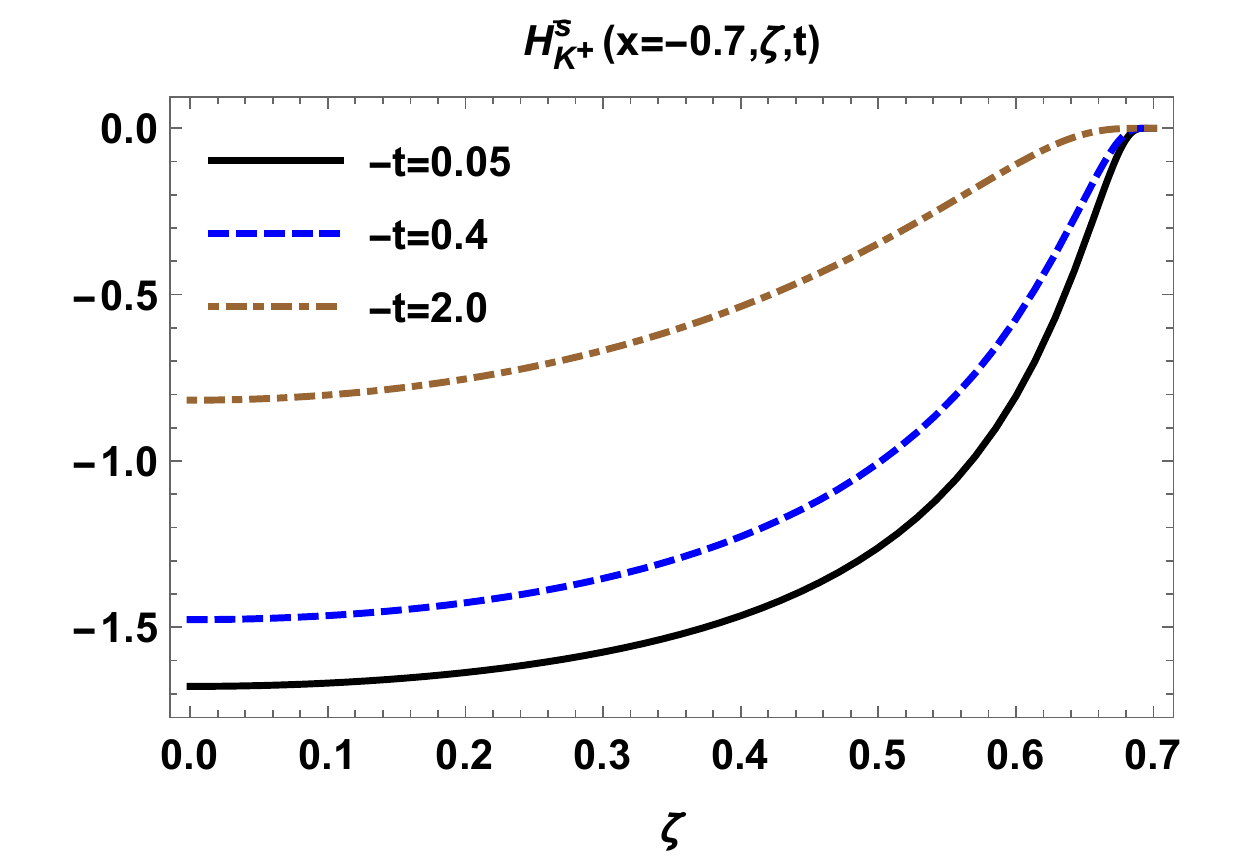}
(b)\includegraphics[width=.45\textwidth]{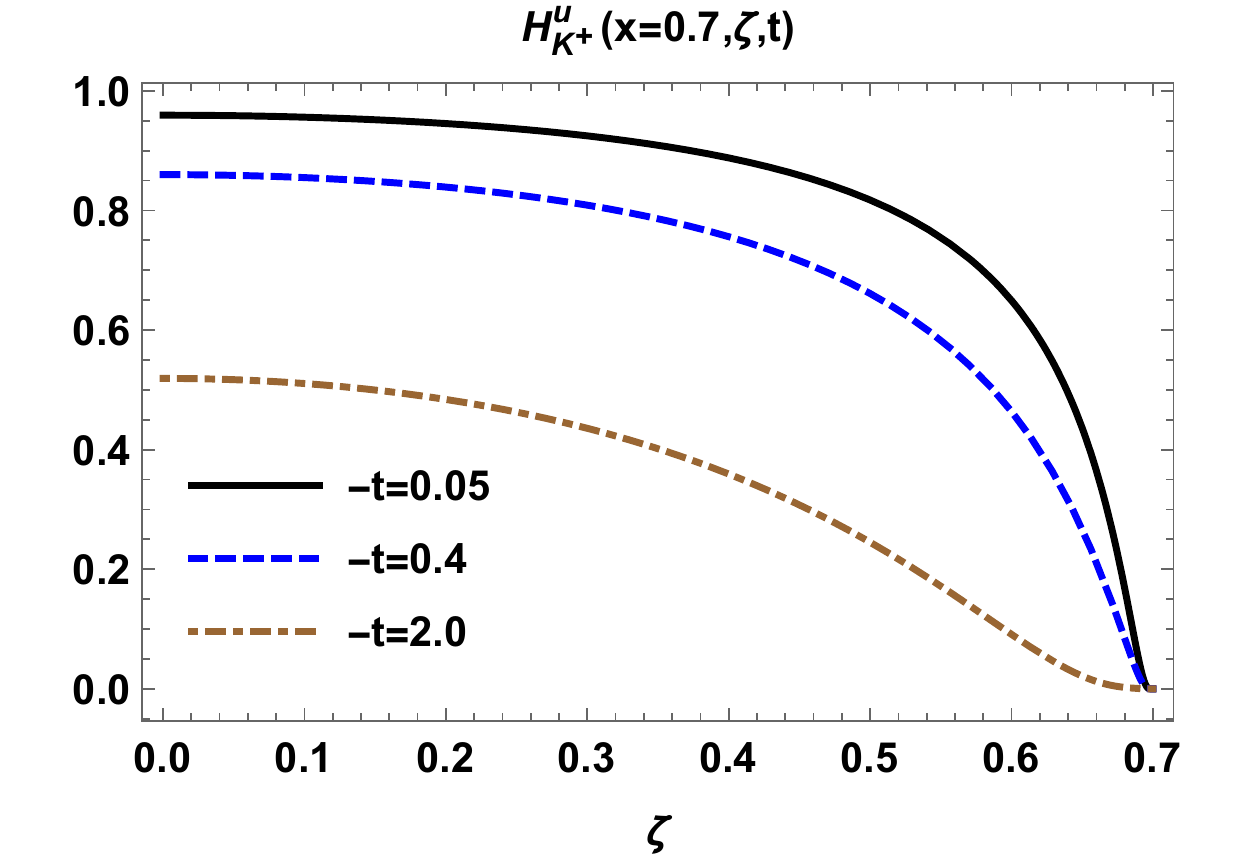}
\end{minipage}
\caption{The variation of unpolarized kaon GPD $H$ with $\zeta$ at different values of $-t=0.05, 0.4,2.0$ $GeV^2$ for (a)  $-x=0.7$ and (b) $x=0.7$ corresponding to $\bar{s}$ and $u$ quark respectively.}
\label{gpd_constant_x}
\end{figure*}

To get a deeper understanding of the relation between the parameters, we present the quark and antiquark distributions of kaon by fixing the value of $x$ and observe it as a function of $-t$ and $\zeta$ in Fig. \ref{gpd_constant_x}(a) and (b). Here, it can be seen that at $\zeta=0$, the distribution is different for different total momentum transferred $-t$  to the kaon from initial to final state. We also observe that in $\bar{s}$ ($u$) quark case, the distribution smoothly rises (falls) when $\zeta$ increases. 

From the discussions of Figs. \ref{gpd_different_t}-\ref{gpd_constant_x}, it can concluded that the distribution $H(x,\zeta,t)$ is zero for $x=\zeta$ and $x=-\zeta$ for the $u$ and $\bar{s}$ quarks respectively. The absence of distribution in this limit is the supportive domain of the model for evaluating the valence quark (antiquark) distributions which is possible at only lower values of $x$. At higher values of $x$, the sea quarks dominate and experiments are being planned to extend the measurements in this $x$ region. Even though the evaluation of total distributions includes the collective effects of  valence and sea quarks, we are ignoring the contribution  coming from the sea quarks in the present work.  
\section{IV. Transverse impact-parameter dependent GPDs }
We have now taken the two-dimensional Fourier transformation of GPD w.r.t. transverse momentum transfer $({\bf \Delta}_\perp)$, i.e. ${\bf \Delta}_\perp \rightarrow {\bf b}_\perp$. The transverse impact parameter dependent quark GPD of kaon in this case can be expressed as  \cite{ipdgpd1}
\begin{eqnarray}
\mathcal{H}^q(x,\zeta,b)=\frac{1}{(2\pi)^2}\int d^2 {\bf D}_\perp e^{-i{\bf D}_\perp.{\bf b}_\perp} H^q(x,\zeta,t).
\end{eqnarray}
Here ${\bf b}_\perp$ describes the impact parameter position in transverse direction, which comes after applying the Fourier tranformation of GPD w.r.t. ${\bf D}_\perp$. The variable ${\bf D}_\perp$ is because of the non-zero skewedness and is related to ${\bf \Delta}_\perp$ as 
\begin{eqnarray}
{\bf D}_\perp=\frac{{\bf P}''_\perp}{1-\zeta}-\frac{{\bf P}'_\perp}{1+\zeta}= \frac{{\bf \Delta}_\perp}{1-\zeta^2}.
\end{eqnarray}
The valence partons of kaon located at transverse impact-parameter position ${\bf b}_\perp$ are probed by assuming the kaon's initial and final state to be positioned at a fixed point but the relative distance between them is changed by the amount $\zeta {\bf b}_\perp$. In the absence of longitudinal momentum transferred to the kaon, i.e. for $\zeta=0$, the Fourier transform of $GPD(x,0,-\Delta^2)$ provides the parton distribution as a function of $x$ and transverse position $\bf b_\perp$ \cite{ipdgpd2}. We have
\begin{eqnarray}
q(x,{\bf b}_\perp)=\int \frac{d^2{\bf \Delta}_\perp}{(2\pi)^2} e^{-i {\bf b}_\perp. {\bf \Delta}_\perp} H^q(x,0,-{\bf \Delta}^2_\perp).
\end{eqnarray}

\begin{figure*}
\centering
\begin{minipage}[c]{1\textwidth}
(a)\includegraphics[width=.4\textwidth]{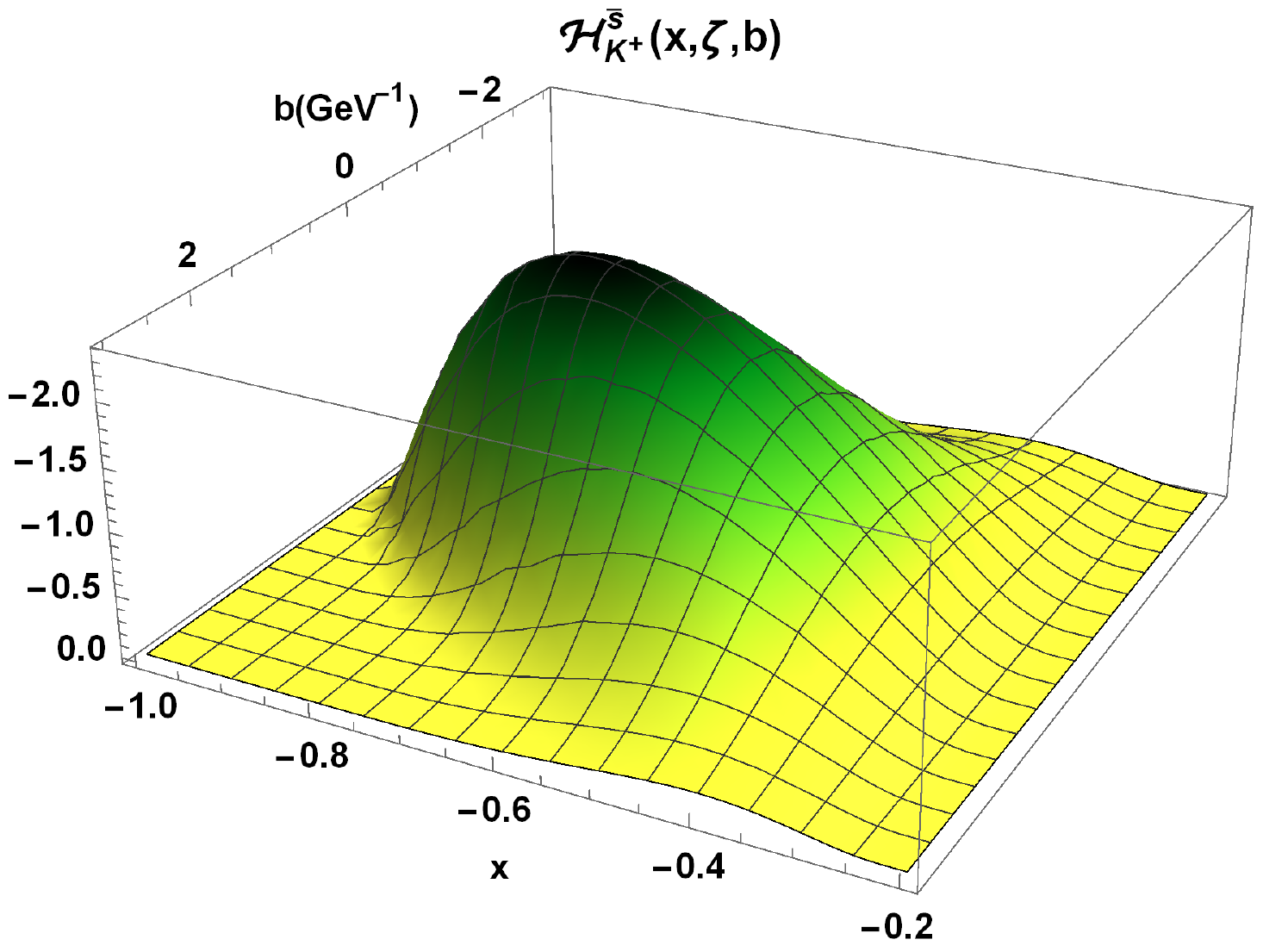}
(b)\includegraphics[width=.4\textwidth]{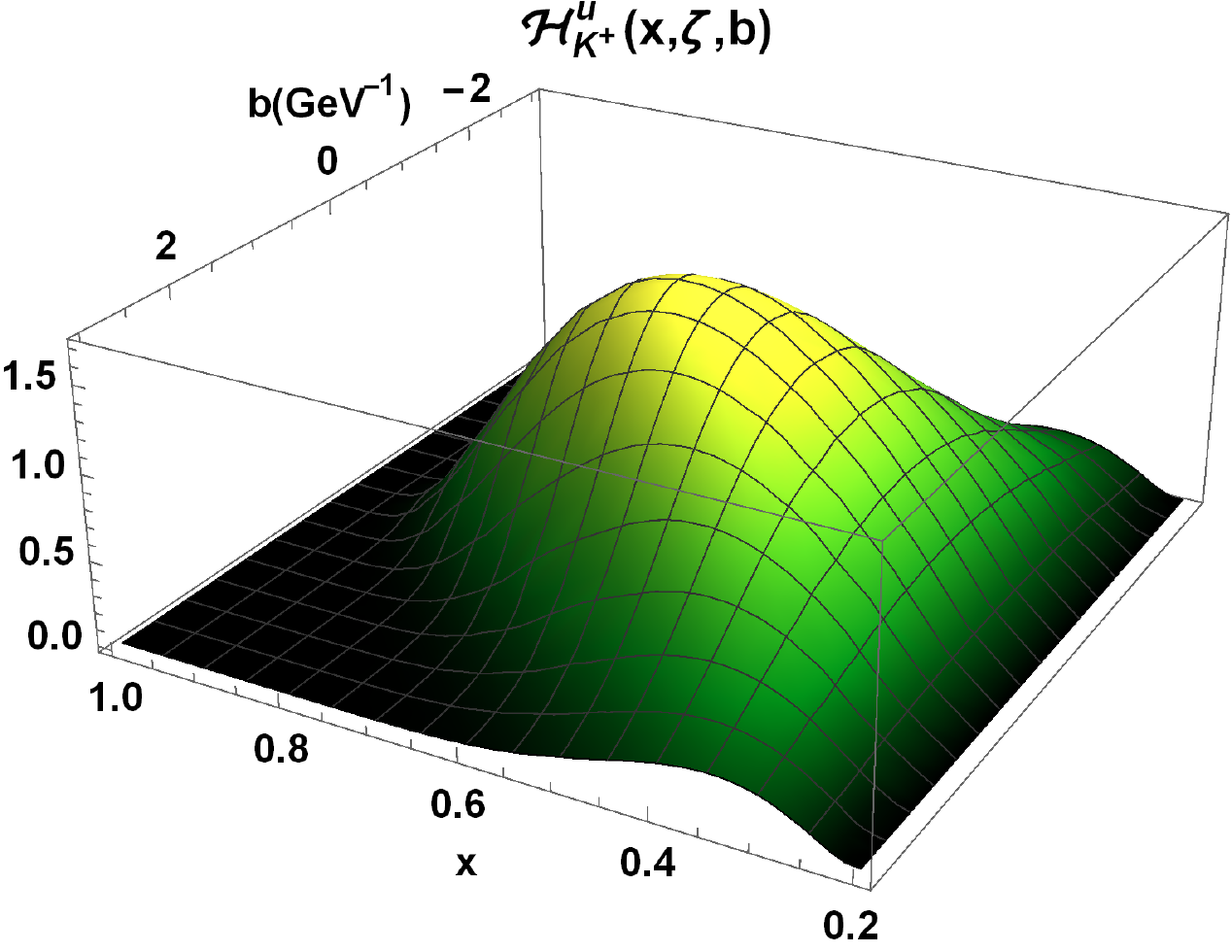}
\end{minipage}
\caption{The 3D plots showing the variation of transverse impact-parameter dependent parton distribution $\mathcal{H}$  of (a) $\bar{s}$ quark and (b) $u$ quark, with respect to $x$ and $b$ for a fixed value of $\zeta=0.2$.}
\label{gpd_x_b}
\end{figure*}

In Fig. \ref{gpd_x_b}(a) and (b), we fix the skewedness $\zeta=0.2$ and present the three-dimensional picture of transverse impact-parameter dependent GPD $\mathcal{H}(x,\zeta,b)$ with respect to longitudinal momentum fraction carried by quark (antiquark) $x(-x)$ and the distance from the transverse centre of momentum  for $\bar{s}$ and $u$ quark. If we let ourself observe $\mathcal{H}$ with respect to only $b$, we find that the distribution is maximum at the center for both quarks. In context of $x$ we observe that, at the point when there is no transverse distance from the centre the momentum of kaon, the distribution peaks have dependence on $x$. The amplitude of the peak changes with the changing values of $x$. 
While there is no distribution of valence partons in kaon at the higher values of $b$, which implies that if we keep on increasing the transverse distance, the distribution become negligible. 
When we compare the distributions of $\bar s$ and $u$ quarks, it can be clearly seen that the magnitude of distribution lies at a slightly higher values of $|x|$ for the case of $\bar s$ as compared to that in the case of $u$.  On the other side, if one concentrates on the variation of $\mathcal{H}$ with respect to $x$, one can see that with the increase in transverse position $b$, the distribution peak shift towards the lower values of $|x|$. 

\begin{figure*}
\centering
\begin{minipage}[c]{1\textwidth}
(a)\includegraphics[width=.4\textwidth]{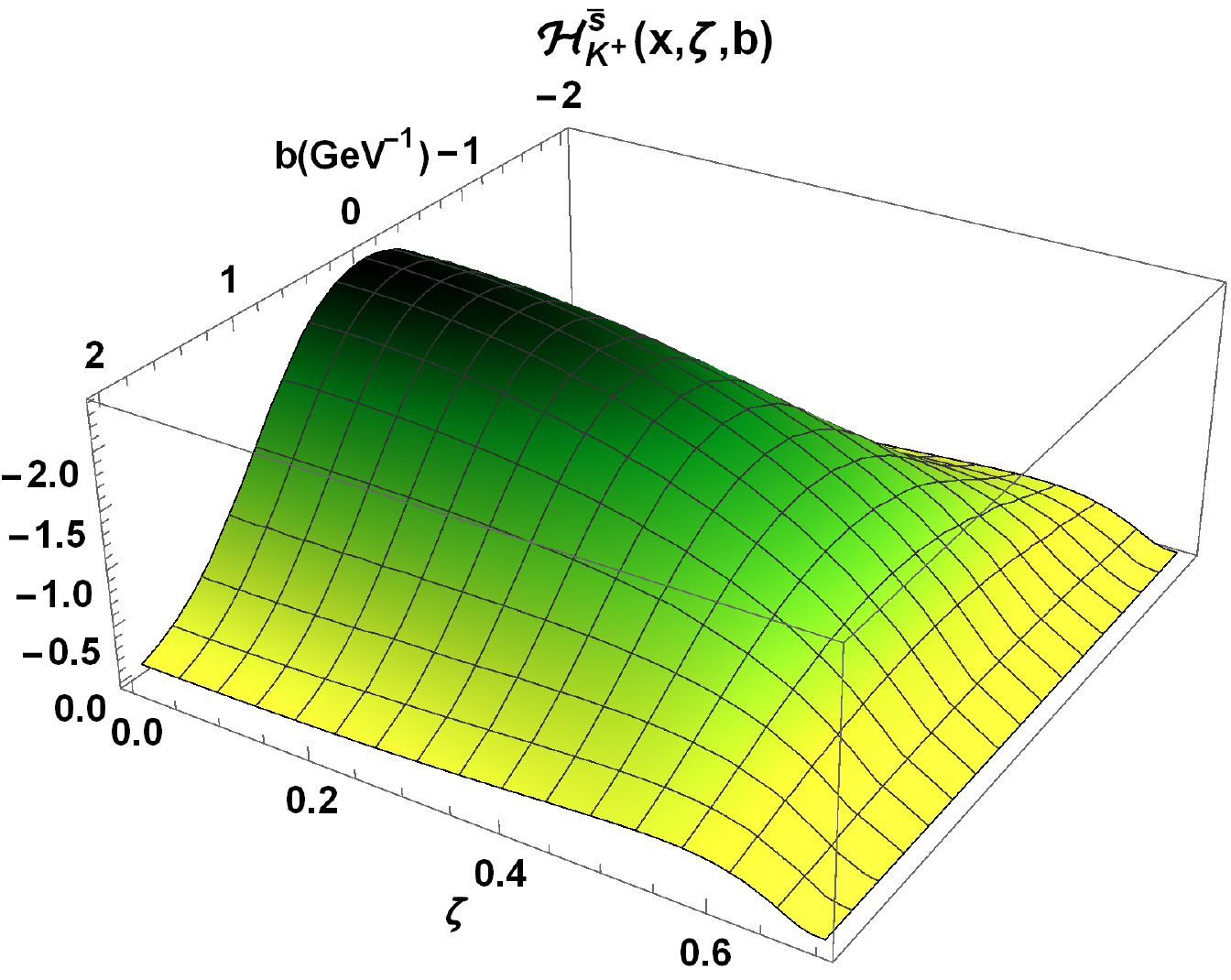}
(b)\includegraphics[width=.4\textwidth]{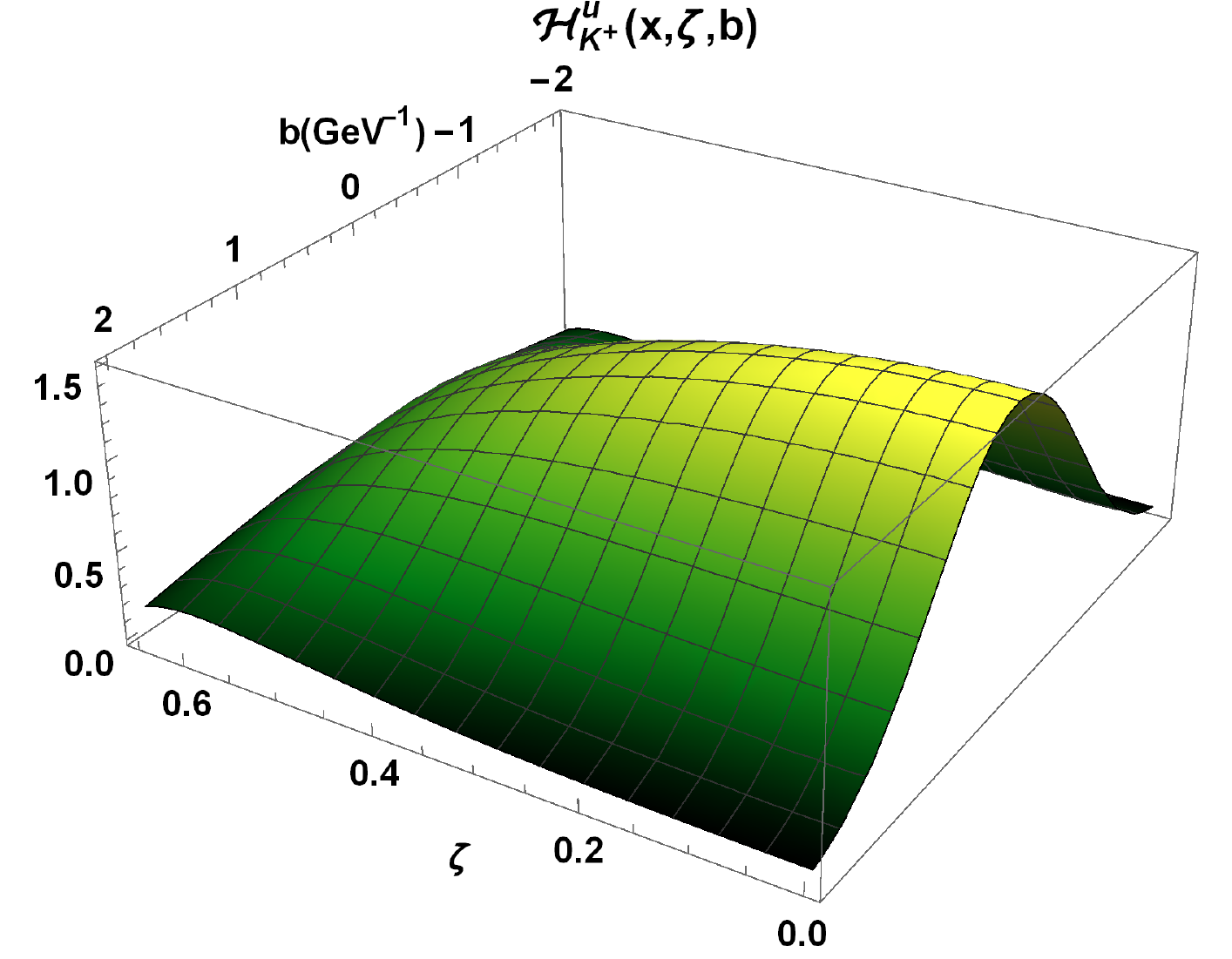}
\end{minipage}
\caption{The 3D plots showing the variation of transverse impact-parameter dependent parton distribution $\mathcal{H}$ with respect to $\zeta$ and $b$ at (a) $-x=0.7$ for $\bar{s}$ quark and  (b) at $x-0.7$ for $u$ quark.}
\label{gpd_zeta_b}
\end{figure*}

Furthermore, we present the distribution $\mathcal{H}(x,\zeta,b)$  as a function of $\zeta$ and ${\bf b}_\perp$ in Figs. \ref{gpd_zeta_b}(a) and(b) by fixing the value of $x$ at $-0.7$ and $0.7$ for $\bar{s}$ and $u$ quarks respectively. We notice that the distribution peaks become wider and the amplitude decreases at the higher values of $\zeta$ for $u$ quark and $\bar{s}$ quark. The spread is maximum, when the momentum transfer in the longitudinal direction is minimum. It implies that if no longitudinal momentum transfer is there, same longitudinal momentum fraction is carried by the active parton in the initial as well as in the final state. This  leads to the maximum spread for $\zeta=0$. 
\section{V. Wigner distributions of $u$ quark and $\bar{s}$ quark in unpolarized kaon}
The quantum phase-space distributions also known as Wigner distributions, describe the five dimensional picture of a hadron.  Specifically,it defines three momentum and two position co-ordinates \cite{wdmodel1, wdmodel2} and is defined as
\begin{eqnarray}
\rho^{[\Gamma]}({\bf b}_\perp,{\bf k}_\perp, x,S)\equiv \int \frac{d^2 {\bf \Delta}_\perp}{(2 \pi)^2} e^{-i {\bf \Delta}_\perp . {\bf b}_\perp} \hat{W}^{[\Gamma]}({\bf \Delta}_\perp, {\bf k}_\perp, x, S).\nonumber\\
\label{wigner}
\end{eqnarray} 
In the Drell-Yan frame $(\Delta^+=0)$, the Wigner operator or correlator $\hat{W}^{[\Gamma]}({\bf \Delta}_\perp, {\bf k}_\perp, x, S)$ at fixed light-cone time $z^+=0$ is defined by
\begin{eqnarray}
&&\hat{W}^{[\Gamma]}(\Delta_\perp, {\bf k_\perp},x;S)=\frac{1}{2} \int \frac{dz^- d^2 z_\perp}{(2 \pi)^3} e^{i k \cdot z} \nonumber\\
&& \bigg\langle{M(P'';S)}\bigg| \bar{\psi}\bigg(-\frac{z}{2}\bigg) \Gamma \mathcal{W}_{[-\frac{z}{2},\frac{z}{2}]}\psi\bigg(\frac{z}{2}\bigg)\bigg|{M(P';S)}\bigg\rangle\Bigg\vert_{z^{+}=0}, \nonumber\\
\label{wigner_correlator}
\end{eqnarray} 
where $\Gamma$ denotes the twist-2 Dirac $\gamma$-matrices $\gamma^+$, $\gamma^+ \gamma_5$, $i \sigma^{j+}\gamma_5$ ($j=1 {\ \rm or \ } 2$, depending on the polarization direction of quark) corresponding to unpolarized, longitudinally-polarized, transversely-polarized parton. The initial (final) momentum state and spin of hadron is described by $P'(P'')$ and $S$ respectively. The symbol $\mathcal{W}_{[-\frac{z}{2},\frac{z}{2}]}$ describes the Wilson line which ensures the SU(3) color gauge invariance of the operator $\hat{W}$. As we have restricted ourselves to the calculations of kaon distributions in the present work, kaon being a pseudoscalar-meson will remain unpolarized throughout the calculations ($S=0$). Therefore, spin will not be included in Eqs (\ref{wigner}) and (\ref{wigner_correlator}). 

By combining the different polarization configrations of quark within the unpolarized hadron (having spin-$0$), the Wigner distributions are defined as follows \cite{wdmodel8}. For the unpolarized quark in the unpolarized hadron, we have
\begin{eqnarray}
\rho_{UU}({\bf b}_\perp,{\bf k}_\perp, x)&=&\rho^{[\gamma^+]}({\bf b}_\perp,{\bf k}_\perp, x),
\label{def_uu}
\end{eqnarray}
for the longitudinally-polarized quark in the unpolarized hadron, we have
\begin{eqnarray}
\rho_{UL}({\bf b}_\perp,{\bf k}_\perp, x)&=&\rho^{[\gamma^+ \gamma_5]}({\bf b}_\perp,{\bf k}_\perp, x),
\label{def_ul}
\end{eqnarray}
for the transversely-polarized quark in the unpolarized hadron, we have
\begin{eqnarray}
\rho^j_{UT}({\bf b}_\perp,{\bf k}_\perp, x)&=&\rho^{[i\sigma^{j+}\gamma_5]}({\bf b}_\perp,{\bf k}_\perp, x).
\label{def_ut}
\end{eqnarray}
By using the Eq. (\ref{meson_eqn}) in Eq. (\ref{wigner_correlator}), we get the Wigner correlation operator $\hat{W}^{[\Gamma]}({\bf \Delta}_{\perp}, {\bf k}_{\perp}, x)$ for $\Gamma = \gamma^+, \gamma^+ \gamma_5,i\sigma^{j+}\gamma_5$ in the overlap form as 
\begin{widetext}
\begin{eqnarray}
\hat{W}^{[\gamma^+]}({\bf \Delta}_\perp, {\bf k}_\perp,x)&=&\frac{1}{16\pi^3} \big[\psi_0^{*\uparrow, \uparrow}(x,{\bf k}_\perp'')\psi^{\uparrow, \uparrow}_0(x,{\bf k}_\perp')
+\psi_0^{*\downarrow, \uparrow}(x,{\bf k}_\perp'')\psi^{\downarrow, \uparrow}_0(x,{\bf k}_\perp')+\psi_0^{*\uparrow, \downarrow}(x,{\bf k}_\perp'')\psi^{\uparrow, \downarrow}_0(x,{\bf k}_\perp')
\nonumber\\
&&+\psi_0^{*\downarrow, \downarrow}(x,{\bf k}_\perp'')\psi^{\downarrow, \downarrow}_0(x,{\bf k}_\perp')\big],
\label{unpolarized_overlap}\\
\hat{W}^{[\gamma^+\gamma_5]}({\bf \Delta}_\perp, {\bf k}_\perp,x)&=&\frac{1}{16\pi^3} \big[\psi_0^{*\uparrow, \uparrow}(x,{\bf k}_\perp'')\psi^{\uparrow, \uparrow}_0(x,{\bf k}_\perp')
-\psi_0^{*\downarrow, \uparrow}(x,{\bf k}_\perp'')\psi^{\downarrow, \uparrow}_0(x,{\bf k}_\perp')+\psi_0^{*\uparrow, \downarrow}(x,{\bf k}_\perp'')\psi^{\uparrow, \downarrow}_0(x,{\bf k}_\perp')
\nonumber\\
&&-\psi_0^{*\downarrow, \downarrow}(x,{\bf k}_\perp'')\psi^{v}_0(x,{\bf k}_\perp')\big],
\label{longi_polarized_overlap}\\
\hat{W}^{[i\sigma^{j+}\gamma_5]}({\bf \Delta}_\perp, {\bf k}_\perp,x)&=&\frac{1}{16\pi^3}\epsilon^{ij}_\perp \big[(-i)^i \psi^{*\uparrow, \uparrow}_0(x, {\bf k}_\perp'')\psi^{\downarrow, \uparrow}_0(x, {\bf k}_\perp')+(i)^i \psi^{*\downarrow, \uparrow}_0(x, {\bf k}_\perp'')\psi^{\uparrow, \uparrow}_0(x, {\bf k}_\perp')\nonumber\\
&&+ (-i)^i \psi^{*\uparrow, \downarrow}_0(x, {\bf k}_\perp'')\psi^{\downarrow, \downarrow}_0(x, {\bf k}_\perp')+(i)^i \psi^{*\downarrow, \downarrow}_0(x, {\bf k}_\perp'')\psi^{\uparrow, \downarrow}_0(x, {\bf k}_\perp') \big].
\label{trans_polarized_overlap}
\end{eqnarray} 
\end{widetext}
\begin{figure*}
\centering
\begin{minipage}[c]{1\textwidth}
(a)\includegraphics[width=.4\textwidth]{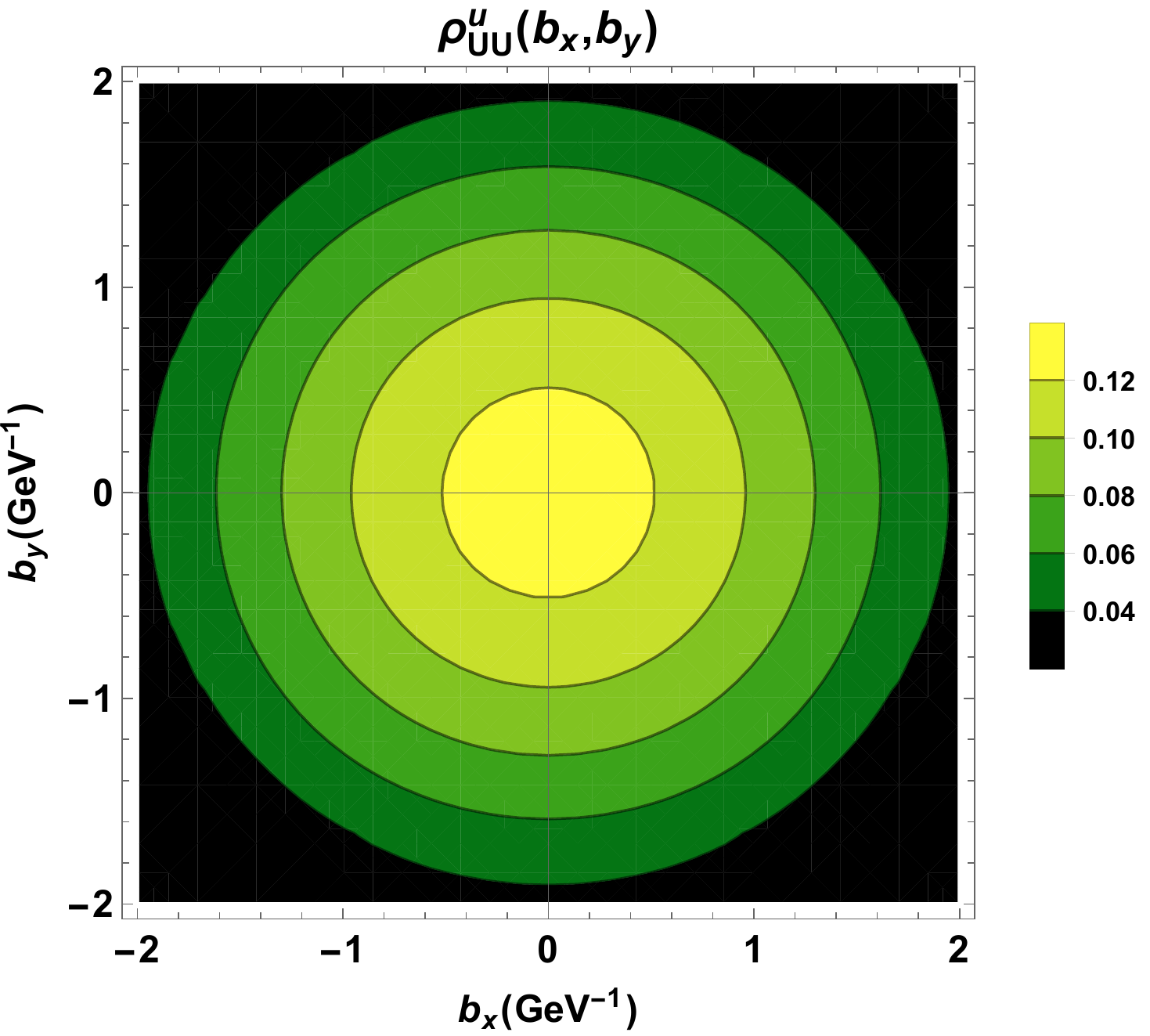}
(b)\includegraphics[width=.4\textwidth]{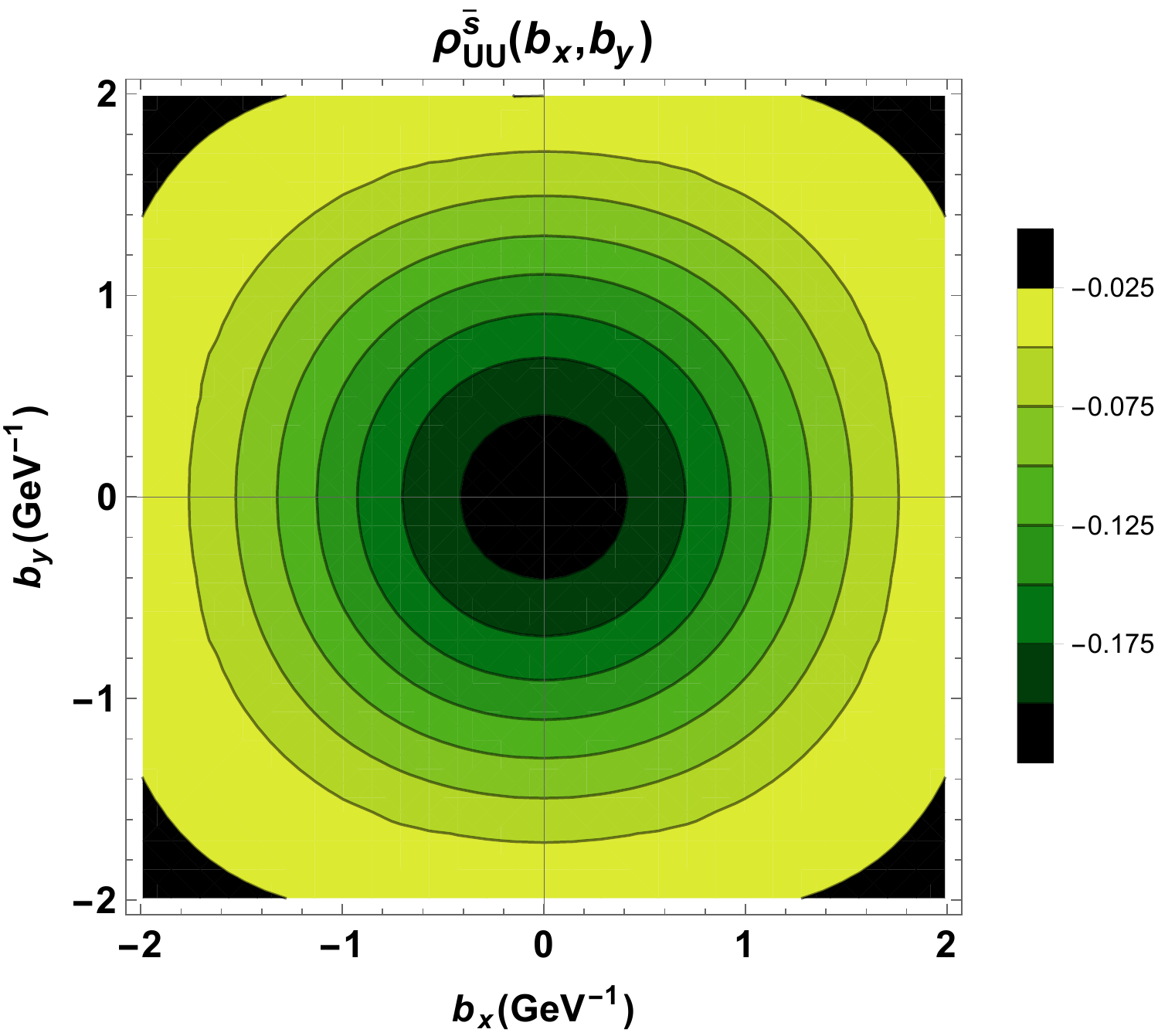}
\end{minipage}
\begin{minipage}[c]{1\textwidth}
(c)\includegraphics[width=.4\textwidth]{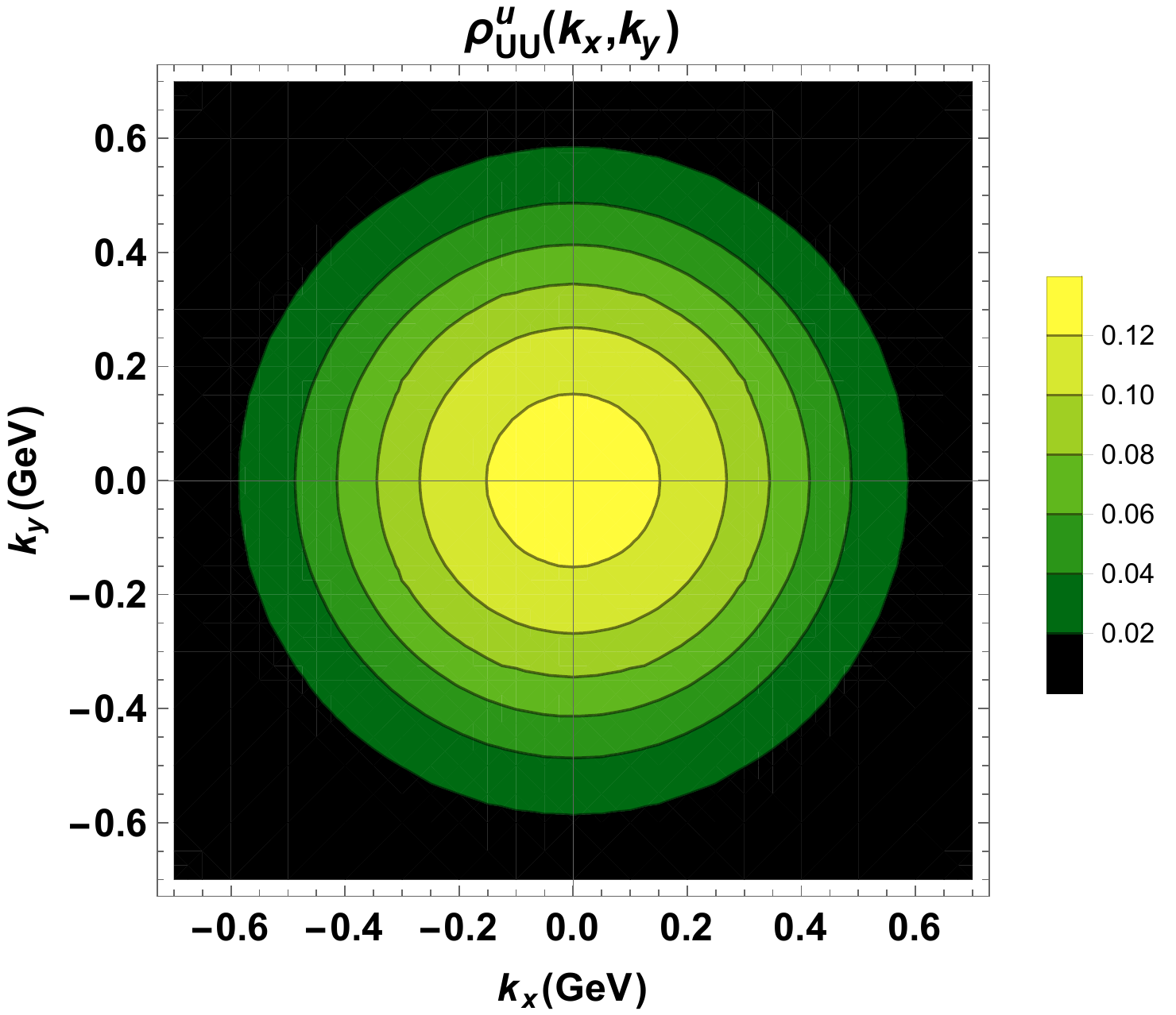}
(d)\includegraphics[width=.4\textwidth]{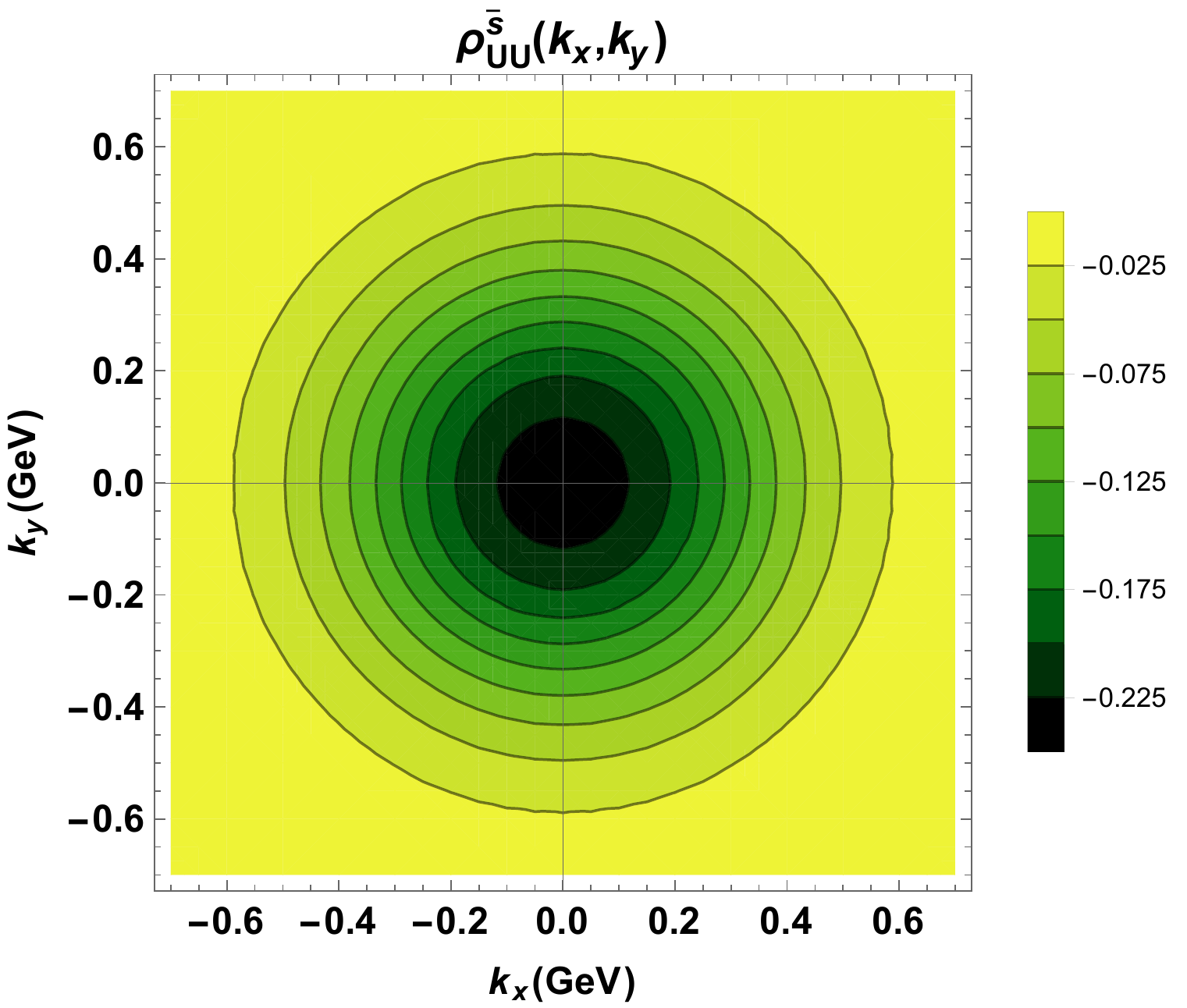}
\end{minipage}
\begin{minipage}[c]{1\textwidth}
(e)\includegraphics[width=.4\textwidth]{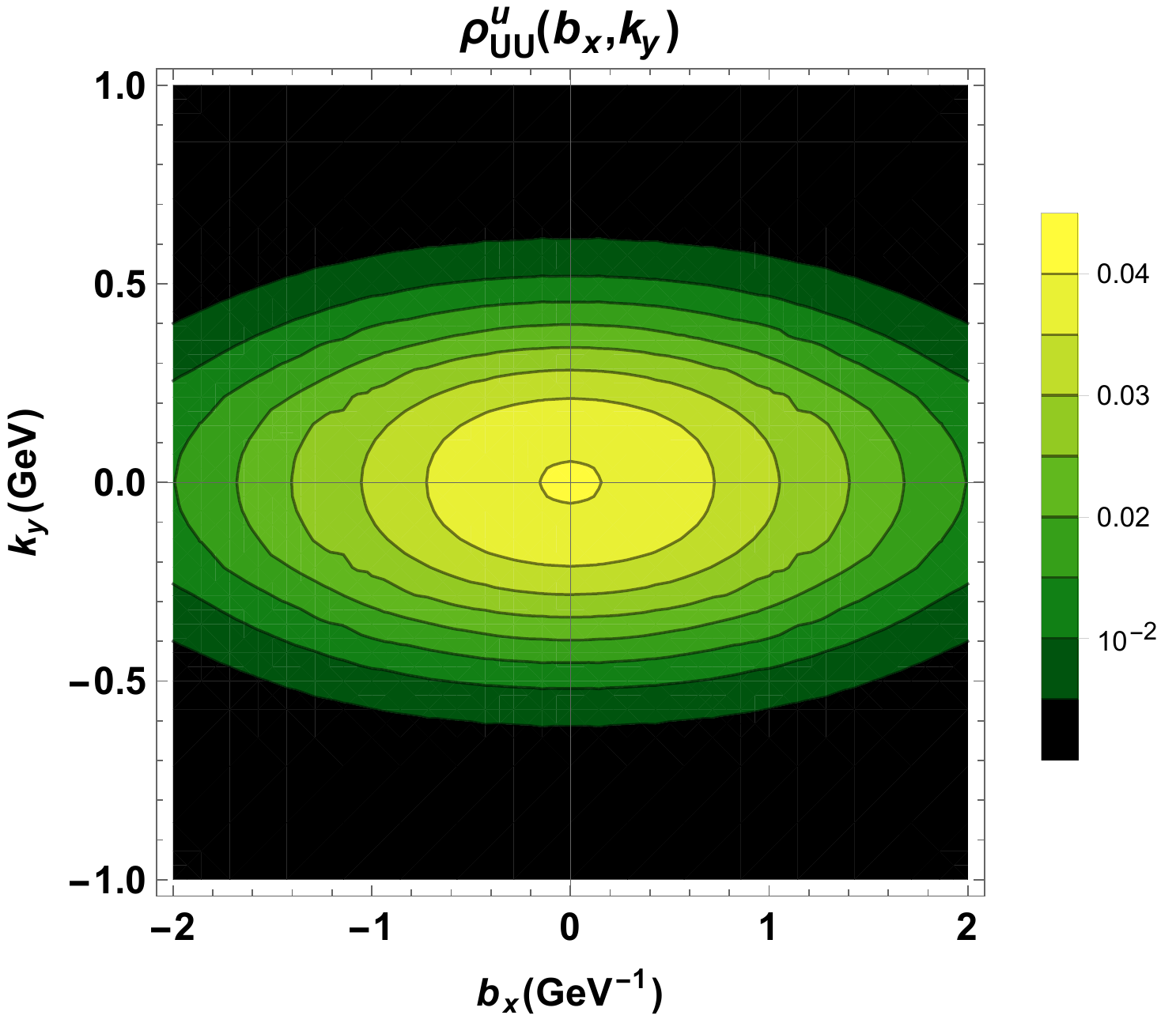}
(f)\includegraphics[width=.4\textwidth]{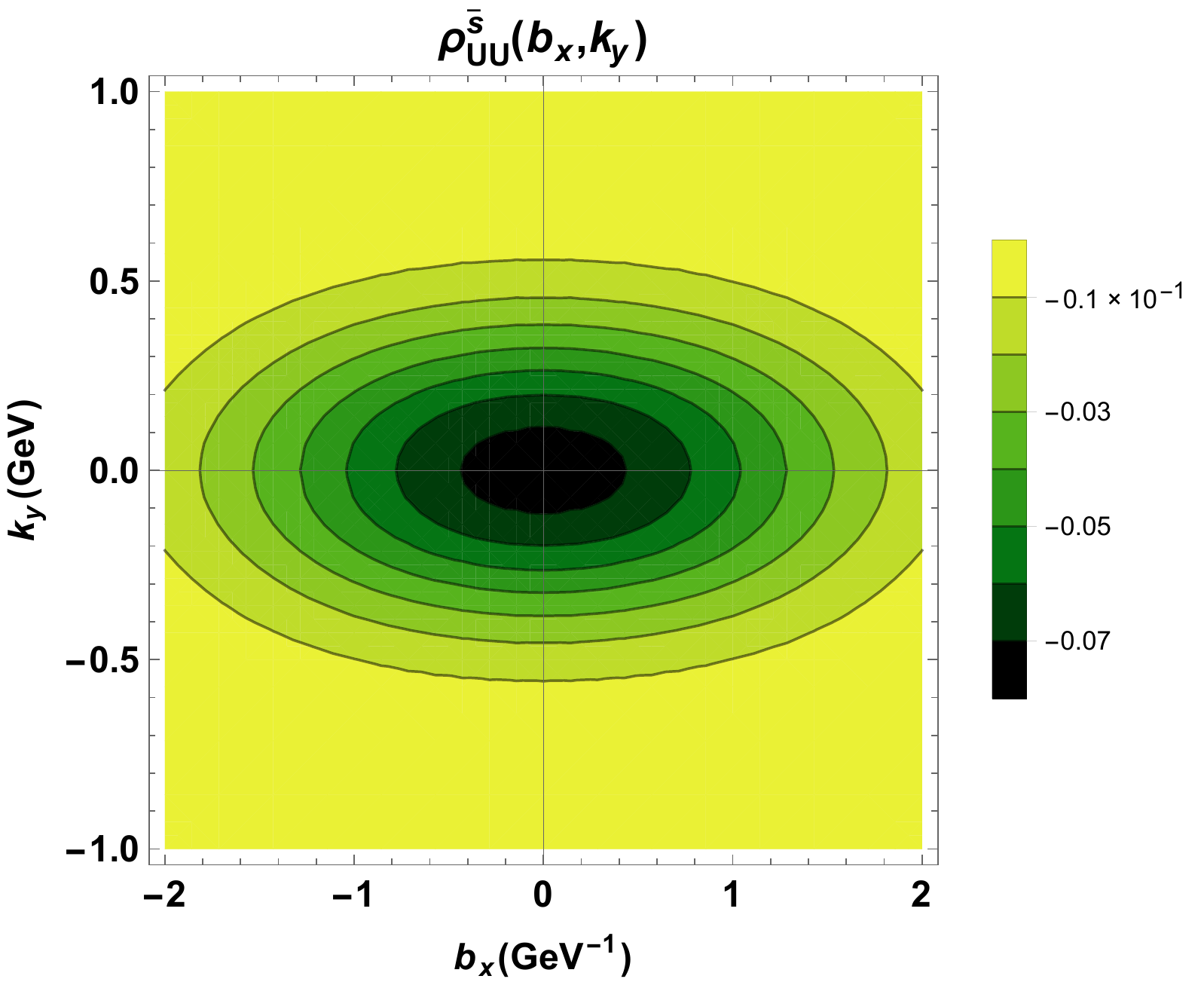}
\end{minipage}
\caption{The unpolarized Wigner distribution $\rho_{UU}$ of (i) $u$ quark (left panel), and (ii) $\bar{s}$ quark (right panel) for kaon in the impact-parameter plane, the transverse momentum plane, and the mixed plane.}
\label{uu}
\end{figure*}
\begin{figure*}
\centering
\begin{minipage}[c]{1\textwidth}
(a)\includegraphics[width=.4\textwidth]{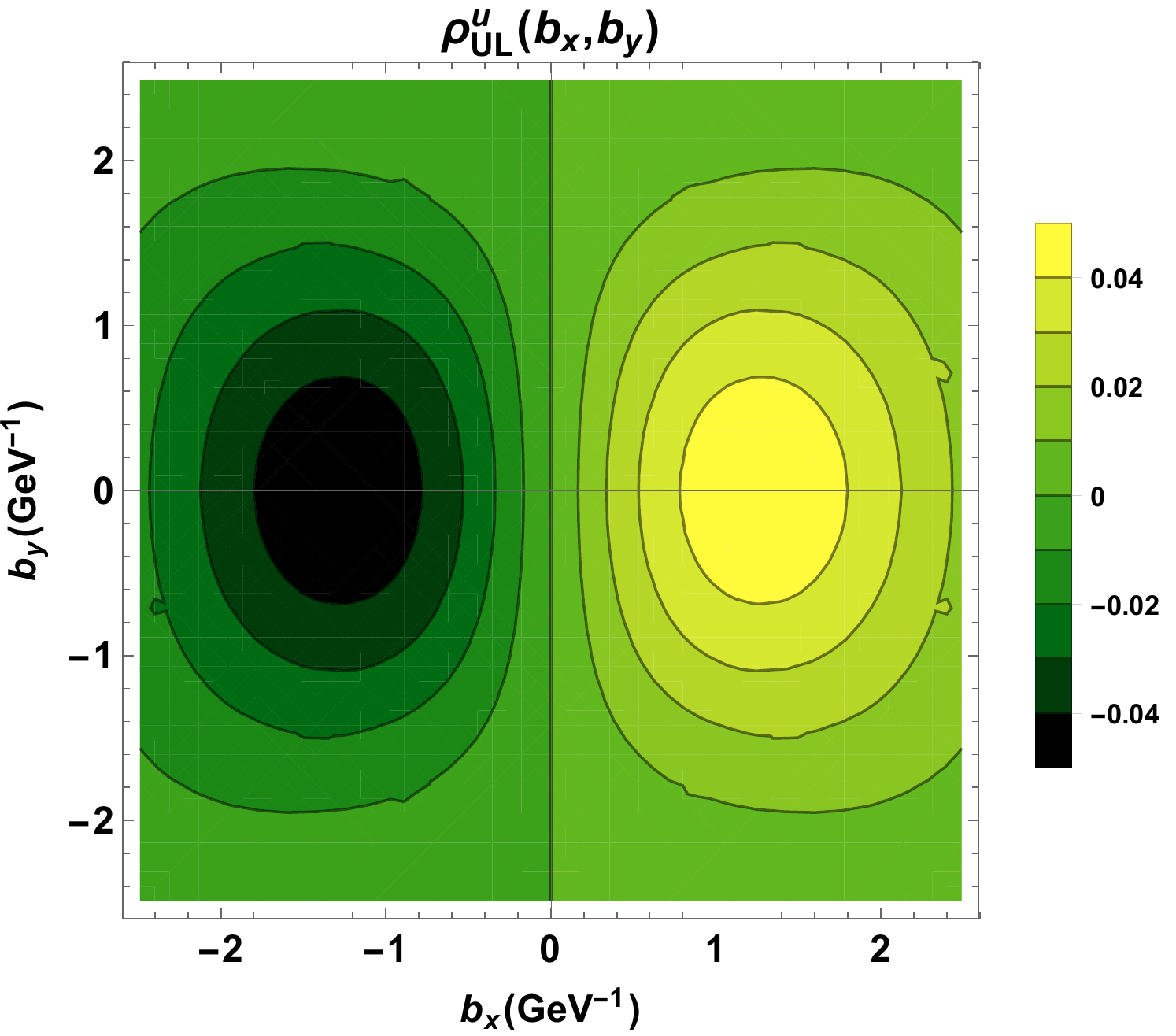}
(b)\includegraphics[width=.4\textwidth]{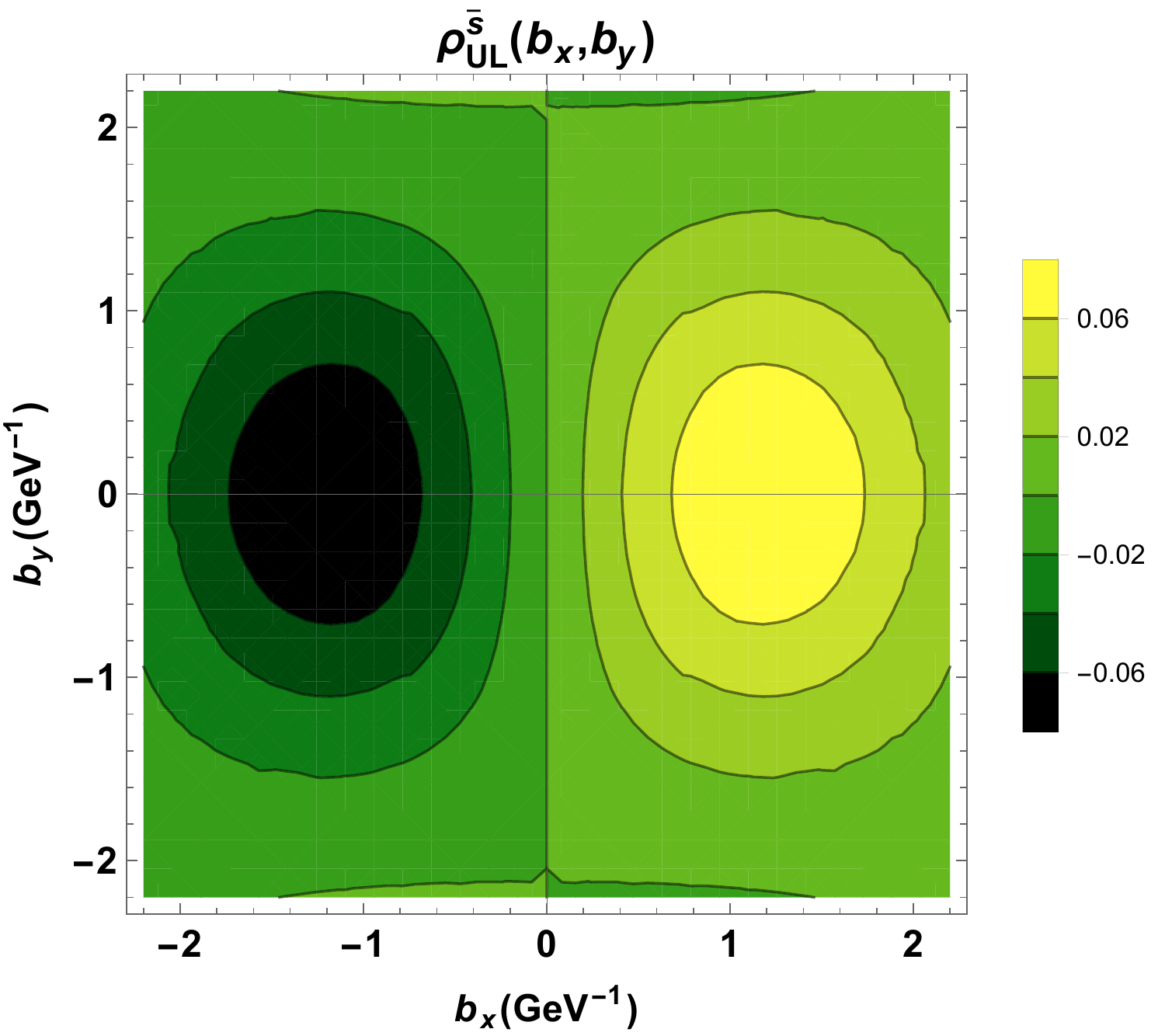}
\end{minipage}
\begin{minipage}[c]{1\textwidth}
(c)\includegraphics[width=.4\textwidth]{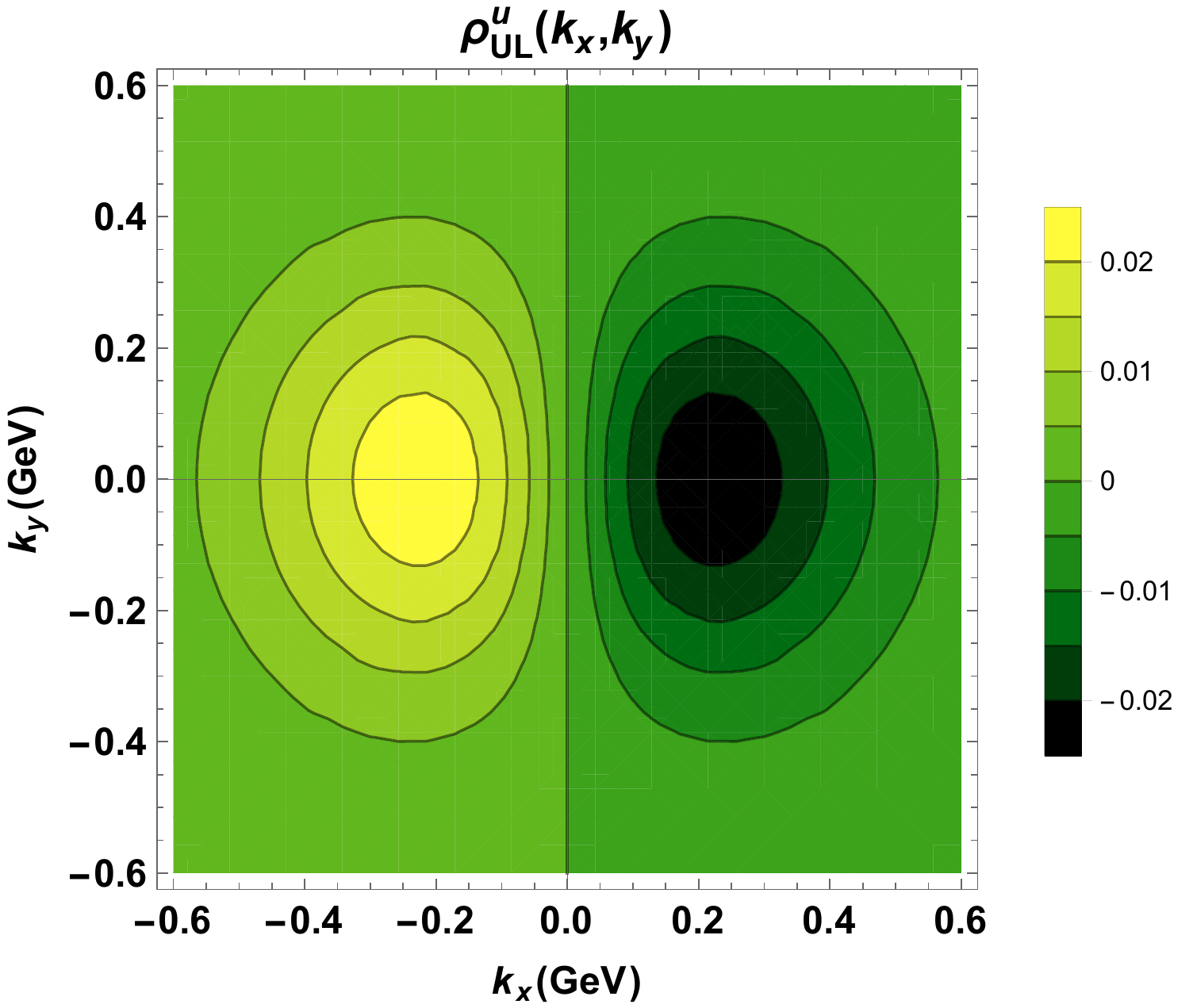}
(d)\includegraphics[width=.4\textwidth]{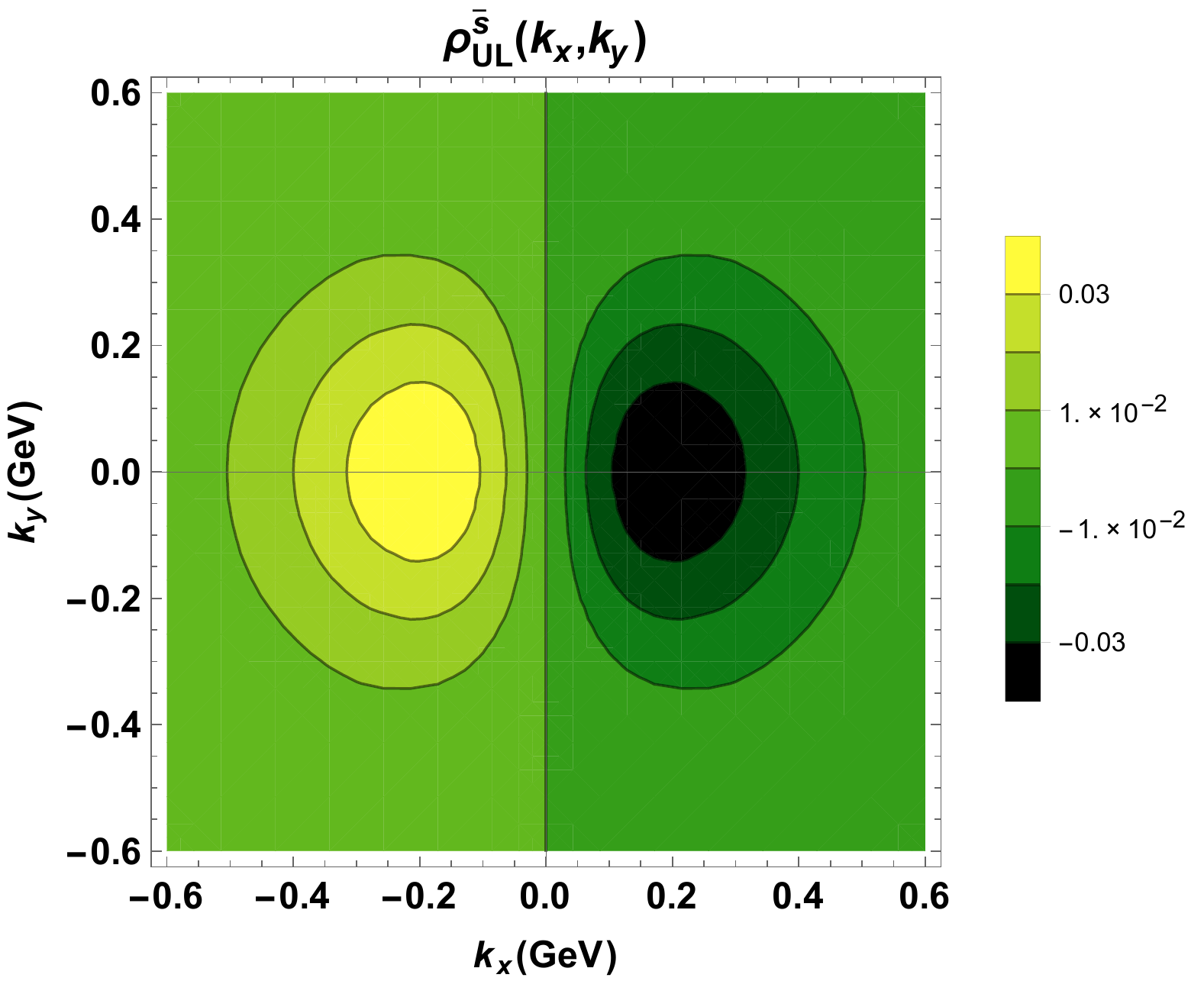}
\end{minipage}
\begin{minipage}[c]{1\textwidth}
(e)\includegraphics[width=.4\textwidth]{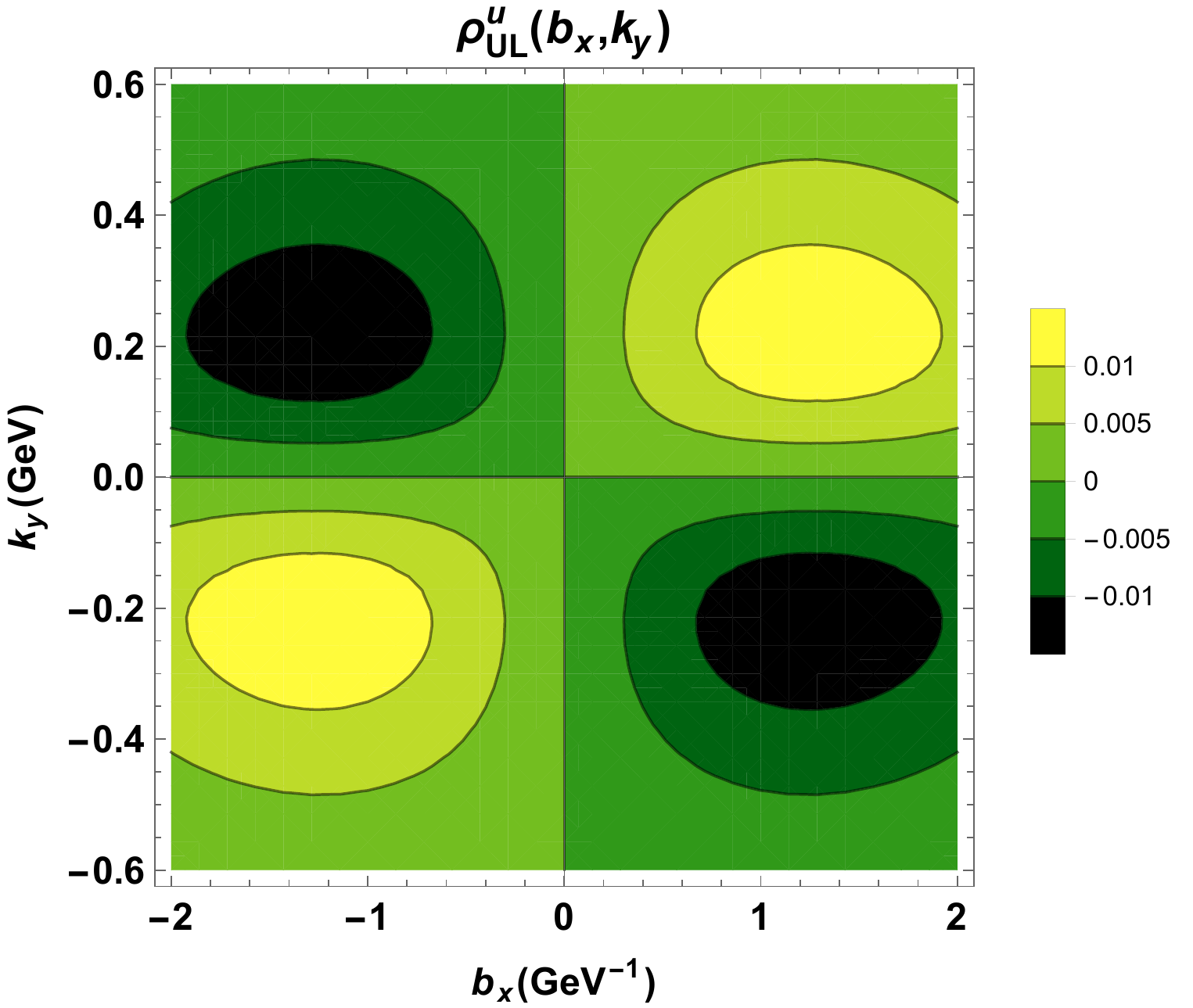}
(f)\includegraphics[width=.4\textwidth]{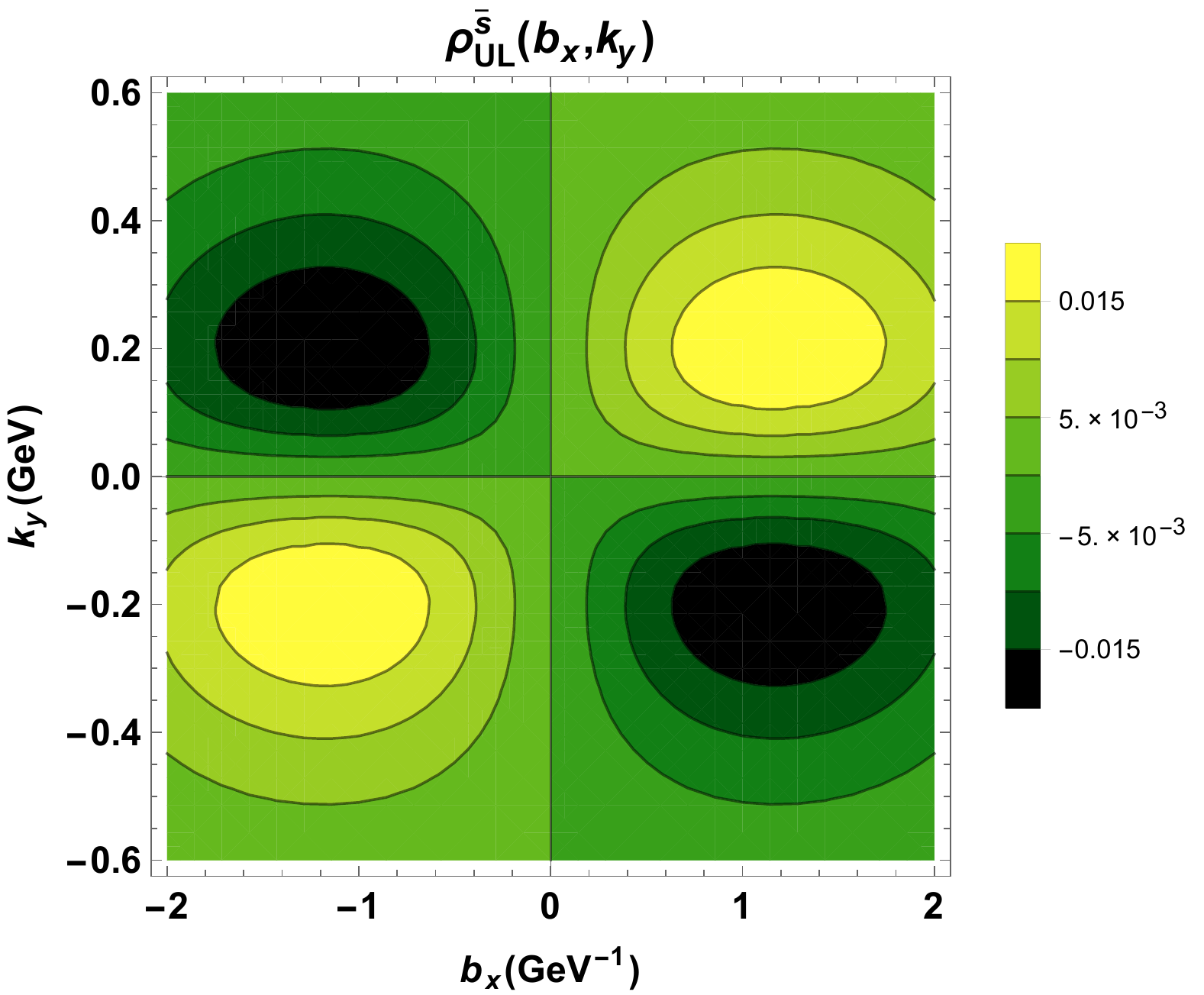}
\end{minipage}
\caption{The unpolarized-longitudinal Wigner distribution $\rho_{UL}$ of (i) $u$ quark (left panel), and (ii) $\bar{s}$ quark (right panel) for kaon in the impact-parameter plane, the transverse momentum plane, and the mixed plane.}
\label{ul}
\end{figure*}
\begin{figure*}
\centering
\begin{minipage}[c]{1\textwidth}
(a)\includegraphics[width=.4\textwidth]{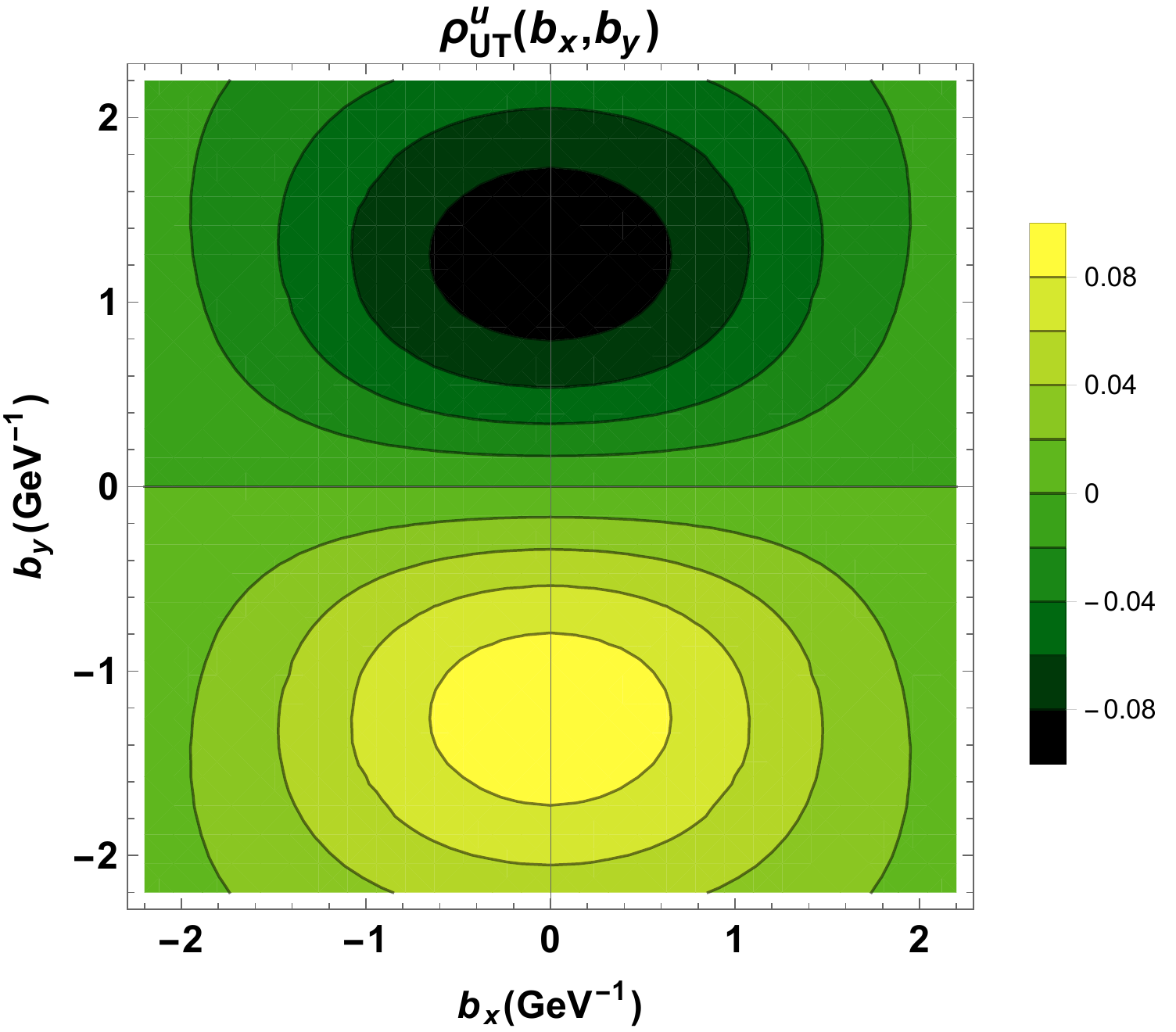}
(b)\includegraphics[width=.4\textwidth]{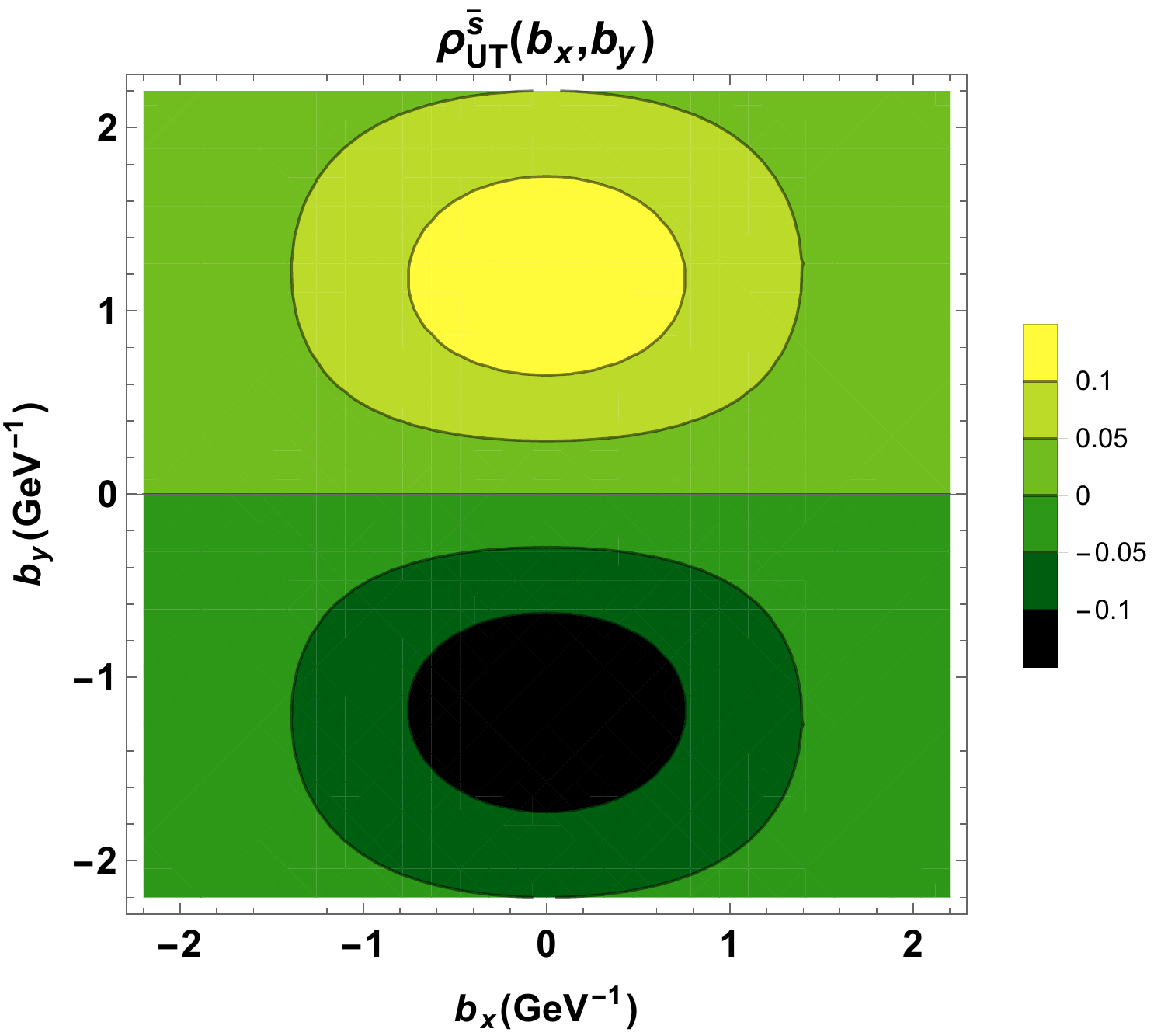}
\end{minipage}
\begin{minipage}[c]{1\textwidth}
(c)\includegraphics[width=.4\textwidth]{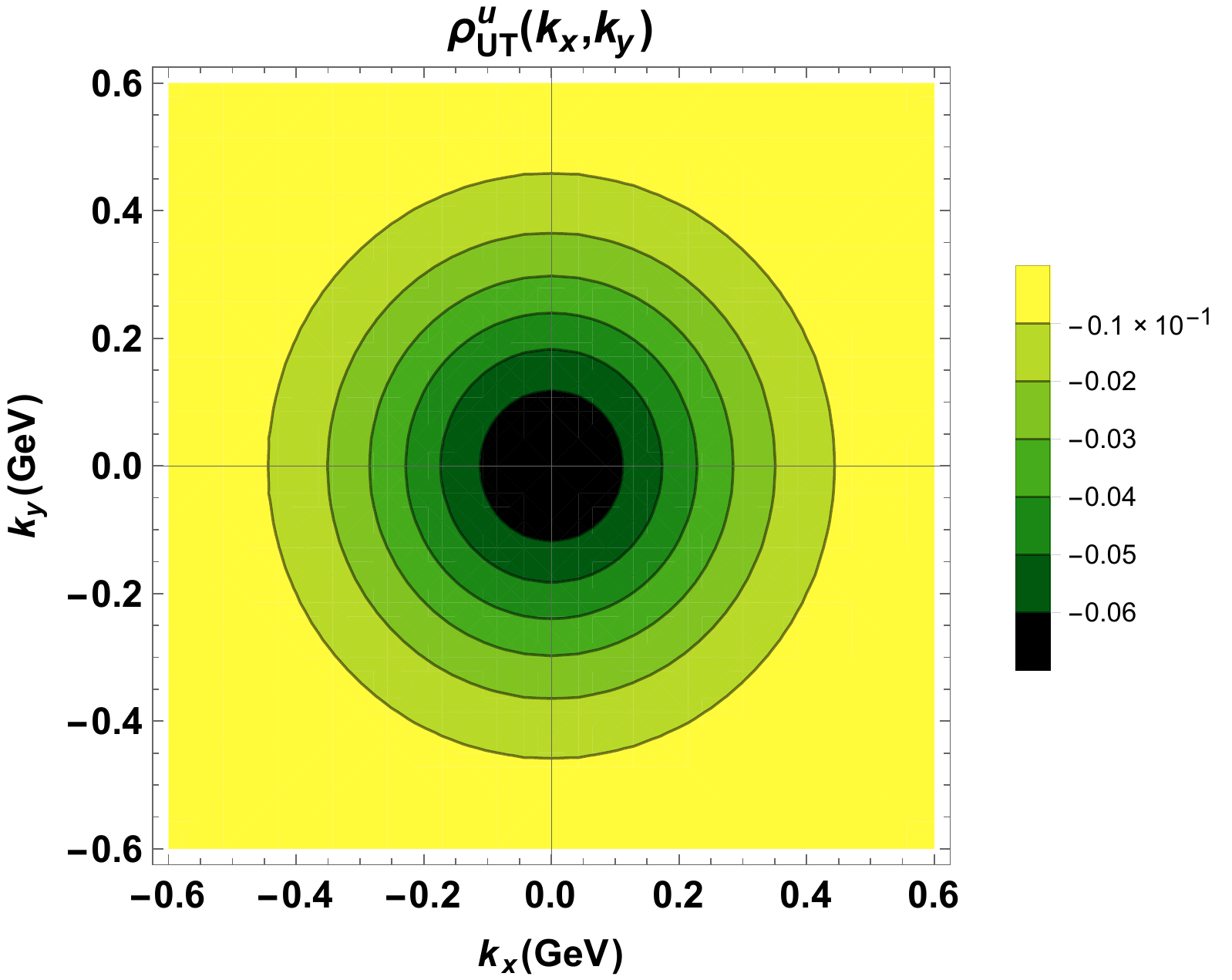}
(d)\includegraphics[width=.4\textwidth]{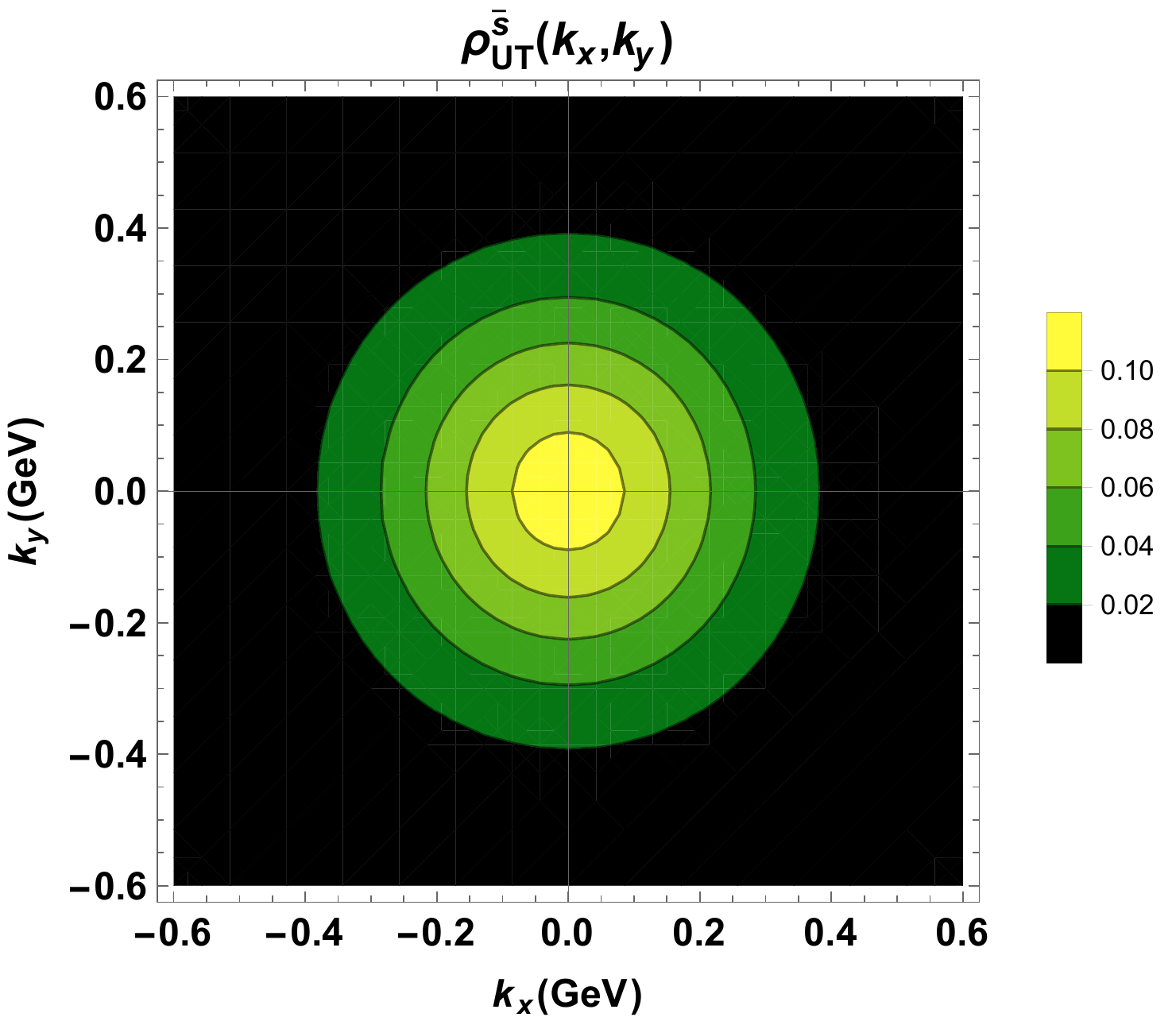}
\end{minipage}
\begin{minipage}[c]{1\textwidth}
(e)\includegraphics[width=.4\textwidth]{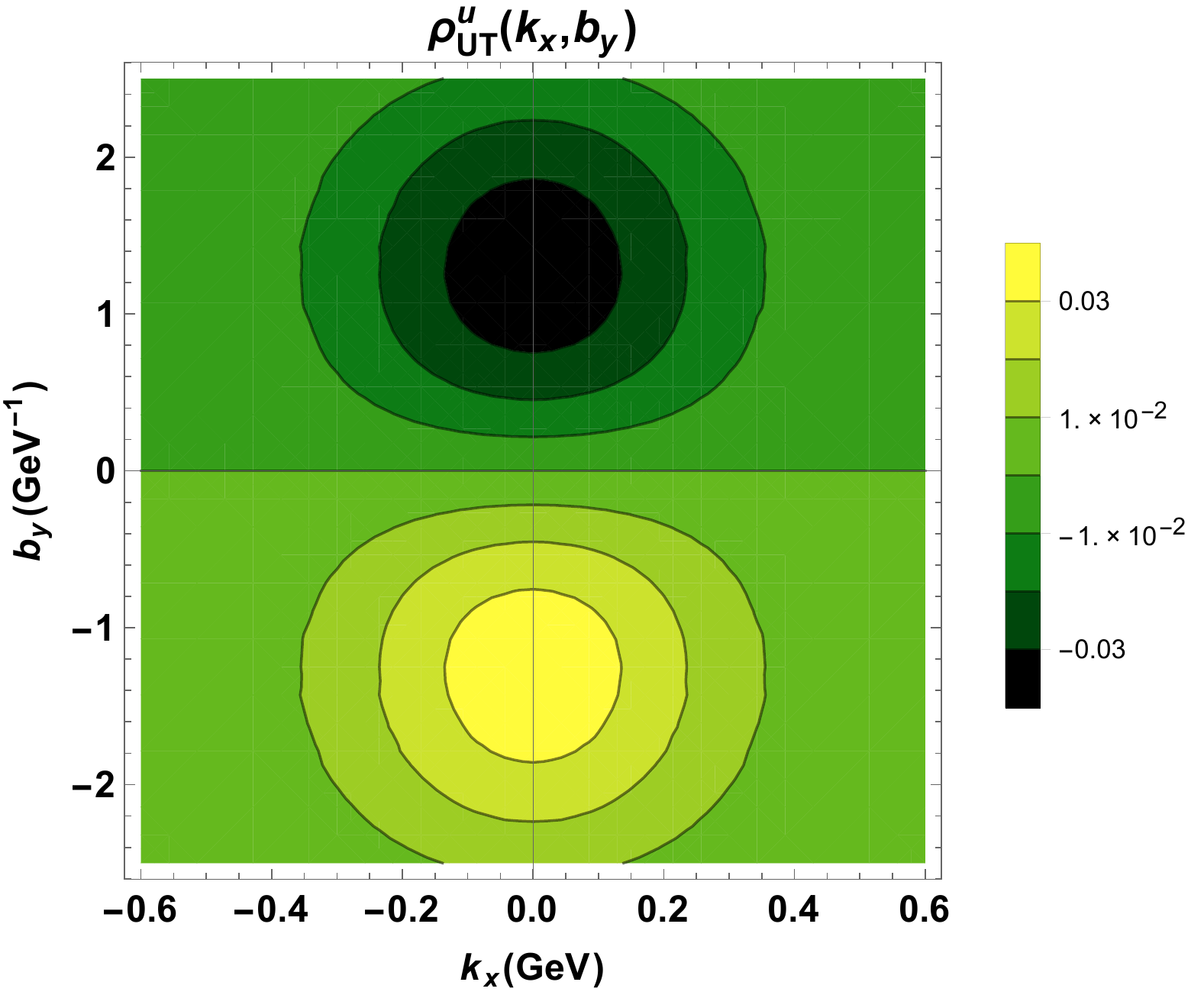}
(f)\includegraphics[width=.4\textwidth]{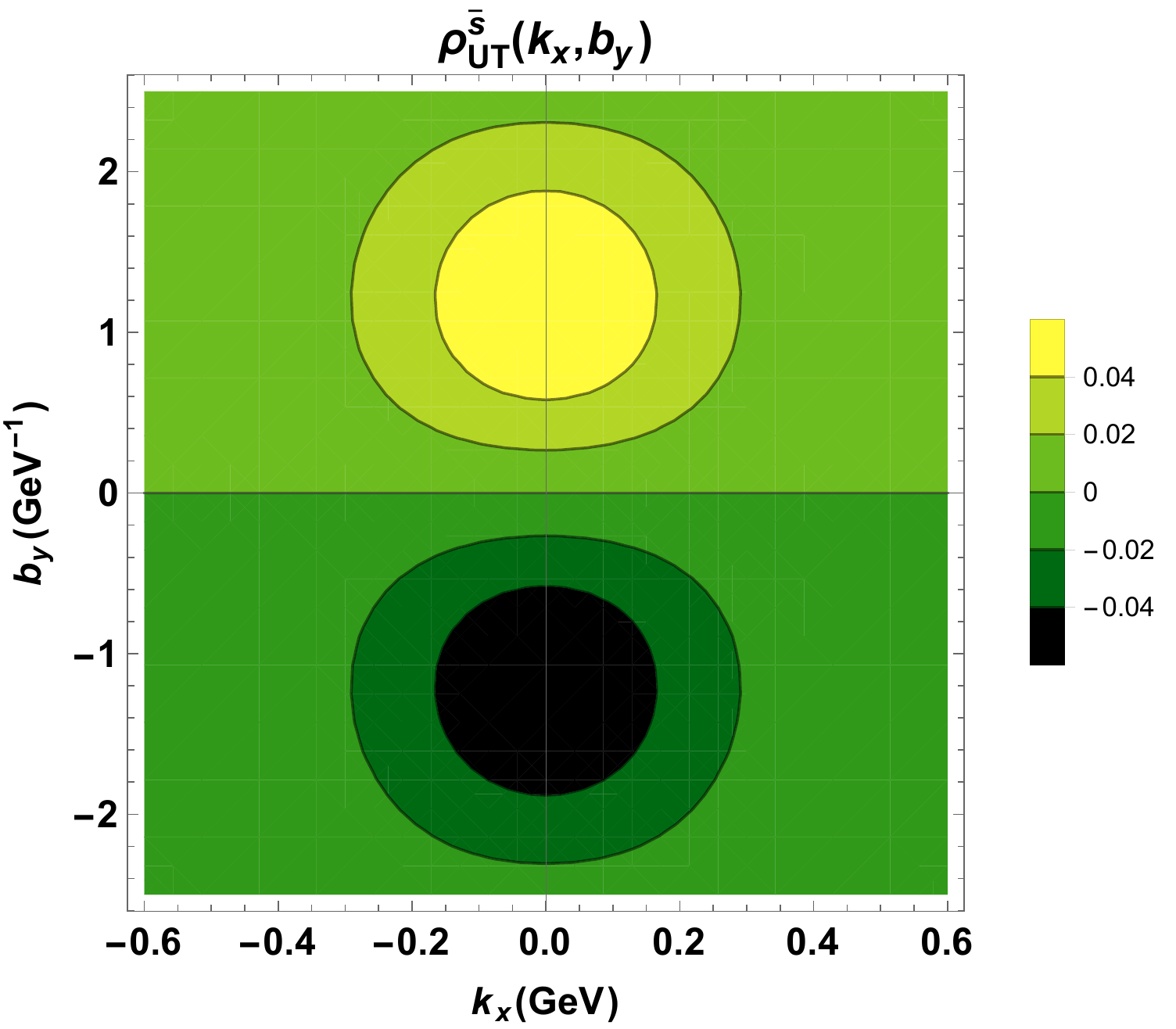}
\end{minipage}
\caption{The unpolarized-transverse Wigner distribution $\rho^j_{UT}$ of (i) $u$ quark (left panel), and (ii) $\bar{s}$ quark (right panel) for kaon in the impact-parameter plane, the transverse momentum plane, and the mixed plane.}
\label{ut}
\end{figure*}
The explicit expressions for quark Wigner distributions in kaon using Eqs. (\ref{wigner}), (\ref{unpolarized_overlap}), (\ref{longi_polarized_overlap}) and (\ref{trans_polarized_overlap}) are given as
\begin{eqnarray}
\rho^{u}_{UU}({\bf b_\perp},{\bf k_\perp},x)&=&\frac{1}{ 16 \pi^3}\int \frac{d\Delta_x d\Delta_y}{(2\pi)^2} {\cos}(\Delta_x b_x+ \Delta_y b_y)\nonumber\\
&&\times \Big[{\bf k}_\perp^2-(1-x)^2 \frac{{\bf \Delta}_\perp^2}{4}\nonumber\\
&&+((1-x)m_1+x m_2)^2\Big] \nonumber\\
&&\times \frac{\varphi_u^{\dagger}(x,{\bf k}''_\perp)\varphi_u (x, {\bf k}'_\perp)}{\sqrt{{\bf k}_\perp''^2+l_u^2}{\sqrt{{\bf k}_\perp'^2+l_u^2}}},\label{uu_u}
\end{eqnarray}
\begin{eqnarray}
\rho^{u}_{UL}({\bf b_\perp},{\bf k_\perp},x)&=&\frac{1}{ 16 \pi^3}\int \frac{d\Delta_x d\Delta_y}{(2\pi)^2} {\sin}(\Delta_x b_x+ \Delta_y b_y) \nonumber\\
&&\times (1-x)(k_y \Delta_x-k_x \Delta_y)\nonumber\\
&&\times \frac{\varphi_u^{\dagger}(x,{\bf k}''_\perp)\varphi_u (x, {\bf k}'_\perp)}{\sqrt{{\bf k}_\perp''^2+l_u^2}{\sqrt{{\bf k}_\perp'^2+l_u^2}}},\label{ul_u}\\
\rho^{u}_{UT}({\bf b_\perp},{\bf k_\perp},x)&=&-\frac{1}{ 16 \pi^3}\int \frac{d\Delta_x d\Delta_y}{(2\pi)^2} {\sin}(\Delta_x b_x+ \Delta_y b_y) \nonumber\\
&&\times ((1-x)m_1+x m_2)(1-x)\Delta_y\nonumber\\
&&\times \frac{\varphi_u^{\dagger}(x,{\bf k}''_\perp)\varphi_u (x, {\bf k}'_\perp)}{\sqrt{{\bf k}_\perp''^2+l_u^2}{\sqrt{{\bf k}_\perp'^2+l_u^2}}}\label{ut_u}.
\end{eqnarray}

The flavor decompositions of valence partons in kaon are associated with each other through the relation
\begin{equation}
\rho^{u}({\bf b}_\perp,{\bf k}_\perp, x, m_1,m_2) = -\rho^{\bar{s}}({\bf b}_\perp, -{\bf k}_\perp, -x,m_2,m_1).
\label{quark_antiquark_relation}
\end{equation}
The explicit expressions for Wigner distributions of antiquark after implementing Eq. (\ref{quark_antiquark_relation}) and the  conditions, we get 
\begin{eqnarray}
\rho^{\bar{s}}_{UU}({\bf b_\perp},{\bf k_\perp},x)&=&-\frac{1}{ 16 \pi^3}\int \frac{d\Delta_x d\Delta_y}{(2\pi)^2} {\cos}(\Delta_x b_x+ \Delta_y b_y)\nonumber\\
&&\times \Big[{\bf k}_\perp^2-(1+x)^2 \frac{{\bf \Delta}_\perp^2}{4}\nonumber\\
&&+((1+x)m_2-x m_1)^2\Big] \nonumber\\
&&\times \frac{\varphi_{\bar{s}}^{\dagger}(x,{\bf k}''_\perp)\varphi_{\bar{s}} (x, {\bf k}'_\perp)}{\sqrt{{\bf k}_\perp''^2+l_{\bar{s}}^2}{\sqrt{{\bf k}_\perp'^2+l_{\bar{s}}^2}}}\label{uu_s},
\end{eqnarray}
\begin{eqnarray}
\rho^{\bar{s}}_{UL}({\bf b_\perp},{\bf k_\perp},x)&=&\frac{1}{ 16 \pi^3}\int \frac{d\Delta_x d\Delta_y}{(2\pi)^2} {\sin}(\Delta_x b_x+ \Delta_y b_y) \nonumber\\
&&\times (1+x)(k_x \Delta_y-k_y \Delta_x)\nonumber\\
&&\times \frac{\varphi_{\bar{s}}^{\dagger}(x,{\bf k}''_\perp)\varphi_{\bar{s}} (x, {\bf k}'_\perp)}{\sqrt{{\bf k}_\perp''^2+l_{\bar{s}}^2}{\sqrt{{\bf k}_\perp'^2+l_{\bar{s}}^2}}},\label{ul_s}\\
\rho^{\bar{s}}_{UT}({\bf b_\perp},{\bf k_\perp},x)&=&\frac{1}{ 16 \pi^3}\int \frac{d\Delta_x d\Delta_y}{(2\pi)^2} {\sin}(\Delta_x b_x+ \Delta_y b_y) \nonumber\\
&&\times ((1+x)m_2-x m_1)(1+x)\Delta_y\nonumber\\
&&\times \frac{\varphi_{\bar{s}}^{\dagger}(x,{\bf k}''_\perp)\varphi_{\bar{s}} (x, {\bf k}'_\perp)}{\sqrt{{\bf k}_\perp''^2+l_{\bar{s}}^2}{\sqrt{{\bf k}_\perp'^2+l_{\bar{s}}^2}}}\label{ut_s}.
\end{eqnarray}
Since the Wigner distributions are interpreted with $\zeta=0$,  the initial and final state momenta and longitudinal momentum fractions can be used by putting $\zeta=0$  in Eqs. (\ref{active_quark_momenta1})-(\ref{conditions}). 

Further, we can obtain the purely transverse Wigner distributions by integrating them over $x$ as
\begin{eqnarray}
\rho_{UX}({\bf b}_\perp,{\bf k}_\perp)\equiv \int dx \ \rho_{UX}({\bf b}_\perp,{\bf k}_\perp,x),
\label{x_integration}
\end{eqnarray}
where $X$ denotes the different polarizations of parton (quark or antiquark) and $U$ stands for kaon being unpolarized. By taking  certain limits, Wigner distributions in impact-parameter plane (${\bf b}_\perp$-plane) and the transverse momentum plane (${\bf k}_\perp$-plane) can be explained. In addition, the distributions in mixed plane, commonly known as mixed probability densities $\rho(b_x,k_y)$ or $\rho(k_x,b_y)$, can also be explained through
\begin{eqnarray}
\int db_y dk_x \rho_{UX}({\bf b}_\perp,{\bf k}_\perp)&=&\rho_{UX}(b_x,k_y),\\
{\rm or}\nonumber\\
\int db_x dk_y \rho_{UX}({\bf b}_\perp,{\bf k}_\perp)&=&\rho_{UX}(k_x,b_y).
\end{eqnarray}
The variables surviving in mixed distribution, which actually show the correlation between the transverse momentum and transverse co-ordinate of quark (antiquark), are not protected by uncertainty principle. Therefore, they are known to describe the probabilistic distributions.

In Figs. \ref{uu}, \ref{ul} and \ref{ut}, we present respectively the results of Wigner distributions of unpolarized, longitudinally-polarized and transversely-polarized $u$ quark as well as $\bar s$ quark in the  unpolarized kaon. In our numerical calculations, the active quark (antiquark) is considered to be the $u$ quark ($\bar{s}$ quark), and the spectator being the $\bar{s}$ quark ($u$ quark). Corresponding to the Eqs. (\ref{uu_u})-(\ref{ut_u}) for $u$ and Eqs. (\ref{uu_s})-(\ref{ut_s}) for $\bar{s}$, we plot the results in the impact-parameter plane, momentum plane and mixed plane. In the impact-parameter plane, we show the the plots of $\rho_{UU}$ (Figs. \ref{uu}(a) and (b)), $\rho_{UL}$ (Figs. \ref{ul}(a) and (b)) and $\rho_{UT}$ (Figs. \ref{ut}(a) and (b)) by taking the fixed transverse momentum as ${\bf k}_\perp=k_\perp \hat{e}_y$ where $k_\perp=0.2$ $GeV$. Similarly, in the transverse momentum plane we plot  $\rho_{UU}$ (Figs. \ref{uu}(c) and (d)), $\rho_{UL}$ (Figs. \ref{ul}(c) and (d)) and $\rho_{UT}$ (Figs. \ref{ut}(c) and (d)) by  choosing the impact-parameter co-ordinate along $\hat{e}_y$, i.e. ${\bf b}_\perp=b_\perp \hat{e}_y$ and $b_\perp =  0.4$ $GeV^{-1}$. Further, for  the mixed plane we plot  $\rho_{UU}$ (Figs. \ref{uu}(e) and (f)), $\rho_{UL}$ (Figs. \ref{ul}(e) and (f)) and $\rho_{UT}$ (Figs. \ref{ut}(e) and (f))

From Fig. \ref{uu}, where the unpolarized Wigner distribution in the impact-parameter plane, momentum plane and mixed plane have been plotted for $u$ quark on the left panel and for $\bar s$ quark on the right panel, we observe that in the  case of $\bar{s}$ the distribution is more concentrated at the center in contrast to the $u$ quark.  The distributions in both cases have opposite behavior. From Figs. \ref{uu}(a) and (b), the distribution in the impact-parameter plane for the case of $\bar{s}$ does not remain circularly symmetric at the higher values of transverse co-ordinates of ${\bf b}_\perp$-plane, whereas in case of $u$ quark, it shows symmetry. This is due to the effect of heavier mass of the active $\bar{s}$ quark. We observe that the probability of rotation of quark (antiquark) to move in either clockwise direction or anti-clockwise direction is same. Further, in Figs. \ref{uu}(c) and \ref{uu}(d), the distributions $\rho_{UU}$ are presented in the momentum plane. The distributions in this case are observed to be circularly symmetric for both $u$ and $\bar{s}$, but the concentrations are in opposite direction.  In the case of distributions in the mixed plane $\rho{(b_x,k_y)}$, the $u$ and $\bar{s}$ quark distributions are axially symmetric. The spread is however, more dispersed in the case $u$ quark as compared to the $\bar{s}$ quark. At the center i.e. at $b_x=k_y=0$, the probability density for $u$ quark is maximum while in case of $\bar{s}$ quark it is minimum. The distribution $\rho_{UU}$ is related to the unpolarized TMD $f_1$ and unpolarized GPD $H$, when integrated upon certain limits.

The results of Wigner distributions for longitudinally-polarized quark (antiquark) in unpolarized kaon viz. $\rho_{UL}$ are shown in Fig. \ref{ul}. In the impact-parameter plane, the distributions show the dipole behavior and the behavior is positive for $b_x>0$ for both quark and antiquark, as shown in Fig. \ref{ul}(a) and (b). However, in the transverse momentum plane, shown in Fig. \ref{ul}(c) and (d), it reverses the direction and $\rho_{UL}$ is positive for $b_x<0$. In the mixed plane ($b_x, k_y$), we observe a quadrupole behaviour of the distributions in both the cases with same polarities as shown in Figs. \ref{ul}(e) and (f). A positive distribution is observed in the region where the product of $b_x$ and $k_y$ is greater, whereas it is negative where the product is negative. For the probability density we have
\begin{eqnarray}
\rho_{UL}(b_x,k_y) \left\{
\begin{array}{@{}ll@{}}
>0 & \text{if}\ b_x*k_y>0 \\
<0 & \text{if}\ b_x*k_y<0
\end{array}\right.
\end{eqnarray}
No TMD or GPD is present corresponding to $\rho_{UL}$. Generally, this Wigner distribution is related to orbital angular momentum (OAM), but for the pseudoscalar meson, the net quark and antiquark spin and OAM are zero. 

We plot the upolarized-transverse distribution $\rho^j_{UT}$ of $u$ and $\bar{s}$ respectively on the left and right panel in Fig. \ref{ut}. The distribution $\rho^j_{UT}$ sheds light on the quark/antiquark distribution when the quark is transversely polarized in an unpolarized kaon. Here $j$ describes the polarization direction of quark (antiquark). We take the quark/antiquark polarization along $x$-axis. We notice a strong correlation between the transverse co-ordinate and the perpendicular polarization direction of quark/antiquark. From Eqs. (\ref{ut_u}) and (\ref{ut_s}), we observe that the distribution vanishes if we consider the quark/antiquark spin direction along the direction of quark/antiquark transverse co-ordinate. The distribution displayed in the impact-parameter plane shows a dipolar behavior for both quark and antiquark but with opposite polarities. This is clear from Figs. \ref{ut}(a) and (b). Further, in the transverse momentum plane (Figs. \ref{ut}(c) and (d)), $\rho^1_{UT}$ is observed to be more focused at the center ($p_x=p_y=0$) in the case of $\bar{s}$ whereas it is more extended to the periphery in case of $u$ quark distribution. The distribution shows a dipolar behaviour in mixed space due to its symmetry in the momentum plane as well as in in the impact-parameter plane. Here, we choose the plane ($k_x,b_y$) instead of ($b_x,k_y$), because it leads to the Dirac Delta function $\delta(\Delta_y)$ in case of $\rho^j_{UT}$, when the Fourier transformation is taken. $\rho^j_{UT}$ relates to the T-odd Boer-Mulder TMD $h^\perp_1$ and T-odd GPD $E_T$ at TMD limit and IPD limit respectively. But we are not able to extract any TMD or GPD in the present work, because we have not considered any  gluon contribution.
\section{VI. Generalized transverse momentum-dependent distributions of $u$  and $\bar{s}$ quark in kaon}
The twist-2 GTMDs related to unpolarized pseudoscalar meson with spin$-0$ are connected with Wigner correlator or operator as \cite{gtmd1}
\begin{eqnarray}
\hat{W}^{[\gamma^+]}&=&F_1,\\
\hat{W}^{[\gamma^+ \gamma_5]}&=&\frac{i \epsilon_\perp^{ij} k_\perp^i\Delta_\perp^j}{M^2}\tilde{G}_1,\\
\hat{W}^{[i\sigma^{j+}\gamma_5]}&=&\frac{i \epsilon_\perp^{ij}k_\perp^i}{M}H_1^k+\frac{i\epsilon_\perp^{ij}\Delta_\perp^i}{M}H_1^\Delta,
\end{eqnarray}
with the anti-symmetric tensor $\epsilon_\perp^{ij}=\epsilon^{-+ij}$, $\epsilon^{0123}=1$ and $\sigma^{ab}=\frac{i}{2}[\gamma^a,\gamma^b]$. All the leading-twist GTMDs are  function of six variables and we have $(x,\zeta,{\bf k}_\perp^2,{\bf k}_\perp.{\bf \Delta}_\perp, {\bf \Delta}_\perp^2)$. There are 4 complex-valued twist-2 GTMDs in the case of spin-$0$ hadron, whereas in case of spin$-\frac{1}{2}$ there are 16. 

The explicit expressions of $u$ quark GTMDs for $\zeta \neq 0$ in kaon are evaluated from  
\begin{eqnarray}
F_1^{(u)}&=&\frac{1}{16\pi^3}\bigg[{\bf k}_\perp^2-(1-x')(1-x'')\frac{{\bf\Delta}_\perp^2}{4}\nonumber\\
&&+\frac{x''-x'}{2}(k_x \Delta_x+k_y \Delta_y)+\mathcal{M}_u'\mathcal{M}_u''\bigg]\nonumber\\
&&\times \frac{\varphi_u^{\dagger}(x'',{\bf k}''_\perp)\varphi_u(x', {\bf k}'_\perp)}{\sqrt{{\bf k}_\perp''^2+{l}_u''^2}{\sqrt{{\bf k}_\perp'^2+{l}_u'^2}}}
\label{f1_u}\\
\tilde{G}^{(u)}_1&=&-\frac{ M^2}{16\pi^3} \frac{(2-x'-x'')}{2} \nonumber\\
&&\times \frac{\varphi_u^{\dagger}(x'',{\bf k}''_\perp)\varphi_u (x', {\bf k}'_\perp)}{\sqrt{{\bf k}_\perp''^2+l_u''^2}{\sqrt{{\bf k}_\perp'^2+l_u'^2}}}
\label{g1_u}\\
H_1^{k(u)}&=&-\frac{M}{16\pi^3} \big[\mathcal{M}_u'-\mathcal{M}_u''\big]\frac{\varphi_u^{\dagger}(x'',{\bf k}''_\perp)\varphi_u(x', {\bf k}'_\perp)}{\sqrt{{\bf k}_\perp''^2+l_u''^2}{\sqrt{{\bf k}_\perp'^2+l_u'^2}}},\nonumber\\
\label{h1k_u}\\
H_1^{\Delta(u)}&=& \frac{M}{16\pi^3}\bigg[\mathcal{M}_u'\frac{(1-x'')}{2}+\mathcal{M}_u''\frac{(1-x')}{2}\bigg]\nonumber\\
&&\times\frac{\varphi_u^{\dagger}(x'',{\bf k}''_\perp)\varphi_u (x', {\bf k}'_\perp)}{\sqrt{{\bf k}_\perp''^2+l_u''^2}{\sqrt{{\bf k}_\perp'^2+l_u'^2}}}
\label{h1delta_u}.
\end{eqnarray}
The $\bar{s}$ quark GTMDs in kaon are related to the $u$ quark distributions as Eqs. (\ref{gpd_flavor_decomposition}) and (\ref{quark_antiquark_relation}). We have
\begin{widetext}
\begin{eqnarray}
F^{u}(x,\zeta,{\bf k}^2_\perp, {\bf k}_\perp.{\bf \Delta}_\perp,{\bf \Delta}_\perp^2,m_1,m_2)=-F^{\bar{s}}(-x,\zeta,{\bf k}^2_\perp, -{\bf k}_\perp.{\bf \Delta}_\perp,{\bf \Delta}_\perp^2,m_2,m_1).
\label{gtmd_quark_anti-quark_relation}
\end{eqnarray}
\end{widetext}
The $\bar{s}$ quark twist-2 GTMDs for $\zeta \neq 0$ are  explicitly evaluated from
\begin{eqnarray}
F_1^{(\bar{s})}&=&-\frac{1}{16\pi^3}\bigg[{\bf k}_\perp^2-(1-x')(1-x'')\frac{{\bf\Delta}_\perp^2}{4}\nonumber\\
&&+\frac{x''-x'}{2}(k_x \Delta_x+k_y \Delta_y)+\mathcal{M}_{\bar{s}}'\mathcal{M}_{\bar{s}}''\bigg]\nonumber\\
&&\times \frac{\varphi_{\bar{s}}^{\dagger}(x'',{\bf k}''_\perp)\varphi_{\bar{s}}(x', {\bf k}'_\perp)}{\sqrt{{\bf k}_\perp''^2+{l}_{\bar{s}}''^2}{\sqrt{{\bf k}_\perp'^2+{l}_{\bar{s}}'^2}}},\label{f1_s}\\
\tilde{G}^{(\bar{s})}_1&=&\frac{M^2}{16\pi^3} \frac{(2-x'-x'')}{2} \nonumber\\
&&\times \frac{\varphi_{\bar{s}}^{\dagger}(x'',{\bf k}''_\perp)\varphi_{\bar{s}}(x', {\bf k}'_\perp)}{\sqrt{{\bf k}_\perp''^2+l_{\bar{s}}''^2}{\sqrt{{\bf k}_\perp'^2+l_{\bar{s}}'^2}}},
\label{g1_s}\\
H_1^{k(\bar{s})}&=&\frac{M}{16\pi^3} \big[\mathcal{M}_{\bar{s}}'-\mathcal{M}_{\bar{s}}''\big]\frac{\varphi_{\bar{s}}^{\dagger}(x'',{\bf k}''_\perp)\varphi_{\bar{s}}(x', {\bf k}'_\perp)}{\sqrt{{\bf k}_\perp''^2+l_{\bar{s}}''^2}{\sqrt{{\bf k}_\perp'^2+l_{\bar{s}}'^2}}},\nonumber\\
\label{h1k_s}\\
H_1^{\Delta(\bar{s})}&=& -\frac{M}{16\pi^3}\bigg[\mathcal{M}_{\bar{s}}'\frac{(1-x'')}{2}+\mathcal{M}_{\bar{s}}''\frac{(1-x')}{2}\bigg]\nonumber\\
&&\times\frac{\varphi_{\bar{s}}^{\dagger}(x'',{\bf k}''_\perp)\varphi_{\bar{s}} (x', {\bf k}'_\perp)}{\sqrt{{\bf k}_\perp''^2+l_{\bar{s}}''^2}{\sqrt{{\bf k}_\perp'^2+l_{\bar{s}}'^2}}}
\label{h1delta_s}.
\end{eqnarray}
The details of the longitudinal momentum fractions and transverse momenta carried by quark (antiquark) have already been given in Section-III in the Eqs. (\ref{active_quark_momenta1})-(\ref{conditions}).

In the present work, we have presented the quark and antiquark GTMDs of kaon with respect to longitudinal momentum fraction carried by quark $x$ and antiquark $-x$. We have studied two cases: (a) the variation of GTMDs w.r.t $x$ for $\zeta \neq 0$ where $\zeta=-\frac{\Delta^+}{2 P^+}$ is the parameter corresponding to the transfer of momentum to the kaon in longitudinal direction, ii)  the variation of GTMDs w.r.t $x$ for $\zeta=0$. It is well known that the so-called \textit{mother distributions} i.e. GTMDs are reducible to the GPDs and TMDs after suitable integrations. For the case when $\zeta \neq 0$, only the GPDs can be extracted since there is no parameter corresponding to longitudinal direction in the case of TMDs (TMDs do not include $\zeta$ contribution). 

\begin{figure*}
\centering
\begin{minipage}[c]{1\textwidth}
(a)\includegraphics[width=.4\textwidth]{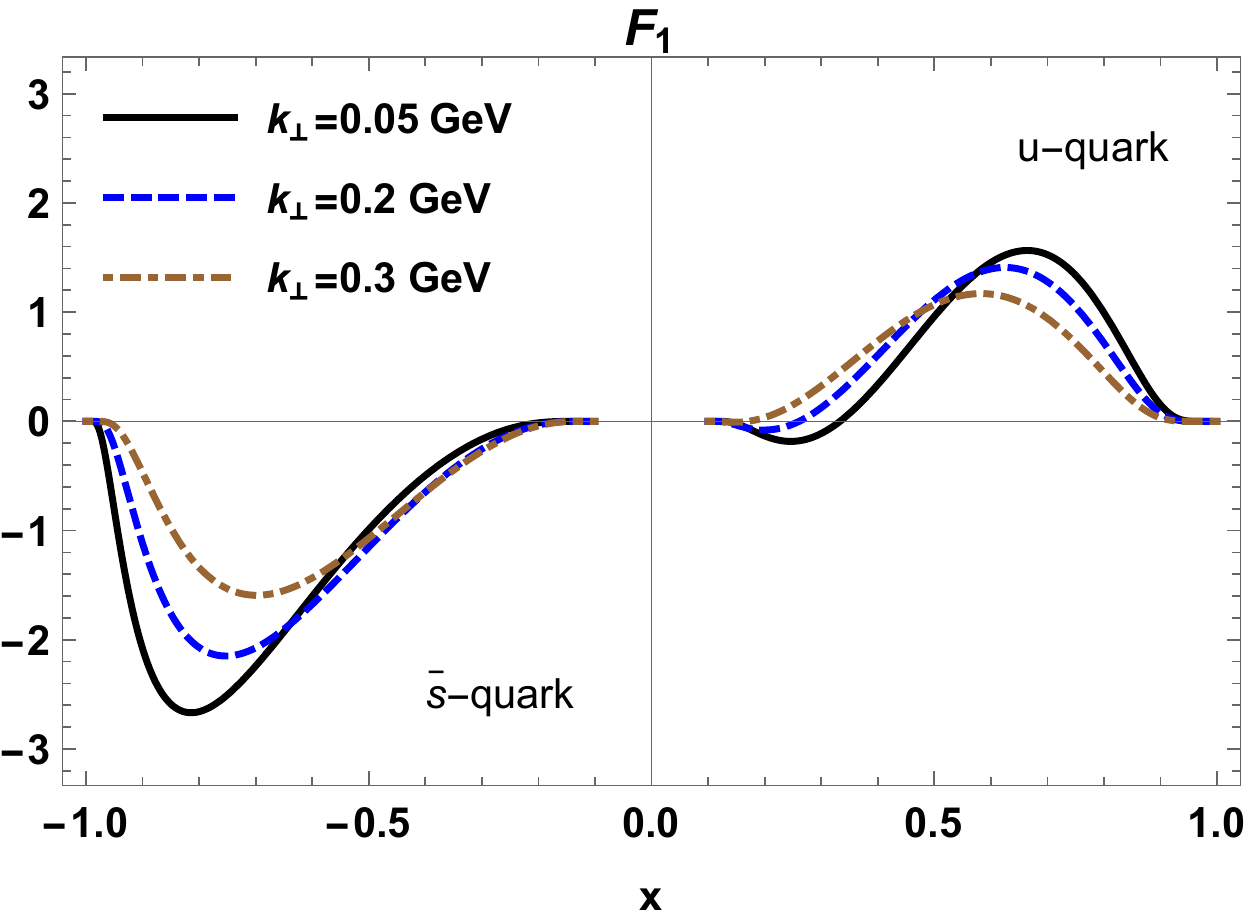}
(b)\includegraphics[width=.4\textwidth]{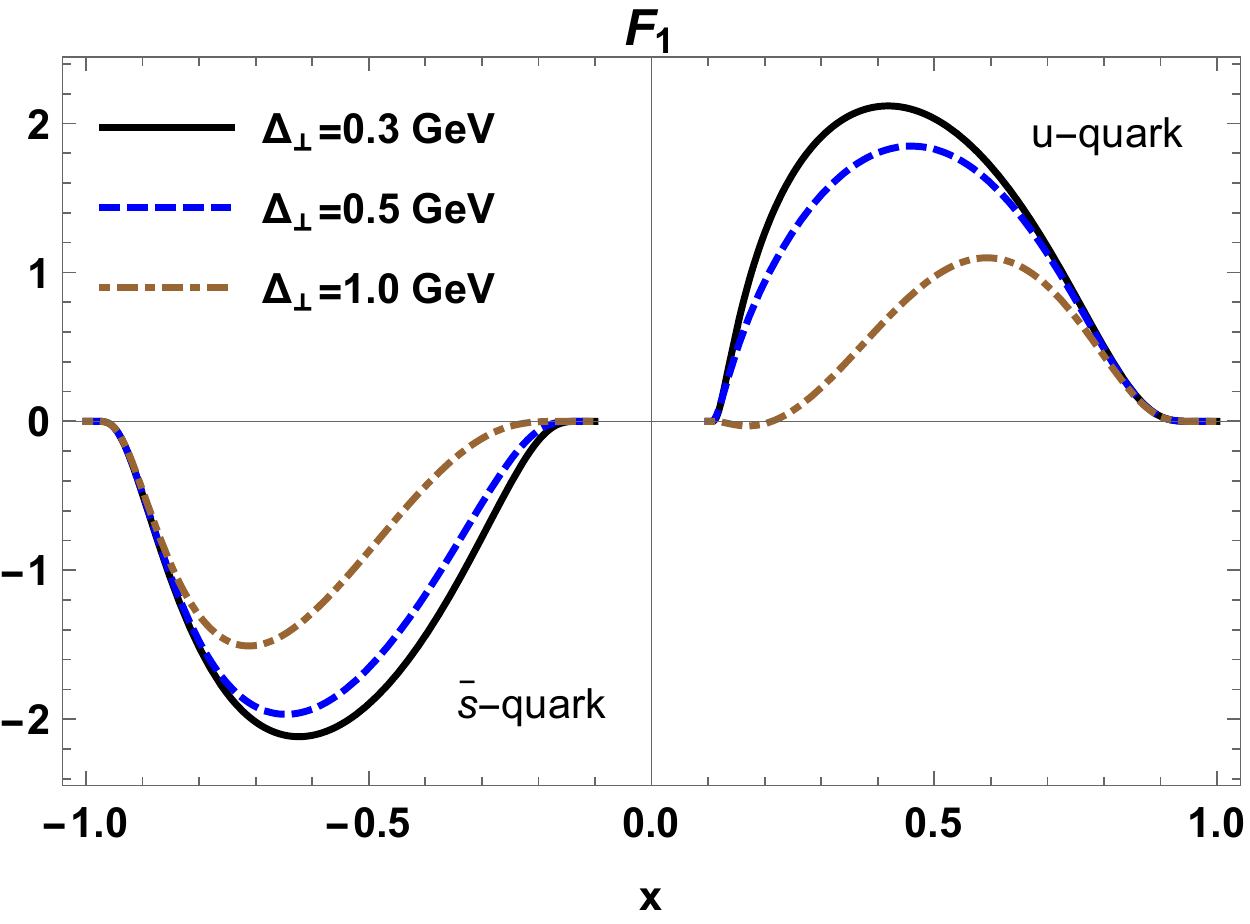}
\end{minipage}
\begin{minipage}[c]{1\textwidth}
(c)\includegraphics[width=.4\textwidth]{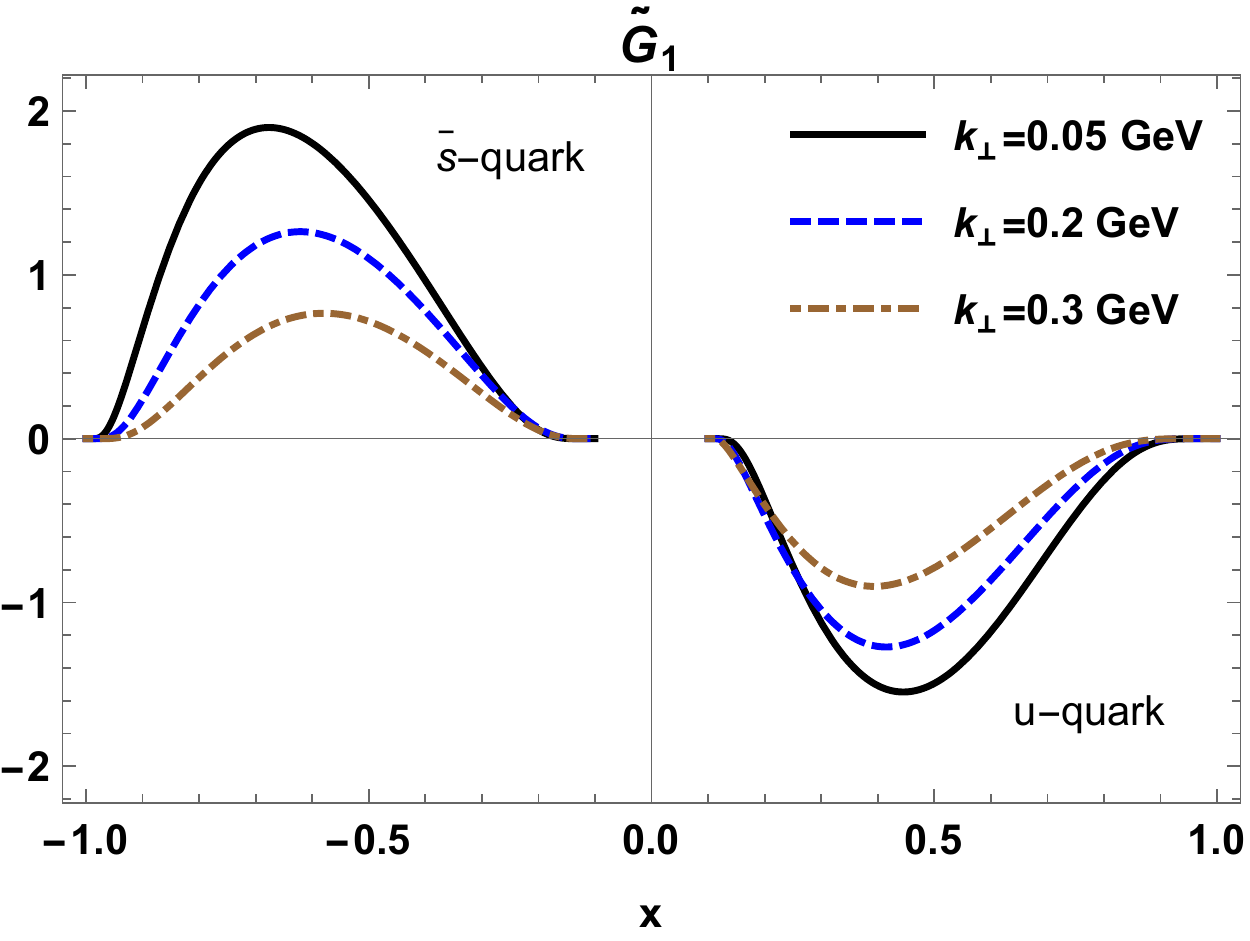}
(d)\includegraphics[width=.4\textwidth]{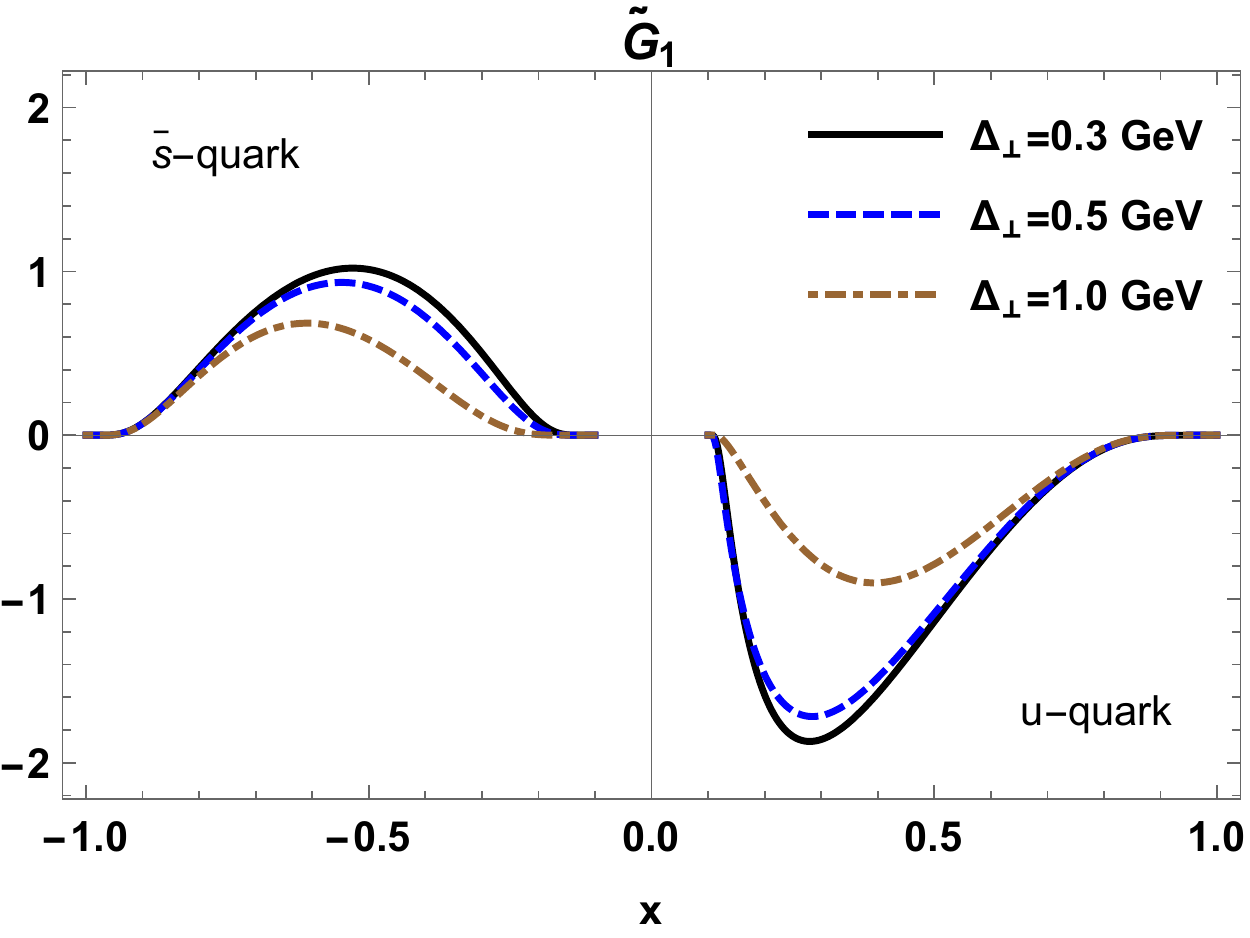}
\end{minipage}
\begin{minipage}[c]{1\textwidth}
(e)\includegraphics[width=.4\textwidth]{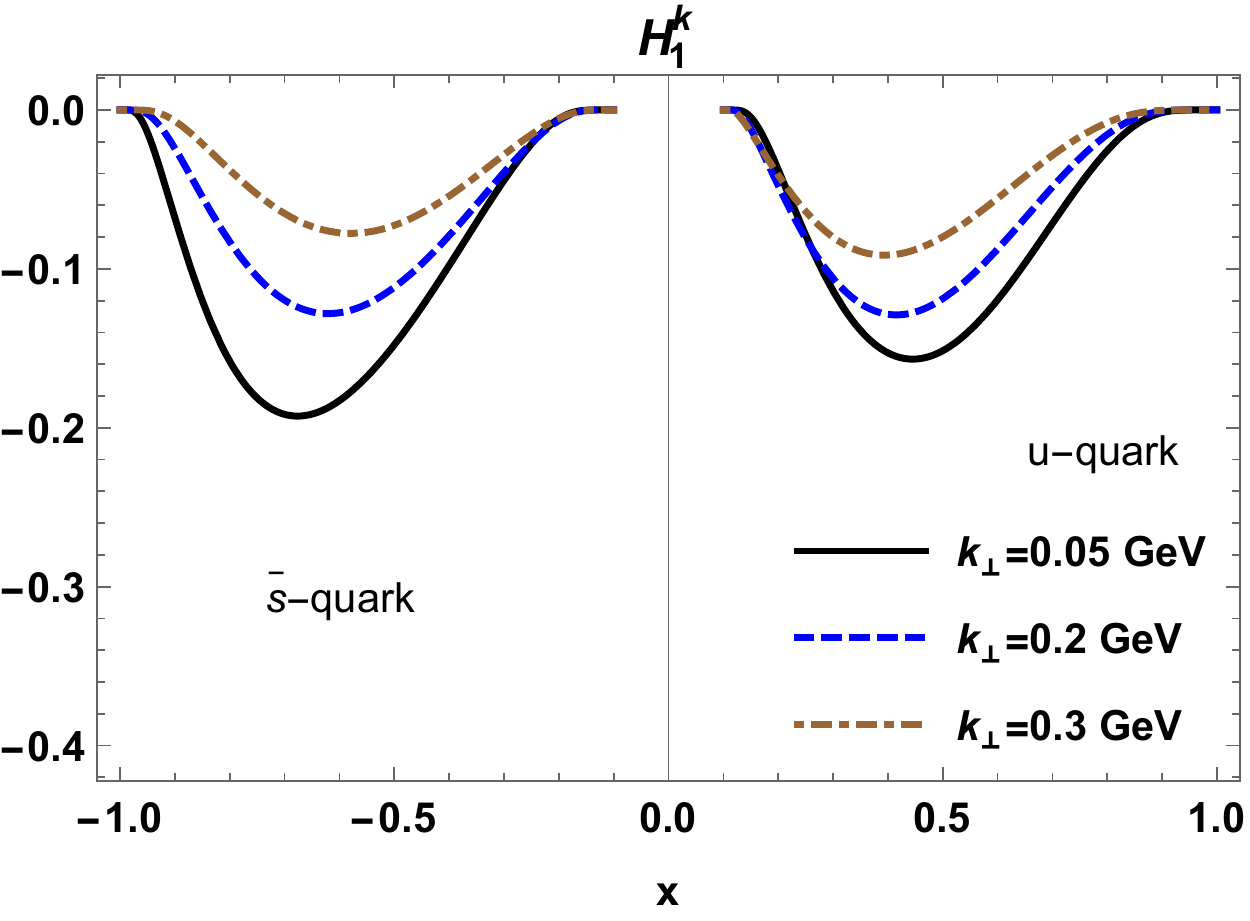}
(f)\includegraphics[width=.4\textwidth]{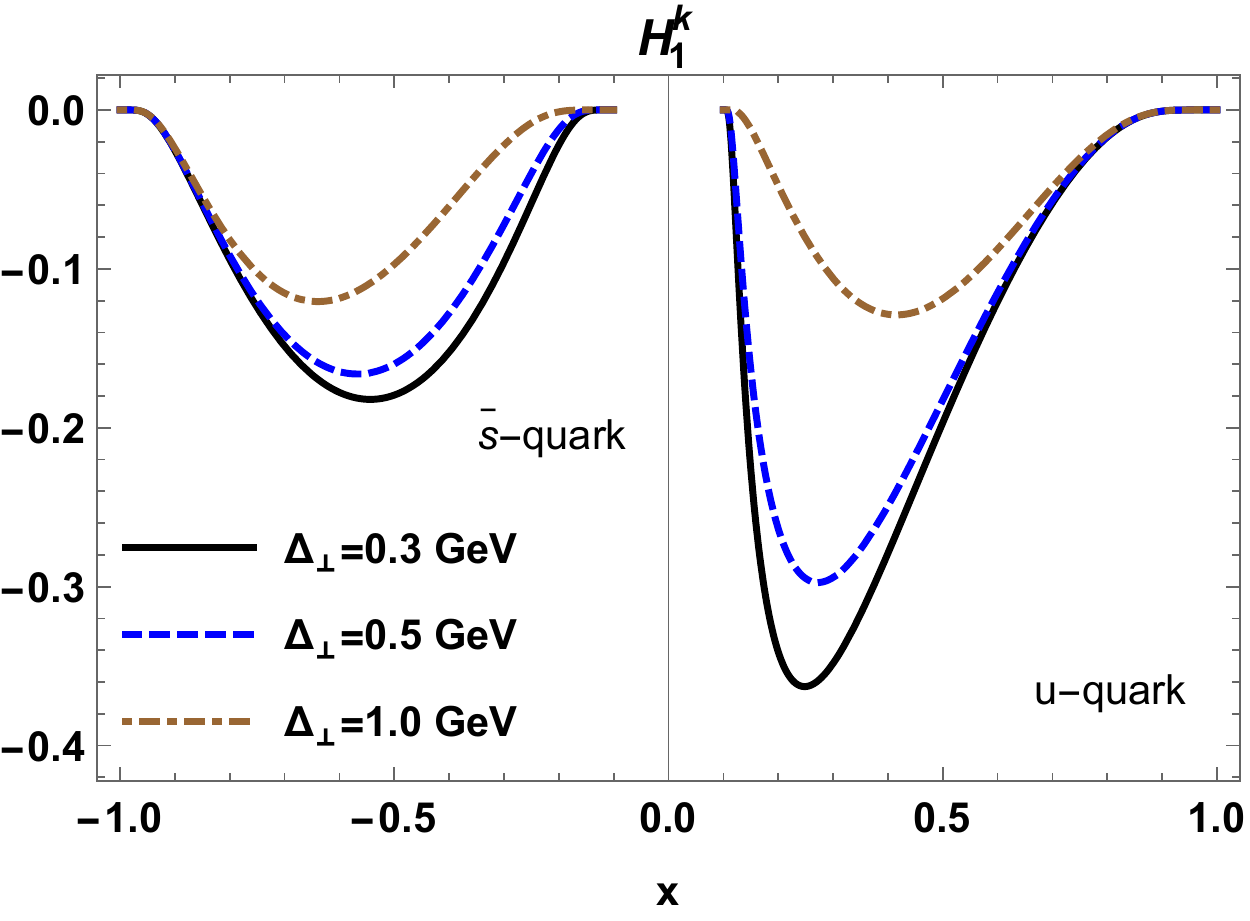}
\end{minipage}
\begin{minipage}[c]{1\textwidth}
(g)\includegraphics[width=.4\textwidth]{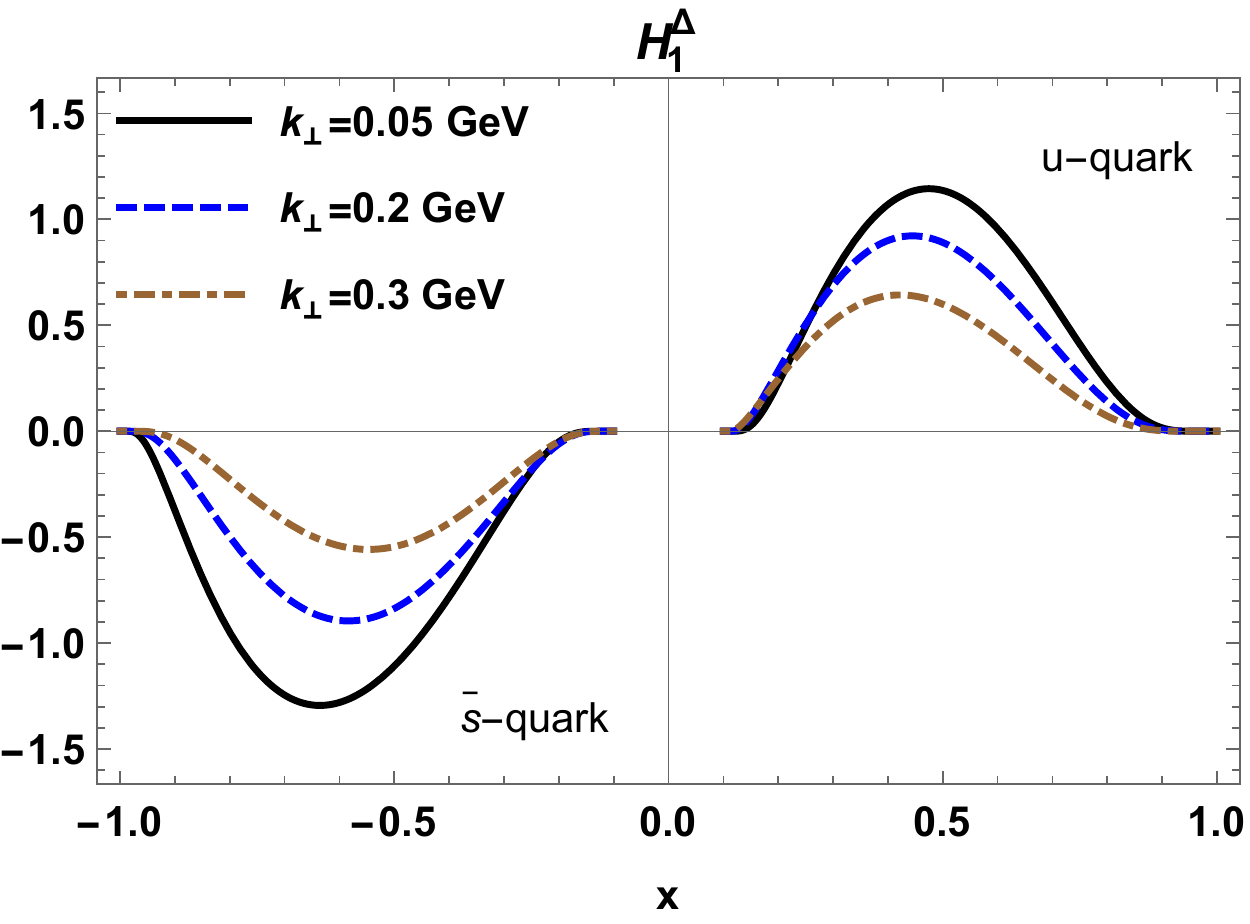}
(h)\includegraphics[width=.4\textwidth]{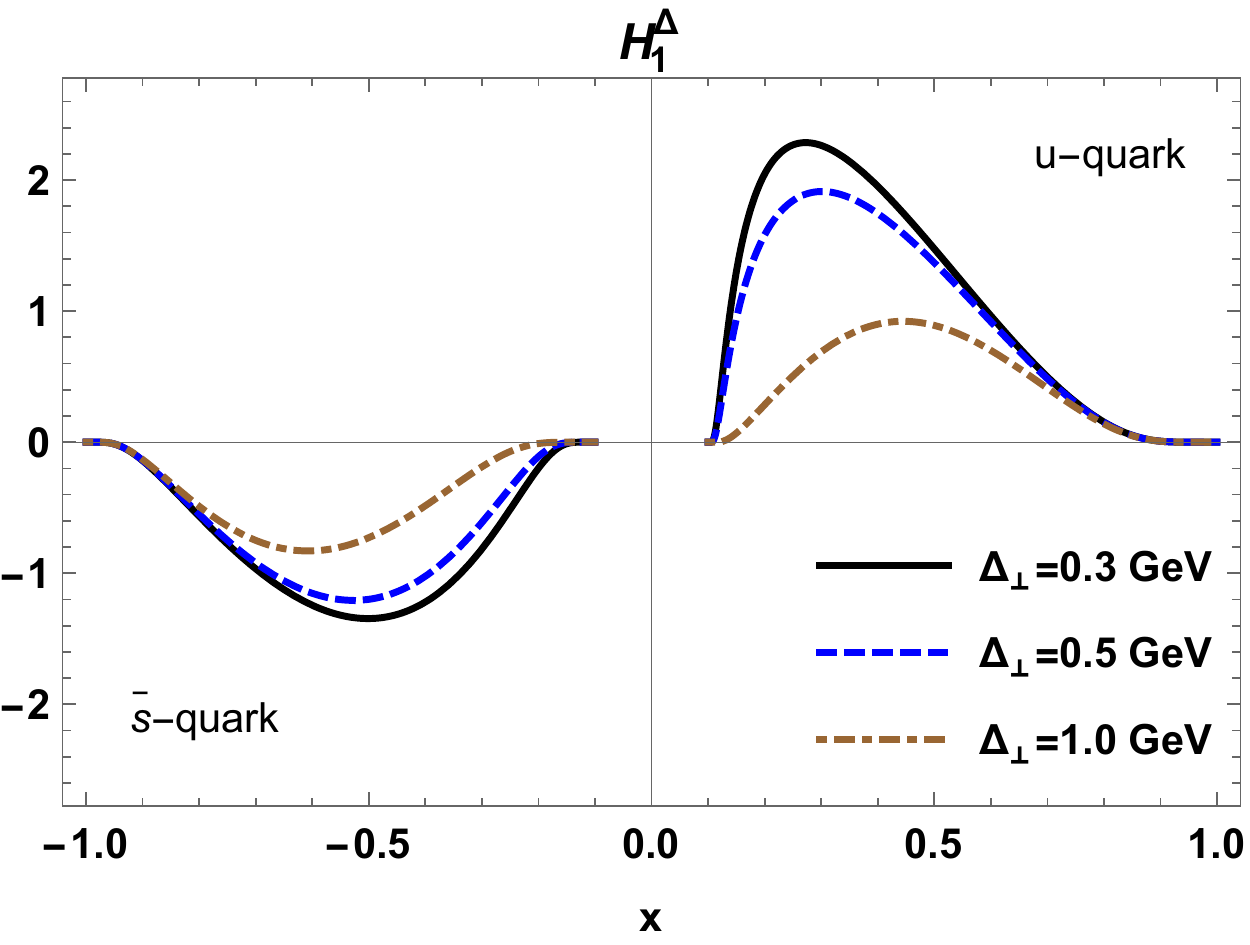}
\end{minipage}
\caption{The plots of GTMDs $F_1(x,\zeta,{\bf k}^2_\perp,{\bf k}_\perp.{\bf \Delta}_\perp,{\bf \Delta}_\perp^2)$, $\tilde{G}_1(x,\zeta,{\bf k}^2_\perp,{\bf k}_\perp.{\bf \Delta}_\perp,{\bf \Delta}_\perp^2)$, $H_1^k(x,\zeta,{\bf k}^2_\perp,{\bf k}_\perp.{\bf \Delta}_\perp,{\bf \Delta}_\perp^2)$ and $H_1^{\Delta}(x,\zeta,{\bf k}^2_\perp,{\bf k}_\perp.{\bf \Delta}_\perp,{\bf \Delta}_\perp^2)$ for $\zeta=0.1$ w.r.t $x$ for $u$ and $\bar{s}$ quarks (i) at different values of ${\bf k}_\perp$ with fixed ${\bf \Delta}_\perp = 1$ $GeV$ (left panel), and (ii) at different values of ${\bf \Delta}_\perp$ with fixed ${\bf k}_\perp=0.2$ $GeV$ (right panel).}
\label{gtmds_zeta}
\end{figure*}

In Fig. \ref{gtmds_zeta}, we have presented the variation of GTMDs of $u$ and $\bar{s}$ quarks in kaon corresponding to the Eqs. (\ref{f1_u})-(\ref{h1delta_u}) and Eqs. (\ref{f1_s})-(\ref{h1delta_s}) respectively. Since, the distributions have the support interval $-1<x<1$, we restrict ourselves in the DGLAP regions $(-1<x<-\zeta)$ for the antiquark  and for the $(\zeta<x<1)$ quark. In the left panel of Fig. \ref{gtmds_zeta}, we have plotted  the $u$ quark and $\bar{s}$ quark distributions $F_1,\tilde{G}_1,H_1^k {\ \rm and \ }H_1^\Delta$ for fixed values of $\zeta=0.1$ and ${\bf \Delta}_\perp=1$ $GeV$ and at different values of ${\bf k}_\perp$ at  ${\bf k}_\perp=0.05,0.2$ and $0.3$ $GeV$. In all the distributions, we observe that for both quark and antiquark the distribution peaks shift towards the lower values of $x$ and the  magnitude also lowers  down with the increasing the quark transverse momentum. For the cases of $F_1$, $\tilde{G}_1$ and $H_1^\Delta$, the effect of  $\bar{s}$ quark is in opposite direction as compared to the $u$ quark (Figs. \ref{gtmds_zeta}(a), (c) and (g)). 
The exception lies in the case of $H_1^k$  (Figs. \ref{gtmds_zeta}(e) and (f)) where the distributions remain negative for both quark and antiquark in kaon. In all the cases, at $x=\zeta$, when the momentum fraction carried by the quark (antiquark) is equal to the momentum transferred to the kaon in longitudinal direction, the distributions vanish. More the longitudinal momentum transfer to the kaon, more will be the shift in the peaks of distributions with increasing quark (antiquark) transverse momentum. On the right panel of Fig. \ref{gtmds_zeta} (Figs. \ref{gtmds_zeta}(b), (d), (f) and (h)), the quark (antiquark) GTMDs in kaon for $\zeta \neq 0$ are displayed at a constant value of ${\bf k}_\perp=0.2$ $GeV$ and by varying ${\bf \Delta}_\perp$. Unlike the distributions behaviour of $u$ and $\bar{s}$ quark by varying ${\bf k}_\perp$ and keeping  ${\bf \Delta}_\perp$ constant, the distribution peaks move towards the higher values of $x$ when the total momentum transferred to the kaon is increased. The magnitude of distribution however decreases when the total momentum transfer to the final state of kaon is more. Here, the magnitude of $\bar{s}$ quark is less as compared to $u$ quark. The difference in quark and antiquark distributions shifts for $\zeta \neq 0$ is due to the heavy on-shell mass of active antiquark as well as the momentum transfer to the kaon in longitudinal direction leading to the dependence of mass of active quark on the distributions when the total momentum transfer to the kaon is fixed.

\begin{figure*}
\centering
\begin{minipage}[c]{1\textwidth}
(a)\includegraphics[width=.4\textwidth]{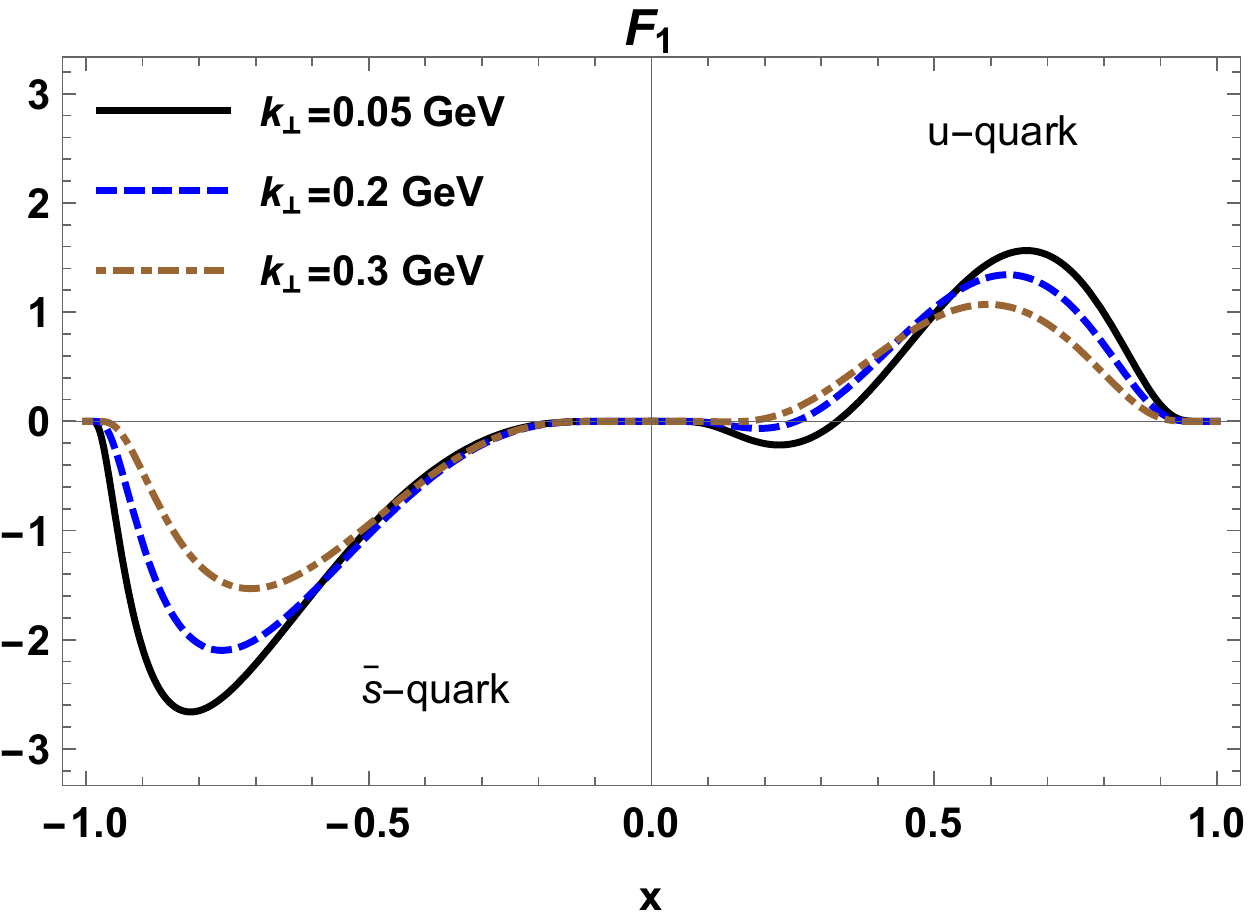}
(b)\includegraphics[width=.4\textwidth]{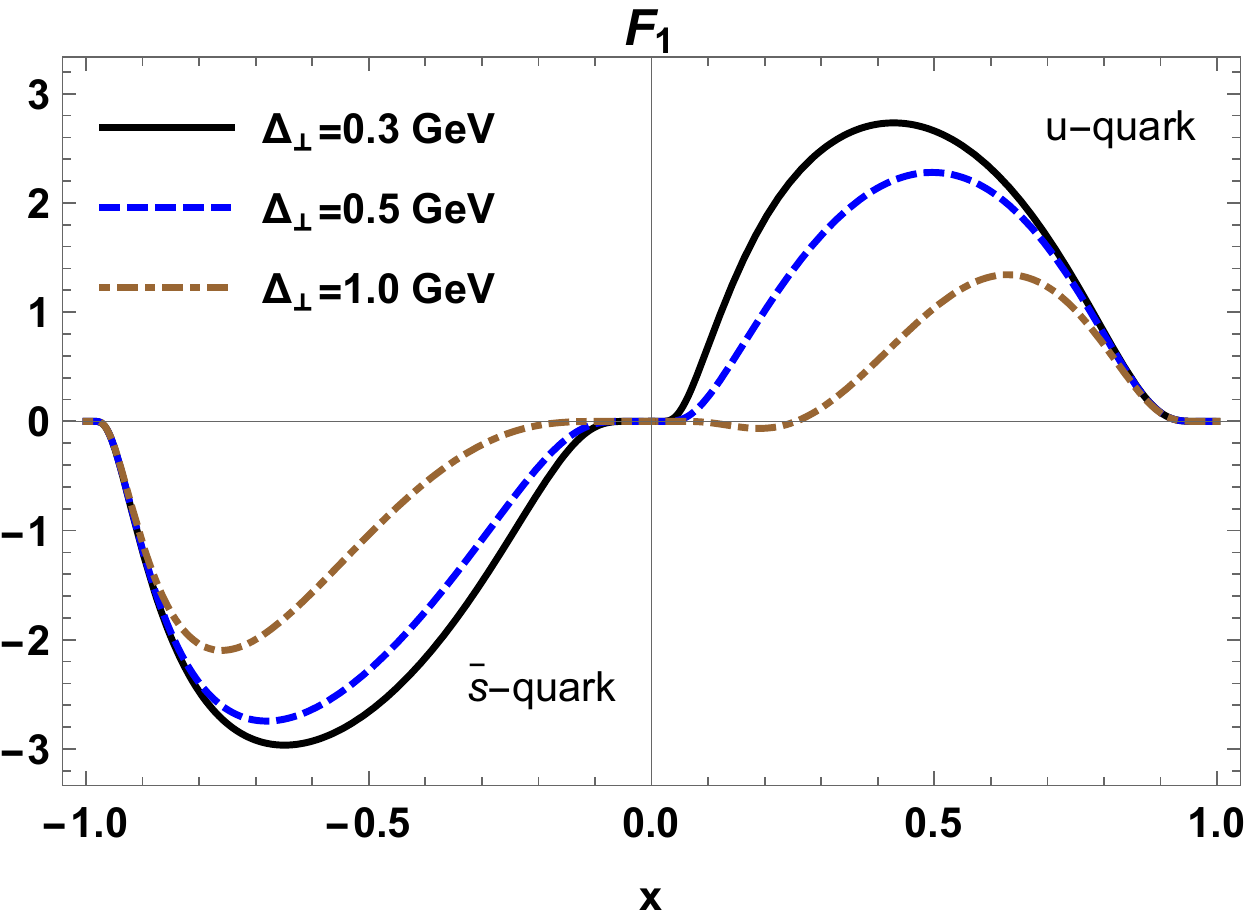}
\end{minipage}
\begin{minipage}[c]{1\textwidth}
(c)\includegraphics[width=.4\textwidth]{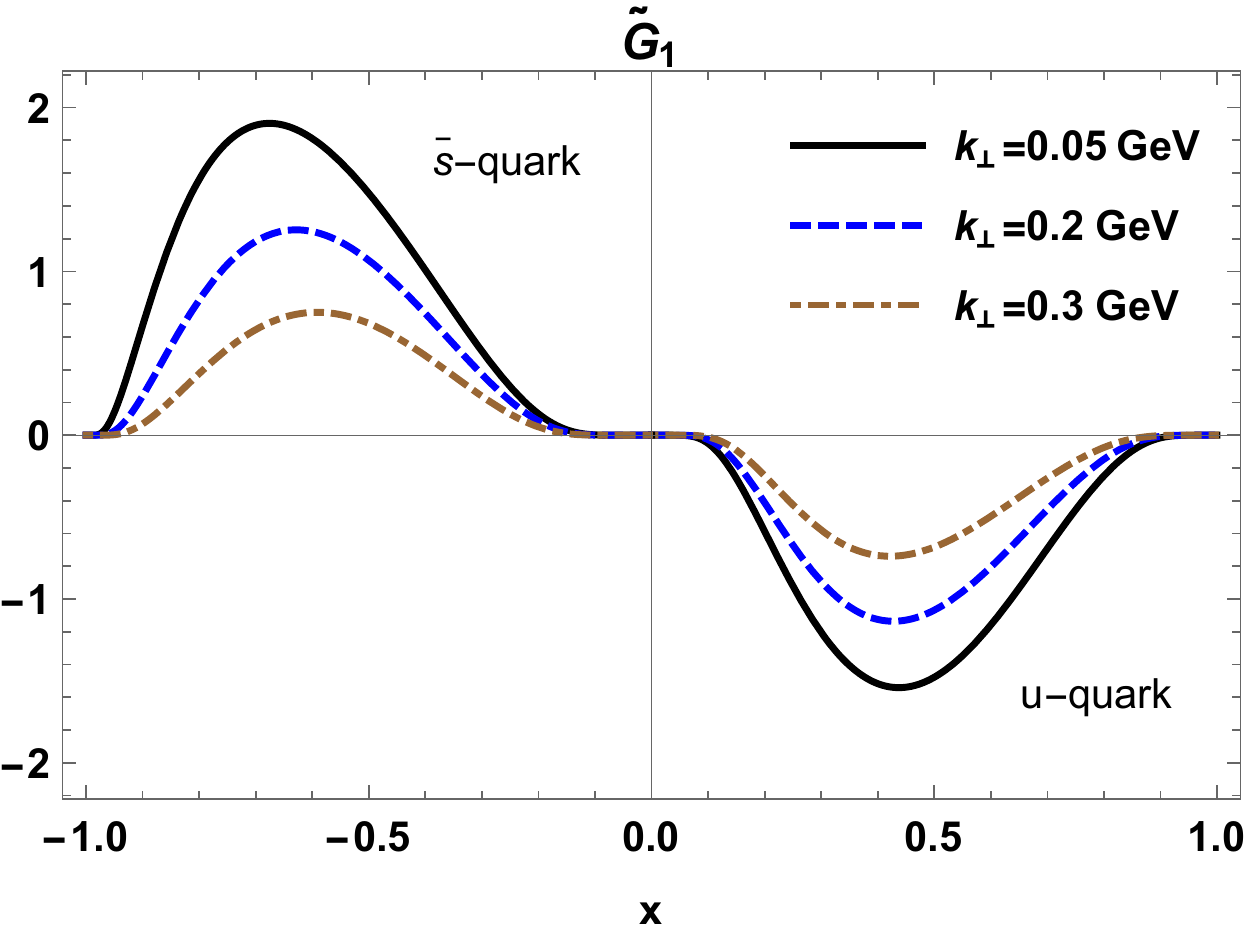}
(d)\includegraphics[width=.4\textwidth]{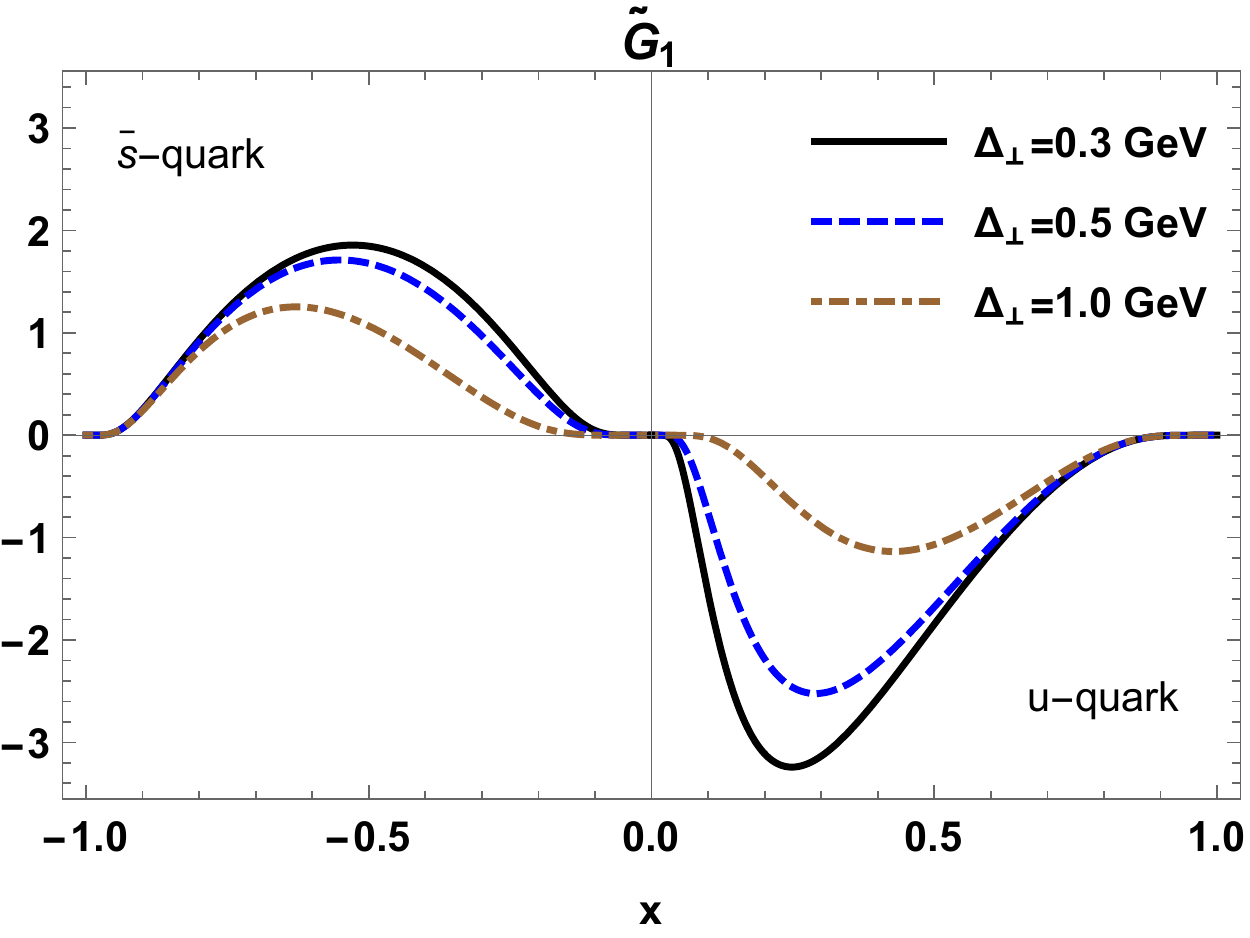}
\end{minipage}
\begin{minipage}[c]{1\textwidth}
(e)\includegraphics[width=.4\textwidth]{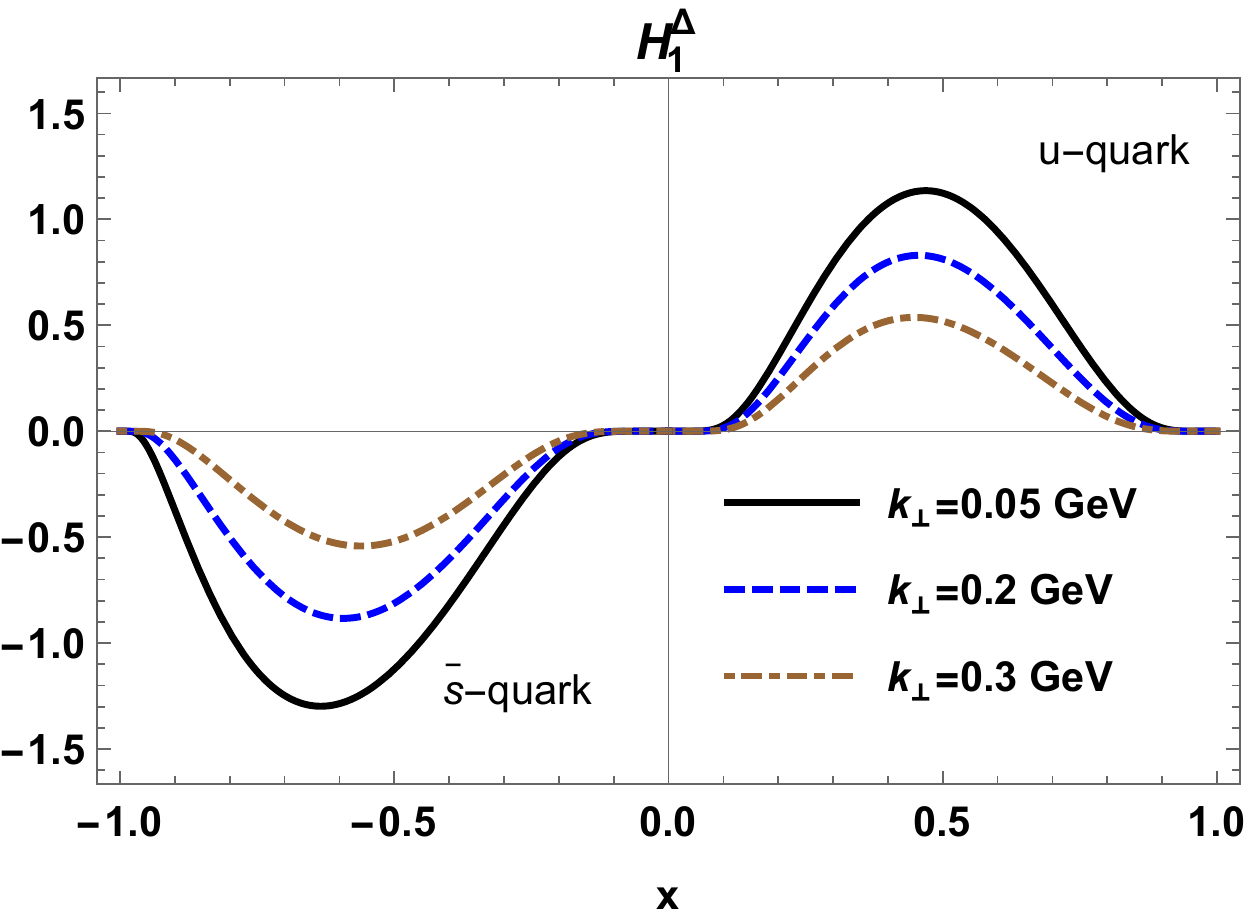}
(f)\includegraphics[width=.4\textwidth]{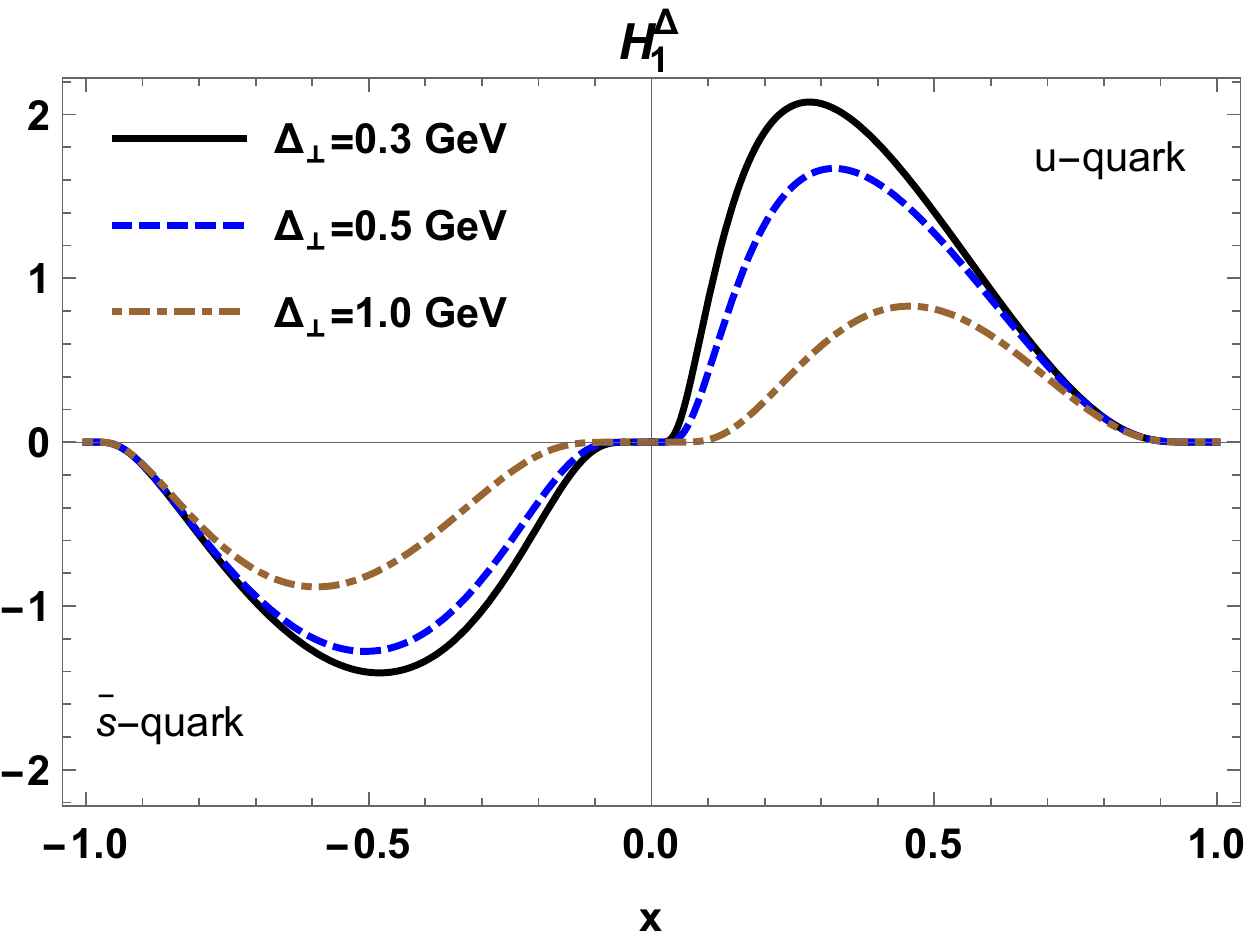}
\end{minipage}
\caption{The plots of GTMDs $F_1(x,0,{\bf k}^2_\perp,{\bf k}_\perp.{\bf \Delta}_\perp,{\bf \Delta}_\perp^2)$, $\tilde{G}_1(x,0,{\bf k}^2_\perp,{\bf k}_\perp.{\bf \Delta}_\perp,{\bf \Delta}_\perp^2)$ and $H_1^{\Delta}(x,0,{\bf k}^2_\perp,{\bf k}_\perp.{\bf \Delta}_\perp,{\bf \Delta}_\perp^2)$ w.r.t $x$ for $u$ and $\bar{s}$ quarks (i) at different values of ${\bf k}_\perp$ with fixed ${\bf \Delta}_\perp=1$ $GeV$ (left panel), and (ii) at different values of ${\bf \Delta}_\perp$ with fixed ${\bf k}_\perp=0.2$ $GeV$ (right panel).}
\label{gtmds_k_delta}
\end{figure*}

In Fig. \ref{gtmds_k_delta}, we show the graphical presentation of GTMDs $F_1, \tilde{G}_1,H_1^\Delta$ for $\zeta=0$. For $\zeta=0$, the complex-valued twist-2 GTMDs reduce to 3 in number as evident from Eqs. (\ref{h1k_u}) and (\ref{h1k_s}). The distribution peaks show the same behaviour for both $u$ and $\bar{s}$ quarks when ${\bf k}_\perp$ is varied  and ${\bf \Delta}_\perp$ is kept constant except for the magnitudes and polarities. This implies that there are no on-shell mass effects when the longitudinal momentum transfer is absent. On the other hand, when the momentum transfer to the kaon is varied by keeping  ${\bf k}_\perp$ constant, the effect on the distributions corresponding to $x$ remains same as for $\zeta \neq 0$. It is known that the GTMDs corresponding to $\zeta=0$ are the Fourier transformations of Wigner distributions. Therefore, the Wigner distribution $\rho_{UU}$ corresponds to the unpolarized GTMD $F_1$, from which one can evaluate the other unpolarized probabilistic distributions. As $\tilde{G}_1$ relates to $\rho_{UL}$, it would lead to the spin and orbital angular momentum correlation, the detailed description of which has been discussed in next Section. 

\section{VII. Spin-orbit correlation}
Following  the study of spin-orbit correlation of the  quark in a proton \cite{wdoam, wdmodel4, spinoam1, spinoam3} and pion \cite{wdmodel8}, we study the spin-orbit correlation of $u$ and $\bar s$ quark in the kaon. The correlation between quark spin and quark OAM is defined in terms of an operator as \cite{wdmodel4, spinoam1}
\begin{widetext}
\begin{eqnarray}
C^q_z(b^-, {\bf b}_\perp,k^+,{\bf k}_\perp)=\frac{1}{2}\int \frac{dz^- d^2{\bf z}_\perp}{(2\pi)^3}e^{ik.z}
\bar{\psi}^q\bigg(b^--\frac{z^-}{2},{\bf b}_\perp\bigg)\gamma^+ \gamma^5 ({\bf b}_\perp \times (-i \overleftrightarrow{\partial}_\perp))\psi^q \bigg(b^-+\frac{z^-}{2},{\bf b}_\perp\bigg).
\end{eqnarray}
\end{widetext}
The quark spin-orbit correlation in terms of Wigner distributions can be written as 
\begin{equation}
C^q_z=\int dx d^2{\bf k}_\perp d^2{\bf b}_\perp ({\bf b}_\perp \times {\bf k}_\perp)_z \rho^{[\gamma^+ \gamma^5]}({\bf b}_\perp,{\bf k}_\perp, x).
\end{equation} 
Following Eq. (\ref{def_ul}), we can write the above equation in terms of Wigner distribution of a longitudinally polarized quark in an unpolarized kaon as
\begin{eqnarray}
C_z^q=\int dx d^2{\bf k}_\perp d^2{\bf b}_\perp({\bf b}_\perp \times {\bf k}_\perp)_z \rho_{UL}({\bf b}_\perp,{\bf k}_\perp, x).
\label{spin-orbit-correlator}
\end{eqnarray}
The unpolarized-longitudinal Wigner distribution is parameterized in terms of GTMD $\tilde{G}_1$ as,
\begin{equation}
\rho_{UL}({\bf b}_\perp,{\bf k}_\perp,x)=\frac{1}{M^2}\epsilon_\perp^{ij} k^i_\perp \frac{\partial}{\partial b_\perp^j} {\mathcal{\tilde{G}}}_1(x,0,{\bf k}_\perp^2,{\bf k}_\perp . {\bf b}_\perp, {\bf b}^2_\perp), 
\end{equation}
where
\begin{eqnarray}
{\mathcal{\tilde{G}}}_1(x,0,{\bf k}_\perp^2,{\bf k}_\perp . {\bf b}_\perp, {\bf b}^2_\perp)&=&\int \frac{d^2 {\bf \Delta}_\perp}{(2\pi)^2} e^{-i {\bf \Delta}_\perp.{\bf b}_\perp}\nonumber\\
&& \tilde{G}_1(x,0,{\bf k}_\perp^2,{\bf k}_\perp . {\bf \Delta}_\perp, {\bf \Delta}^2_\perp). \nonumber\\
\end{eqnarray}
In terms of GTMDs, the definition of quark spin-orbit correlator is given as
\begin{eqnarray}
C_z^q=\int dx d^2{\bf k}_\perp \frac{{\bf k}^2_\perp}{M^2} \tilde{G}_1(x,0,{\bf k}_\perp,0,0).
\end{eqnarray} 
The case $C_z>0$ favors the alignment of quark spin and OAM. Otherwise, for $C_z<0$, it leads to the anti-alignment of quark spin and OAM. 

\begin{figure*}
\includegraphics[width=.4\textwidth]{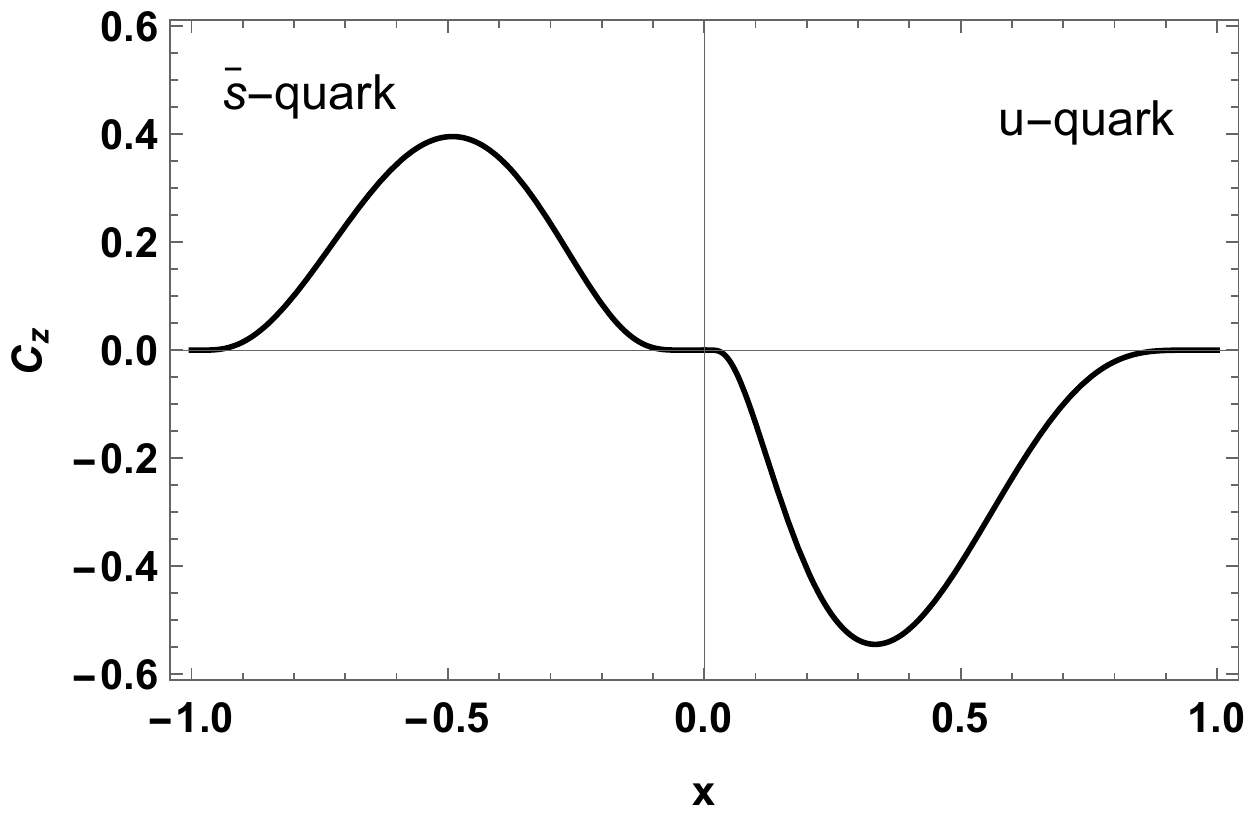}
\caption{The spin-orbit correlation $(C_z)$ of $u$ quark and $\bar{s}$ quark in kaon with respect to $x$.}
\label{k_oam}
\end{figure*}

In Fig. \ref{k_oam}, we have plotted the correlation $C_z$ w.r.t the longitudinal momentum fraction $x$. Here we study how the correlation between the spin and OAM varies with $x$. There seems to be a strong correlation between the $\bar{s}$ quark spin and OAM near the central values of the longitudinal momentum fraction carried by $\bar{s}$ quark i.e. $-x$. However, for the $u$ quark, the correlation peak is negative and is observed at lower values of quark longitudinal momentum fraction i.e. $x$. When $C_z$ is integrated upon $x$, the value comes out to be $C_z^{\bar{s}}=0.176$ and $C_z^u=-0.234$, which implies that the $\bar{s}$ quark OAM is parallel to the $\bar{s}$ quark spin and $u$ quark OAM is anti-parallel to the $u$ quark spin. If we compare these results with the previous studies for the case of proton, it is found that the spin-orbit correlations for $u$  and $d$ quarks are negative and anti-aligned in both the cases in the light-front quark-diquark model inspired from AdS/QCD \cite{wdmodel4, spinoam1}. In the light cone constituent quark model (LCCQM) and light-cone chiral quark-soliton model ($\chi$QSM), $C_z$ is positive for both quarks in the case of proton \cite{wdoam}. Recently, the spin-OAM correlation has been discussed for pion in light-cone quark model, (in Ref. \cite{wdmodel8}) where they have $C_z^q=-0.159$ with the quark spin being anti-parallel to the quark OAM. Our results are in line with these observations. 
\section{VIII. Summary and Conclusions}
In this paper, we have studied the various dimensional distributions of $u$ and $\bar{s}$ quarks in kaon using the light-cone quark model. We evaluate three-dimensional probabilistic generalized parton distributions for the valence $u$ and $\bar{s}$ quarks by considering the presence of longitudinal momentum transferred to the kaon's final state ($\zeta \neq 0$). Also, the quasi-probabilistic distributions: Wigner distributions and the mother distributions GTMDs are also discussed for case of $u$ and $\bar{s}$ quarks in kaon. The momentum-space wavefunction used in the present work is associated with the BHL prescription. The parameters used in the present work are able to give experimentally consistent results of electromagnetic form factors. As the support interval for the distributions is $-1<x<1$, we chose the DGLAP regions $-1<x<-\zeta$ and $\zeta<x<1$ respectively for the antiquark and quark. Here, $x$ defines the longitudinal momentum fraction carried by the parton. For the calculations of $u$ quark distribution, the $\bar{s}$ quark is considered as spectator, and vice versa. 

For the case of unpolarized GPD of quark and antiquark for non-zero skewedness, we study the distribution as a function of  variable $x$ by considering different values of $-t$ in one case and with different values of $\zeta$ in the other case. A shift in the distribution peak is observed along higher magnitudes of $x$ when there is an increase in the total momentum transfer to the kaon. The height of the peak however decreases. We have also presented the  variation of GPD as a function of $x$ but at a fixed $-t=0.5$ and  different values of $\zeta$. We observe a  high peak at lower value $\zeta$. Therefore, if the momentum transfer along longitudinal direction is less, the spread is found to be maximum. Further, we discussed the quark and antiquark distributions at $x=0.7$ and $x=-0.7$ as a function of $\zeta$ for different values of $-t$. It is observed that at $\zeta=0$ the distribution is different for different total momentum transferred to the final state of kaon. In all the cases we observe that the $\bar s$ quark amplitudes are comparatively  large and in opposite direction to the case of  $u$ quark. 

The relation of impact-parameter dependent GPDs are derived by taking the Fourier transformation of unpolarized GPD and they have been studied as a function of $x$ and transverse distance from the center of kaon, i.e. ${\bf b}_\perp$. We observe the absence of distribution of valence quarks when the transverse distance from the center of kaon is large. On the other hand, it is maximum when the transverse distance is small. Another important observation for the distribution of $u$ and $\bar s$ quarks is that the spread moves towards the lower $x$ for higher values of ${\bf b}_\perp$.  The distribution peak of $\bar{s}$ quark is localized at higher values of $x$ as compared to $u$ quark which is because of the heavier active quark mass in case of $bar{s}$ quark. Further, the 3D distribution  for transverse distance distribution as a function of $\zeta$ and ${\bf b}_\perp$ at constant value of longitudinal momentum fraction $x$ (for quark) and $-x$ (for antiquark) is maximum when no momentum is transferred to the final state of kaon for both quarks. For $\zeta=0$, the IPDGPD converts to give the parton distribution $q(x,{\bf b}_\perp)$. The spread is more when the antiquark longitudinal momentum is higher towards negative polarities and in middle of transverse impact-parameter. While for quark, the spread is near the centre w.r.t. quark longitudinal momentum fraction.

Further, the transverse Wigner distributions $\rho({\bf b}_\perp, {\bf k}_\perp)$ for unpolarized kaon, with the unpolarized, longitudinally-polarized an transversely-polarized composites, i.e. $u$ quark and $\bar s$ quark have been presented graphically in transverse impact-parameter plane, transverse momentum plane and mixed plane. For $\rho_{UU}$, the distribution comes out to be circularly symmetric about the center in ${\bf b}_\perp$ and ${\bf k}_\perp$ planes. However, in mixed plane, the axially-symmetric distribution is observed. The $\rho_{UL}$ displays a dipolar distribution in ${\bf b}$ plane as well as in ${\bf k}_\perp$ plane. The quadrupole distribution is observed in mixed plane which is related to the spin-orbital angular momentum correlation of partons in kaon. The polarization direction was taken along $x$-axis to evaluate $\rho^{j}_{UT}$ i.e. $j=1$. A dipolar distribution was observed in transverse impact-parameter plane and mixed plane whereas a circularly symmetric distribution in transverse momentum plane focused at the $k_x=k_y=0$. The spread is more concentrated in case of $\bar{s}$ which is because of  the heavy mass of strange quark in comparison of the up quark. The polarities seem to be opposite for both as the relation consumes the negative sign for strange quark distributions. The exception lies in case of $\rho_{UL}$. The strong correlation between the partons and kaon are well indicated by the phase-space distributions in the context of transverse momentum and transverse impact-parameter co-ordinates. The probabilistic distributions are possible to extract from the Wigner distributions upon certain limits.

Furthermore, GTMDs related to Fourier transformation of Wigner distributions for $\zeta=0$ and entitled as \textit{mother distributions}, help in extracting the corresponding GPDs and TMDs by applying some limits. We have investigated the relation of unpolarized kaon GTMDs with $x$ at different values of ${\bf k}_\perp$ and ${\bf \Delta}_\perp$ for $u$ and $\bar{s}$ quarks. There are 4 GTMDs for the case of kaon: $F_1$, $\tilde{G}_1$, $H_1^k$ and $H_1^{\Delta}$. In all the observations, we get to know that by increasing the quark (antiquark) transverse momentum ${\bf k}_\perp$, the distribution peaks move backward towards the lower values of $x$ with a decrease in their magnitudes. While observing the distributions corresponding to $x$ at different ${\bf \Delta}_\perp$, the trend becomes opposite. By increasing the momentum transfer ${\bf \Delta}_\perp$, the distribution peaks move forward towards the higher values of $x$ while the magnitudes decrease. The $u$ quark and $\bar{s}$ quark distributions have opposite polarities except for the case of $H_1^k$. The distribution $H_1^k$ become zero for $\zeta=0$. GTMD $F_1$ is related to Wigner distribution $\rho_{UU}$ through the Fourier transformation and further at certain limits, GPD $H$ and TMD $f_1$ come into picture. There is no GPD or TMD corresponding to $\rho_{UL}$ and $\tilde{G}_1$ but they are instead connected to the spin-orbit correlation $C_z$. The values come out to be $C_z=0.176$ for $\bar{s}$ quark and $C_z=-0.234$ for $u$ quark. From this it can be concluded that $\bar{s}$ quark's spin and OAM are anti-aligned whereas $u$ quark's spin and OAM are aligned to each other. Also no GPD and TMD related to $\rho^j_{UT}$ and $H_1^k$ and $H_1^{\Delta}$  are present in this model which is because of the absence of any gluon interaction.

To conclude, the results presented in the mss.  can perhaps be substantiated further by future measurements for the kaon where the quark  GTMDs can be extracted through the exclusive double Drell-Yan process where the final state must carry the two virtual photons. The GPDs can be accessible through DVCS  and DVMP processes.

\acknowledgements
S.K. would like to thank M. Diehl for helpful discussions. H.D. would like to thank the Department of Science and Technology (Ref No. EMR/2017/001549) Government of India for financial support.

\end{document}